\documentclass[twocolumn,journal]{IEEEtran}
\usepackage[T1]{fontenc}
\usepackage[latin9]{inputenc}
\usepackage{float}
\usepackage{textcomp}
\usepackage{mathrsfs}
\usepackage{amsmath}
\usepackage{amssymb}
\usepackage{stackrel}
\usepackage{graphicx}
\usepackage[unicode=true,
 bookmarks=true,bookmarksnumbered=true,bookmarksopen=true,bookmarksopenlevel=1,
 breaklinks=false,pdfborder={0 0 0},pdfborderstyle={},backref=false,colorlinks=false]
 {hyperref}
\hypersetup{pdftitle={Your Title},
 pdfauthor={Your Name},
 pdfpagelayout=OneColumn, pdfnewwindow=true, pdfstartview=XYZ, plainpages=false}

\makeatletter

\floatstyle{ruled}
\newfloat{algorithm}{tbp}{loa}
\providecommand{\algorithmname}{Algorithm}
\floatname{algorithm}{\protect\algorithmname}

\usepackage[caption=false,font=footnotesize]{subfig}

\makeatother

\begin{document}
\title{Joint Waveform and Beamforming Optimization for MIMO Wireless Power
Transfer}
\author{Shanpu~Shen,~\IEEEmembership{Member,~IEEE,} and~Bruno~Clerckx,~\IEEEmembership{Senior~Member,~IEEE}\thanks{Manuscript received; This work was supported in part by the EPSRC
of U.K. under Grant EP/P003885/1 and EP/R511547/1. \textit{(Corresponding
author: Shanpu Shen.)}}\thanks{The authors are with the Department of Electrical and Electronic Engineering,
Imperial College London, London SW7 2AZ, U.K. (e-mail: s.shen@imperial.ac.uk;
b.clerckx@imperial.ac.uk).}}
\maketitle
\begin{abstract}
In this paper, we study a multi-sine multiple-input multiple-output
(MIMO) wireless power transfer (WPT) system with the objective to
increase the output DC power. We jointly optimize the multi-sine waveform
and beamforming accounting for the rectenna nonlinearity, and consider
two combining schemes for the rectennas at the receiver, namely DC
and RF combinings. For DC combining, the waveform and transmit beamforming
are optimized, as a function of the channel state information (CSI).
For RF combining, the optimal transmit and receive beamformings are
provided in closed form and the waveform is optimized. We also consider
a practical RF combining circuit using phase shifter and RF power
combiner and optimize the waveform, transmit beamforming, and analog
receive beamforming adaptive to the CSI. Two types of performance
evaluations, based on the nonlinear rectenna model and accurate and
realistic circuit simulations, are provided. The evaluations demonstrate
that the joint waveform and beamforming design can increase the output
DC power by leveraging the beamforming gain, the frequency diversity
gain, and the rectenna nonlinearity. It also shows that the joint
waveform and beamforming design provides a higher output DC power
than the beamforming-only design with a relative gain of 180\% in
a two-transmit antenna sixteen-sinewave two-receive antenna setup.
\end{abstract}


\section{Introduction}

\IEEEPARstart{W}{ireless} power transfer (WPT) has a long history
and nowadays attracts more and more attentions as a promising energy
harvesting technology. Near-field WPT via inductive coupling has been
utilized for charging cell phones, medical implants, and electrical
vehicles, but the limitation is that it can only transfer power in
a very short distance. In contrast with the near-field WPT, far-field
WPT via radio frequency (RF) enables a long-distance power transfer
to energize numerous devices in the Internet of Things (IoT) \cite{2017_TOC_WPT_YZeng_Bruno_RZhang}.
The far-field WPT utilizes a dedicated source to radiate RF energy
through a wireless channel and a rectifying antenna (rectenna) at
the receiver to receive and convert this energy into DC power. In
contrast to batteries that need to be replaced and recharged periodically,
far-field WPT provides a more reliable, controllable, user-friendly,
and cost-effective way to power the devices in the IoT. Furthermore,
the far-field WPT can be extended to simultaneous wireless information
and power transfer (SWIPT) resulting in significant gains in terms
of spectral efficiency and energy efficiency by superposing information
and power transfer \cite{2014_ComMag_SWIPT}. The major challenge
of far-field WPT is to increase the output DC power of rectenna without
increasing the transmit power. To solve this challenge, a large amount
of the technical efforts in the literature have been devoted to designing
efficient rectennas \cite{2019TAP_WPT_WeiLin}, \cite{2019TMTT_HuchengSun_GridArray}.

Designing efficient WPT signals and waveforms can also increase the
output DC power \cite{2017_TOC_WPT_YZeng_Bruno_RZhang}. Multi-sine
signal excitation \cite{2009_IntConfRFID} has been shown through
RF measurements to increase the RF-to-DC conversion efficiency and
therefore increase the output DC power. However, the main limitation
of the method is not only the lack of a systematic approach to design
waveforms, but also the fact that they operate without Channel State
Information (CSI) at the Transmitter (CSIT) and Receiver (CSIR). The
first systematic analysis, design and optimization of waveforms for
WPT was conducted in \cite{2016_TSP_WPT_Bruno_Waveform}. Those waveforms
are adaptive to the CSI and jointly leverage the beamforming gain,
the frequency selectivity of the channel, and the rectenna nonlinearity
to maximize the output DC power. Since then, further enhancements
have been made to waveform optimization adaptive to CSI with the objective
to reduce the design complexity and extend to large scale multi-antenna
multi-sine WPT \cite{2017_TSP_WPT_Bruno_Yang_Large}-\nocite{2017_AWPL_WPT_Bruno_LowComplexity}\nocite{2017_IEEE_SPAWC_Waveform}\cite{2020_WCNC_WPT},
to account for limited feedback \cite{2018_TWC_WPT_Bruno_HYang_LimitedFeedback},
to energize multiple devices (multi-user setting) \cite{2017_TSP_WPT_Bruno_Yang_Large},
\cite{2019_JSAC_Waveform_MultipleDevice}, to transfer information
and power simultaneously \cite{2018_TSP_WIPT_Bruno_WIPT}-\nocite{2019_TOC_SWIPT_nonZeroInput}\cite{2019_TWC_WIPT_AsymmetricModulation},
and to enable efficient wireless powered communications \cite{2017_CL_WPBackscatterComm},
\cite{2019_TWC_MuWPScatteredComm}.

In addition to designing efficient rectenna and waveform, using multiple
rectennas, also known as multiport rectennas, at the receiver to form
a multiple-input multiple-output (MIMO) WPT system can effectively
increase the output DC power. This contrasts with all prior works
\cite{2016_TSP_WPT_Bruno_Waveform}-\nocite{2017_TSP_WPT_Bruno_Yang_Large}\nocite{2017_AWPL_WPT_Bruno_LowComplexity}\nocite{2017_IEEE_SPAWC_Waveform}\nocite{2018_TWC_WPT_Bruno_HYang_LimitedFeedback}\nocite{2019_JSAC_Waveform_MultipleDevice}\nocite{2018_TSP_WIPT_Bruno_WIPT}\nocite{2019_TOC_SWIPT_nonZeroInput}\nocite{2019_TWC_WIPT_AsymmetricModulation}\nocite{2017_CL_WPBackscatterComm}\cite{2019_TWC_MuWPScatteredComm}
that assumed a single rectenna per device. Interestingly, multiport
rectennas have been designed in \cite{ShanpuShen2016_TAP_Impedancematching}-\nocite{ShanpuShen2017_AWPL_DPTB}\nocite{ShanpuShen2018_EuCap_QPDP}\nocite{ShanpuShen2017_TAP_EHPIXEL}\nocite{ShanpuShen2019_TMTT_Freqdepend}\cite{ShanpuShen2019_TIE_Hybrid_Combining}
for ambient RF energy harvesting, which is similar to WPT but does
not have a controllable and dedicated transmitter. It was shown that
using multiport rectennas can linearly increase the output DC power
with the number of rectennas at the receiver. DC combining and RF
combining for the multiple rectennas at the receiver have been investigated
in \cite{2011AWPL_EH_InvestRectArray}, but the investigation is at
the level of RF circuit design and does not consider the impact on
communication and signal designs including adaptive waveform and beamforming
optimization.

Systematic studies of MIMO WPT systems were conducted in \cite{2016_TWC_RZhang_MIMOWPT_Limitedfeed}-\nocite{2019_SPL_MIMO_WPT}\cite{ShanpuShen2020_TWC_MIMO_WPT_SingleTone}.
In \cite{2016_TWC_RZhang_MIMOWPT_Limitedfeed}, a general design framework
for channel acquisition was proposed for MIMO WPT systems with limited
feedback, but the limitation is that it does not consider 1) the rectenna
nonlinearity, 2) RF combining, and 3) waveform design, which are all
beneficial to increase the output DC power. In \cite{2019_SPL_MIMO_WPT},
a generic receiver architecture for MIMO WPT systems was proposed.
The generic receiver architecture leverages the rectenna nonlinearity
by using a sigmoidal function-based rectenna model to maximize the
output DC power, but the limitation is that it only focuses beamforming
design and does not consider the waveform design. More recently, in
\cite{ShanpuShen2020_TWC_MIMO_WPT_SingleTone}, a systematic beamforming
design and optimization for MIMO WPT with DC and RF combinings was
proposed. As a result of the rectenna nonlinearity (modeled based
on a Taylor expansion of the diode I-V characteristics), \cite{ShanpuShen2020_TWC_MIMO_WPT_SingleTone}
has shown that RF combining provides significant performance benefits
over DC combining, though results in a more complex architecture where
transmit and receive beamformings have to be jointly optimized. Though
\cite{ShanpuShen2020_TWC_MIMO_WPT_SingleTone} sheds some new light
on beamforming design for MIMO WPT with various combining, its design
is limited by the use of continuous sinewave. In view of the recent
results in \cite{ShanpuShen2020_TWC_MIMO_WPT_SingleTone} and past
results on waveform design for WPT \cite{2016_TSP_WPT_Bruno_Waveform},
\cite{2017_TSP_WPT_Bruno_Yang_Large}, we go one step further and
ask ourselves how to design an even more efficient WPT architecture
by jointly optimizing the waveform and beamforming in a MIMO WPT.

In this paper, we consider the joint design of waveform and beamforming
for MIMO WPT systems, accounting for the rectenna nonlinearity to
increase the output DC power. This is the first paper to jointly optimize
the waveform and beamforming for MIMO WPT systems. The contributions
of the paper are summarized as follow.

\textit{First}, we analyze a multi-sine MIMO WPT architecture with
joint waveform and beamforming optimization, accounting for the rectenna
nonlinearity. Two combining schemes, DC and RF combinings, for the
multiple rectennas at the receiver are considered.

\textit{Second}, for DC combining, assuming perfect CSIT and leveraging
the nonlinear rectenna model, we jointly optimize the waveform and
transmit beamforming in the multi-sine MIMO WPT system with the objective
to maximize the total output DC power of all rectennas. The waveform
and beamforming are optimized with guarantee of converging to a stationary
point by using successive convex approximation (SCA) and semidefinite
relaxation (SDR).

\textit{Third}, for RF combining, assuming perfect CSIT and CSIR and
leveraging the nonlinear rectenna model, we optimize the waveform
and transmit and receive beamformings in the multi-sine MIMO WPT system
with the objective to maximize the output DC power. The optimal transmit
and receive beamformings are provided in closed form, while the waveform
optimization is formulated as a nonconvex posynomial maximization
problem and solved with guarantee of converging to a stationary point
by using SCA.

\textit{Fourth}, a practical RF combining circuit consisting of phase
shifters and an RF power combiner is considered for the multi-sine
MIMO WPT system. Assuming perfect CSIT and CSIR and leveraging the
nonlinear rectenna model, the waveform, transmit beamforming, and
analog receive beamforming are jointly optimized with the objective
to maximize the output DC power. The optimization is solved with guarantee
of converging to a stationary point by using SCA.

\textit{Fifth}, the joint waveform and beamforming optimization for
DC and RF combinings are shown to increase the output DC power through
an accurate and realistic circuit evaluation. Comparison with the
waveform and beamforming design optimized with the linear rectenna
model is also provided to show the crucial role played by the rectenna
nonlinearity in WPT. In addition, compared with the beamforming only
design proposed by \cite{ShanpuShen2020_TWC_MIMO_WPT_SingleTone},
the proposed joint waveform and beamforming design is shown to have
a relative gain in terms of the output DC power which can exceed 100\%
when there are sixteen sinewaves and can reach to 180\% in a two-transmit
antenna sixteen-sinewave two-receive antenna setup. Moreover, it is
shown that RF combining provides a higher output DC power than DC
combining since it can leverage the rectenna nonlinearity more efficiently.

In contrast with \cite{ShanpuShen2020_TWC_MIMO_WPT_SingleTone} which
assumes a continuous sinewave and only considers beamforming design,
this paper tackles the joint waveform and beamforming design. Such
a joint waveform and beamforming design leads to a significantly enhanced
performance. However, it is important to note that the joint waveform
and beamforming design also brings new challenges including: 1) The
MIMO WPT system model is more complex when the multi-sine waveform
is considered. 2) The optimization objective function, i.e. the output
DC power, has a much more complex expression. Due to the rectenna
nonlinearity, the received RF signals at different frequencies are
coupled with each other in the expression of the output DC power,
which makes the optimization a NP-hard problem. 3) The optimization
involves more variables across frequency (for waveform) and space
(for beamforming). These variables cannot be uncoupled in the optimization
and have to be jointly optimized, which increase the optimization
complexity. Furthermore, it is also important to highlight that the
algorithms for beamforming only design in \cite{ShanpuShen2020_TWC_MIMO_WPT_SingleTone}
cannot guarantee finding a stationary point due to using SDR (only
numerically guarantee for tested channels), however, the algorithms
proposed in this paper not only generalize the beamforming only design
but also has an advancement that it can theoretically guarantee finding
a stationary point.

\textit{Organization}: Section II introduces the multi-sine MIMO WPT
system model and Section III briefly revisits the nonlinear rectenna
model. Section IV and Section V tackle the joint waveform and beamforming
optimization for DC and RF combinings, respectively. Section VI evaluates
the performance and Section VII concludes the work.

\textit{Notations}: Bold lower and upper case letters stand for vectors
and matrices, respectively. A symbol not in bold font represents a
scalar. $\mathscr{\mathcal{E}}\left\{ \cdot\right\} $ refers to the
expectation/averaging operator. $\Re\left\{ x\right\} $ and $\left|x\right|$
refer to the real part and modulus of a complex number $x$, respectively.
$\left\Vert \mathbf{x}\right\Vert $ and $\left[\mathbf{x}\right]_{i}$
refer to the $l_{2}$-norm and $i$th element of a vector $\mathbf{x}$,
respectively. $\mathrm{arg}\left(\mathbf{x}\right)$ refers to a vector
with each element being the phase of the corresponding element in
a vector $\mathbf{x}$. $\mathbf{X}^{T}$, $\mathbf{X}^{H}$, $\mathrm{Tr}\left(\mathbf{X}\right)$,
and $\mathrm{rank}\left(\mathbf{X}\right)$ refer to the transpose,
conjugate transpose, trace, and rank of a matrix $\mathbf{X}$, respectively.
$\mathbf{X}\succeq0$ means that $\mathbf{X}$ is positive semidefinite.
$\mathbf{0}$ denotes an all-zero vector. log is in base $e$.

\section{Multi-sine MIMO WPT Model}

We consider a point-to-point multi-sine MIMO WPT system. There are
$M$ antennas at the transmitter and $Q$ antennas at the receiver.
A multi-sine waveform over $N$ angular frequencies $\omega_{1}$,
$\omega_{2}$, ..., $\omega_{N}$ is transmitted. The multi-sine waveform
transmitted by the $m$th transmit antenna is given by 
\begin{equation}
x_{m}\left(t\right)=\Re\left\{ \sum_{n=1}^{N}s_{m,n}e^{j\omega_{n}t}\right\} 
\end{equation}
where $s_{m,n}$ is a complex weight accounting for the magnitude
and phase of the $n$th sinewave on the $m$th transmit antenna. We
group $s_{m,n}$ into a $M$-dimensional vector $\mathbf{s}_{n}=\left[s_{1,n},s_{2,n},\ldots,s_{M,n}\right]^{T}$.
We further group $\mathbf{s}_{n}$ $\forall n$ into a $MN$-dimensional
vector $\mathbf{s}=\left[\mathbf{s}_{1}^{T},\mathbf{s}_{2}^{T},\ldots,\mathbf{s}_{N}^{T}\right]^{T}$.
The transmitter is subject to a transmit power constraint given by
\begin{equation}
\frac{1}{2}\left\Vert \mathbf{s}\right\Vert ^{2}\leq P\label{eq:power constraint}
\end{equation}
where $P$ denotes the transmit power. We group $x_{m}\left(t\right)$
into $\mathbf{x}\left(t\right)=\left[x_{1}\left(t\right),x_{2}\left(t\right),...,x_{M}\left(t\right)\right]^{T}$
and it can be rewritten as
\begin{equation}
\mathbf{x}\left(t\right)=\Re\left\{ \sum_{n=1}^{N}\mathbf{s}_{n}e^{j\omega_{n}t}\right\} .
\end{equation}

The multi-sine waveform transmitted by the multiple transmit antennas
propagate through a wireless channel. The received signal at the $q$th
receive antenna can be expressed as
\begin{align}
y_{q}\left(t\right) & =\Re\left\{ \sum_{n=1}^{N}\mathbf{h}_{q,n}\mathbf{s}_{n}e^{j\omega_{n}t}\right\} 
\end{align}
where $\mathbf{h}_{q,n}=\left[h_{q,1,n},h_{q,2,n},...,h_{q,M,n}\right]$
with $h_{q,m,n}$ referring to the complex channel gain between the
$m$th transmit antenna and the $q$th receive antenna at the $n$th
angular frequency. We collect all $\mathbf{h}_{q,n}$ into a matrix
$\mathbf{H}_{n}=\left[\mathbf{h}_{1,n}^{T},\mathbf{h}_{2,n}^{T},...,\mathbf{h}_{Q,n}^{T}\right]^{T}$
where $\mathbf{H}_{n}$ represents the $Q\times M$ channel matrix
at the $n$th angular frequency of the multi-sine MIMO WPT system.
We assume that the channel matrix $\mathbf{H}_{n}$ is perfectly known
to the transmitter and the receiver.

In this paper, we optimize $\mathbf{s}$ with the transmit power constraint
\eqref{eq:power constraint} to maximize the output DC power of multi-sine
MIMO WPT systems. At the $n$th sinewave, $\mathbf{s}_{n}$ characterizes
the beamforming for the multiple transmit antennas, so that optimizing
$\mathbf{s}$ includes the beamforming optimization. On the other
hand, at the $m$th transmit antenna, $s_{m,1}$, $s_{m,2}$, ...,
$s_{m,N}$ characterize the multi-sine waveform $x_{m}\left(t\right)$,
so that optimizing $\mathbf{s}$ includes the waveform optimization.
It should be noted that the waveform optimized in this paper is multi-sine
waveform at fixed $N$ angular frequencies, not waveform with arbitrary
spectrum. Therefore, optimizing $\mathbf{s}$ means the joint waveform
and beamforming optimization.

\section{Rectenna Model}

We briefly revisit a rectenna model derived in the past literature
\cite{2017_TSP_WPT_Bruno_Yang_Large}. The model accounts for the
rectenna nonlinearity through the higher order terms in the Taylor
expansion of the diode I-V characteristics while having a simple and
tractable expression\footnote{There is another nonlinear rectenna model proposed in \cite{2015_CL_Nonlinaer_SWIPT},
however, the model in \cite{2015_CL_Nonlinaer_SWIPT} cannot be used
for waveform optimization. A detailed comparison between the model
in \cite{2017_TSP_WPT_Bruno_Yang_Large} and the model in \cite{2015_CL_Nonlinaer_SWIPT}
is provided in \cite{2018_TSP_WIPT_Bruno_WIPT}, \cite{2019_JSAC_WIPT_Bruno_RZhang_RSchober_DIKim_HVPoor}.}.

\begin{figure}[t]
\begin{centering}
\includegraphics[scale=0.16]{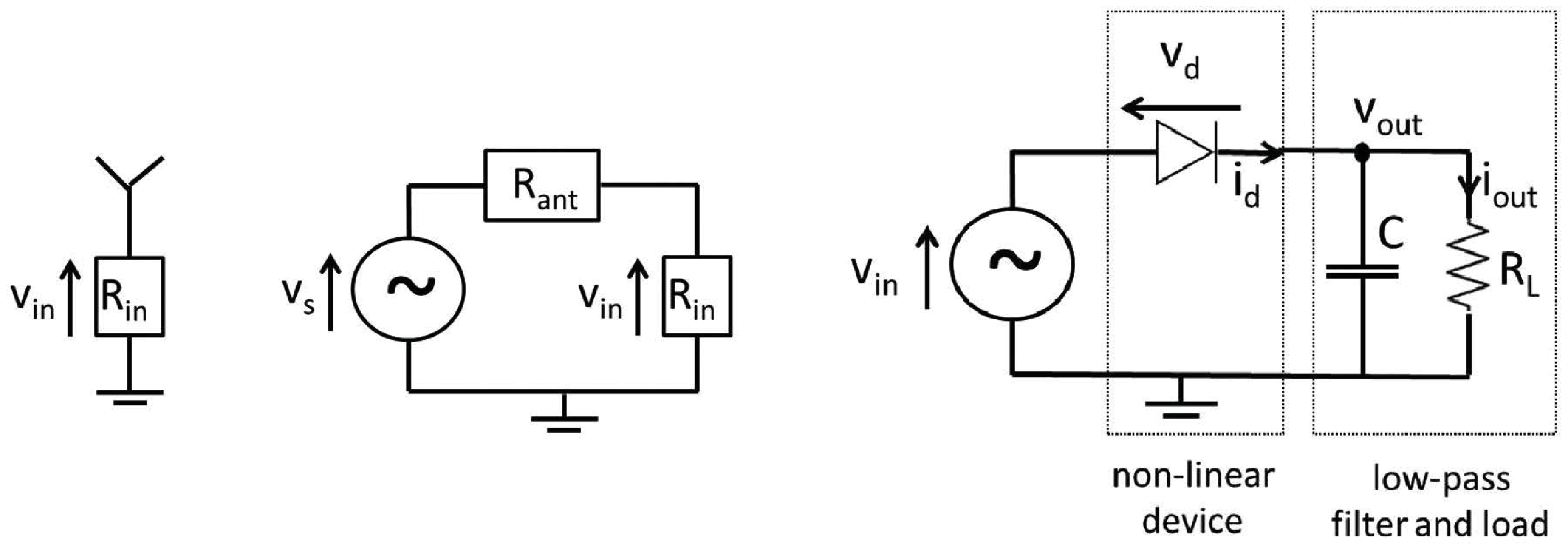}
\par\end{centering}
\caption{\label{fig:Antenna-equivalent-circuit}Antenna equivalent circuit
(left) and a single diode rectifier (right).}
\end{figure}

Consider a rectifier with input impedance $R_{\mathrm{in}}$ connected
to a receive antenna as shown in Fig. \ref{fig:Antenna-equivalent-circuit}.
The signal $y\left(t\right)$ impinging on the antenna has an average
power $P_{\mathrm{av}}=\mathscr{\mathcal{E}}\left\{ y\left(t\right)^{2}\right\} $.
The receive antenna is assumed lossless and modeled as an equivalent
voltage source $v_{s}\left(t\right)$ in series with an impedance
$R_{\mathrm{ant}}=50\,\Omega$ as shown in Fig. \ref{fig:Antenna-equivalent-circuit}.
With perfect matching ($R_{\mathrm{in}}=R_{\mathrm{ant}}$), the input
voltage of the rectifier $v_{\mathrm{in}}\left(t\right)$ can be related
to the received signal $y\left(t\right)$ by $v_{\mathrm{in}}\left(t\right)=y\left(t\right)\sqrt{R_{\mathrm{ant}}}$.
A rectifier is always made of a nonlinear rectifying component such
as a diode followed by a low pass filter with a load \cite{ShanpuShen2017_AWPL_DPTB},
\cite{ShanpuShen2017_TAP_EHPIXEL}, \cite{ShanpuShen2019_TMTT_Freqdepend},
as shown in Fig. \ref{fig:Antenna-equivalent-circuit}. The current
$i_{d}\left(t\right)$ flowing through an ideal diode (neglecting
its series resistance) relates to the voltage drop across the diode
$v_{d}\left(t\right)$ = $v_{\mathrm{in}}\left(t\right)$ \textminus{}
$v_{\mathrm{out}}\left(t\right)$ as $i_{d}\left(t\right)=i_{s}\left(e^{\frac{v_{d}\left(t\right)}{n_{i}v_{t}}}-1\right)$
where $i_{s}$ is the reverse bias saturation current, $v_{t}$ is
the thermal voltage, $n_{i}$ is the ideality factor (assumed equal
to 1.05).

In \cite{2017_TSP_WPT_Bruno_Yang_Large}, by assuming zero output
DC current and taking Taylor expansion at zero quiescent point to
the $n_{0}$th-order term, the output DC voltage of the rectifier
$v_{\mathrm{out}}$ is approximated as
\begin{align}
v_{\mathrm{out}} & =\sum_{i\,\mathrm{even},\,i\geq2}^{n_{0}}\beta_{i}\mathscr{\mathcal{E}}\left\{ y\left(t\right)^{i}\right\} \label{eq:VoutRectennaModel}
\end{align}
where $\beta_{i}=\frac{R_{\mathrm{ant}}^{i/2}}{i!\left(n_{i}v_{t}\right)^{i-1}}$.
There is no odd (first, third, and fifth, etc) order terms in the
output DC voltage \eqref{eq:VoutRectennaModel} because $\mathscr{\mathcal{E}}\left\{ y\left(t\right)^{i}\right\} =0$
for $i$ odd, i.e. odd order terms have zero mean, which has also
been shown in \cite{2016_TSP_WPT_Bruno_Waveform}. In the following
Sections, we mainly consider the truncation order $n_{0}=4$ since
$n_{0}=4$ is a good choice \cite{2017_TSP_WPT_Bruno_Yang_Large}.

\section{Joint Waveform and Beamforming Optimization with DC Combining}

\begin{figure}[t]
\begin{centering}
\includegraphics[width=8.5cm]{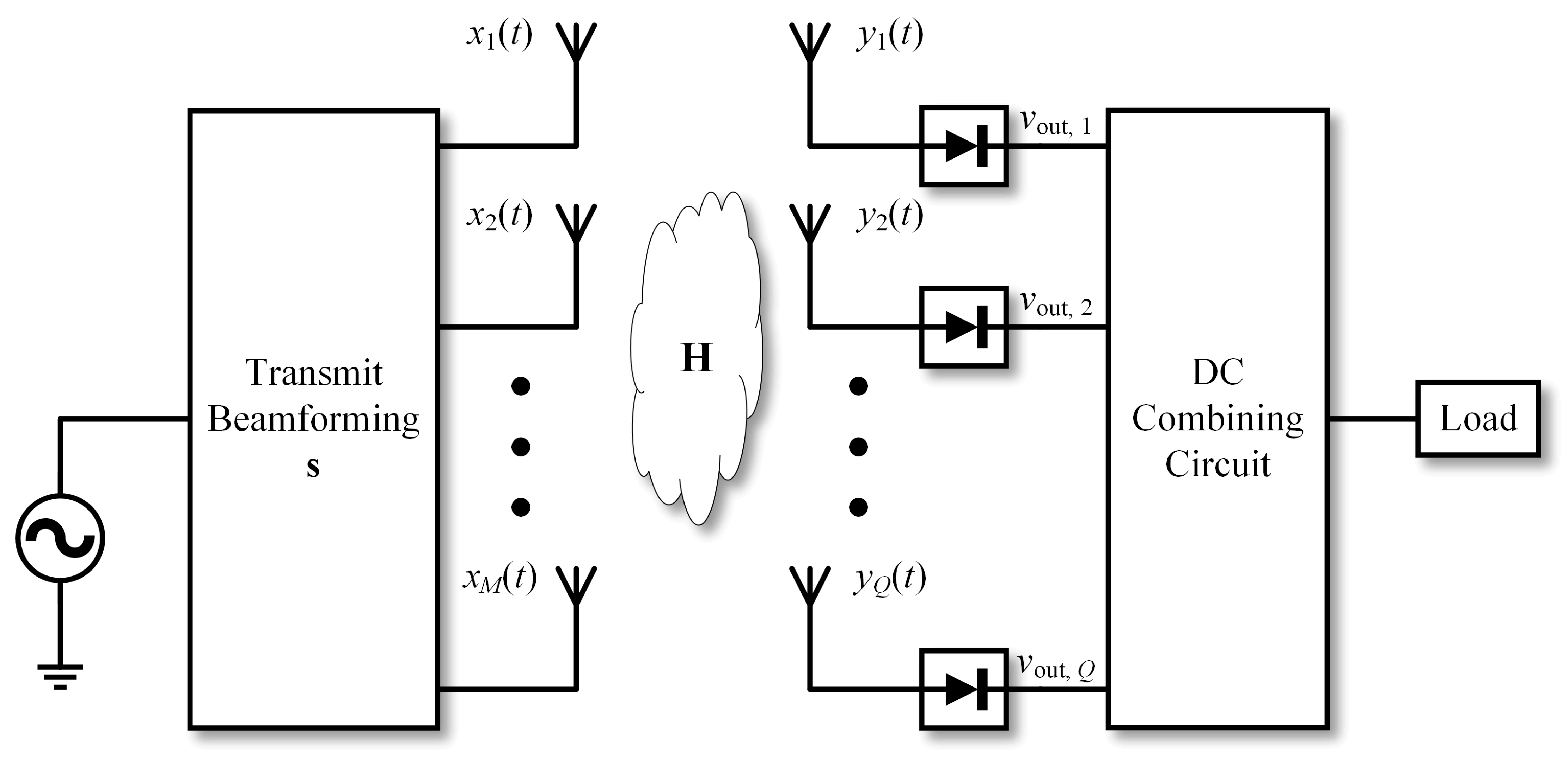}
\par\end{centering}
\caption{\label{fig:DC combinging}Schematic of the multi-sine MIMO WPT system
with DC combining in the receiver.}
\end{figure}

\begin{figure}[t]
\begin{centering}
\includegraphics[width=8.5cm]{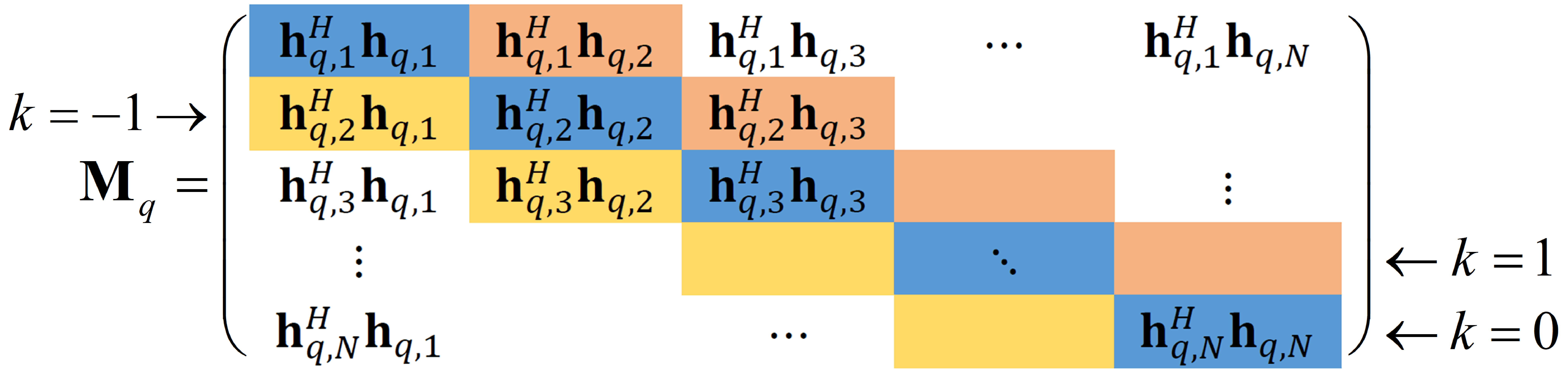}
\par\end{centering}
\caption{\label{fig:Mq}$\mathbf{M}_{q,1}$ is the above matrix only maintaining
the block diagonal (whose index is $k=1$) in pink, while all the
other blocks are set as zero matrices.}
\end{figure}

Consider the DC combining scheme for the multiple receive antenna
system as shown in Fig. \ref{fig:DC combinging}. Each receive antenna
is connected to a rectifier so that the RF signal received by each
antenna is individually rectified. Using the nonlinear rectenna model
\eqref{eq:VoutRectennaModel} with the truncation order $n_{0}=4$,
the output DC voltage of the $q$th rectifier (connected to the $q$th
receive antenna) is given by
\begin{align}
v_{\mathrm{out},q} & =\beta_{2}\mathscr{\mathcal{E}}\left\{ y_{q}\left(t\right)^{2}\right\} +\beta_{4}\mathscr{\mathcal{E}}\left\{ y_{q}\left(t\right)^{4}\right\} ,
\end{align}
where $\mathscr{\mathcal{E}}\left\{ y_{q}\left(t\right)^{2}\right\} $
and $\mathscr{\mathcal{E}}\left\{ y_{q}\left(t\right)^{4}\right\} $
are given by
\begin{align}
\mathscr{\mathcal{E}}\left\{ y_{q}\left(t\right)^{2}\right\}  & =\frac{1}{2}\sum_{n=1}^{N}\mathbf{s}_{n}^{H}\mathbf{h}_{q,n}^{H}\mathbf{h}_{q,n}\mathbf{s}_{n},\\
\mathscr{\mathcal{E}}\left\{ y_{q}\left(t\right)^{4}\right\}  & =\frac{3}{8}\sum_{\substack{n_{1},n_{2},n_{3},n_{4}\\
n_{1}+n_{2}=n_{3}+n_{4}
}
}\left(\mathbf{s}_{n_{3}}^{H}\mathbf{h}_{q,n_{3}}^{H}\mathbf{h}_{q,n_{1}}\mathbf{s}_{n_{1}}\right.\nonumber \\
 & \left.\qquad\qquad\qquad\quad\;\;\,\cdot\mathbf{s}_{n_{4}}^{H}\mathbf{h}_{q,n_{4}}^{H}\mathbf{h}_{q,n_{2}}\mathbf{s}_{n_{2}}\right).
\end{align}
We can rewrite $v_{\mathrm{out},q}$ in a more compact form by introducing
$MN$-by-$MN$ matrices $\mathbf{M}_{q}$ and $\mathbf{M}_{q,k}$.
$\mathbf{M}_{q}$ is defined by $\mathbf{M}_{q}\triangleq\mathbf{h}_{q}^{H}\mathbf{h}_{q}$
with $\mathbf{h}_{q}=\left[\mathbf{h}_{q,1},\mathbf{h}_{q,2},\ldots,\mathbf{h}_{q,N}\right]$.
As shown in Fig. \ref{fig:Mq}, $k\in\left\{ 1,2,\ldots,N-1\right\} $
is the index of the $k$th block diagonal above the main block diagonal
(whose index $k=0$) of $\mathbf{M}_{q}$, while $k\in\left\{ -\left(N-1\right),\ldots,-2,-1\right\} $
is the index of the $\left|k\right|$th block diagonal below the main
block diagonal. Given a certain $k$, $\mathbf{M}_{q,k}$ is generated
by retaining the $k$th block diagonal of $\mathbf{M}_{q}$ but setting
all the other blocks as zero matrices. For $k\neq0$, the non-Hermitian
matrix $\mathbf{M}_{q,-k}=\mathbf{M}_{q,k}^{H}$, while $\mathbf{M}_{q,0}\succeq0$.
Thus, $v_{\mathrm{out},q}$ can be rewritten as
\begin{align}
v_{\mathrm{out},q} & =\frac{1}{2}\beta_{2}\mathbf{s}^{H}\mathbf{M}_{q,0}\mathbf{s}+\frac{3}{8}\beta_{4}\mathbf{s}^{H}\mathbf{M}_{q,0}\mathbf{s}\left(\mathbf{s}^{H}\mathbf{M}_{q,0}\mathbf{s}\right)^{H}\nonumber \\
 & \quad+\frac{3}{4}\beta_{4}\sum_{k=1}^{N-1}\mathbf{s}^{H}\mathbf{M}_{q,k}\mathbf{s}\left(\mathbf{s}^{H}\mathbf{M}_{q,k}\mathbf{s}\right)^{H}.
\end{align}

The output DC power of all rectifiers are combined together by a DC
combining circuit such as MIMO switching DC-DC converter \cite{KKKKI}
as shown in Fig. \ref{fig:DC combinging}. The total output DC power
is then given by $P_{\mathrm{out}}=\sum_{q=1}^{Q}v_{\mathrm{out},q}^{2}/R_{L}$
where we assume each rectifier has the same load $R_{L}$. Therefore,
we aim to maximize the total output DC power subject to the transmit
power constraint, which can be formulated as
\begin{align}
\underset{\mathbf{s}}{\mathrm{max}}\;\;\;\left\{ \sum_{q=1}^{Q}\frac{v_{\mathrm{out},q}^{2}}{R_{L}}\colon\frac{1}{2}\left\Vert \mathbf{s}\right\Vert ^{2}\leq P\right\} .\label{eq:OP1-MaxPoutStP-1}
\end{align}
The objective function \eqref{eq:OP1-MaxPoutStP-1} is an octic polynomial,
which in general makes problems \eqref{eq:OP1-MaxPoutStP-1} NP-hard.
To tackle the octic polynomial, auxiliary variables $t_{q,k}=\mathbf{s}^{H}\mathbf{M}_{q,k}\mathbf{s}$,
for $q=1,\ldots,Q$ and $k=0,\ldots,N-1$, are introduced so that
the octic objective function \eqref{eq:OP1-MaxPoutStP-1} can be reduced
to quartic polynomial and $v_{\mathrm{out},q}$ can be rewritten as
\begin{align}
v_{\mathrm{out},q} & =\frac{1}{2}\beta_{2}t_{q,0}+\frac{3}{8}\beta_{4}t_{q,0}t_{q,0}^{*}+\frac{3}{4}\beta_{4}\sum_{k=1}^{N-1}t_{q,k}t_{q,k}^{*},
\end{align}
which can be further expressed in a more compact form as $v_{\mathrm{out},q}=\frac{1}{2}\beta_{2}t_{q,0}+\mathbf{t}_{q}^{H}\mathbf{A}_{0}\mathbf{t}_{q}$
with $\mathbf{t}_{q}=\left[t_{q,0},t_{q,1},\ldots,t_{q,N-1}\right]^{T}$
and $\mathbf{A}_{0}=\mathrm{diag}\left\{ \frac{3}{8}\beta_{4},\frac{3}{4}\beta_{4},\ldots,\frac{3}{4}\beta_{4}\right\} \succeq0$.
However, for $k\neq0$, $\mathbf{M}_{q,k}$ is not Hermitian so that
the term $\mathbf{s}^{H}\mathbf{M}_{q,k}\mathbf{s}$ is essentially
a bilinear function, which may also lead to a NP-hard problem. To
address this, we introduce an auxiliary rank-1 positive semidefinite
matrix variable $\mathbf{X}=\mathbf{s}\mathbf{s}^{H}$ to linearize
the term such that $t_{q,k}=\mathrm{Tr}\left(\mathbf{M}_{q,k}\mathbf{X}\right)$
for $q=1,\ldots,Q$ and $k=0,\ldots,N-1$. Therefore, we can equivalently
rewrite the problem \eqref{eq:OP1-MaxPoutStP-1} as
\begin{align}
\underset{t_{q,k},\mathbf{\mathbf{X}}\succeq0}{\mathrm{max}}\;\;\; & \sum_{q=1}^{Q}v_{\mathrm{out},q}^{2}\label{eq:OP2-object}\\
\mathsf{\mathrm{s.t.}}\;\;\,\;\;\; & t_{q,k}=\mathrm{Tr}\left(\mathbf{M}_{q,k}\mathbf{X}\right),\:\forall q,k,\label{eq:OPT2-constraint-1}\\
 & \mathrm{Tr}\left(\mathbf{X}\right)\leq2P,\label{eq:OPT2-constraint-2}\\
 & \mathrm{rank}\left(\mathbf{X}\right)=1.\label{eq:OPT2-constraint-4}
\end{align}

The rank constraint \eqref{eq:OPT2-constraint-4}, however, makes
the problem \eqref{eq:OP2-object}-\eqref{eq:OPT2-constraint-4} NP-hard
in general. To handle this, we use SDR to relax the rank constraint
\eqref{eq:OPT2-constraint-4} and then solve the relaxed problem \eqref{eq:OP2-object}-\eqref{eq:OPT2-constraint-2}.
$v_{\mathrm{out},q}$ is convex with respect to $t_{q,k}$ and $v_{\mathrm{out},q}\geq0$,
so the objective function $\sum_{q=1}^{Q}v_{\mathrm{out},q}^{2}$
is convex with respect to $t_{q,k}$ according to the compositions
rules \cite{boyd2004convex}. Therefore, the relaxed problem \eqref{eq:OP2-object}-\eqref{eq:OPT2-constraint-2}
is essentially maximizing a convex function subject to convex constraints,
but unfortunately it is still not a convex problem.

To solve the nonconvex relaxed problem \eqref{eq:OP2-object}-\eqref{eq:OPT2-constraint-2},
we use SCA to approximate the convex objective function as a linear
function and iteratively solve the approximated problem. Particularly,
at iteration $i$, the convex objective function $\sum_{q=1}^{Q}v_{\mathrm{out},q}^{2}$
is approximated at $\mathbf{t}_{q}^{\left(i-1\right)}$, which is
the optimal $\mathbf{t}_{q}$ solved at iteration $\left(i-1\right)$,
as a linear function by its first-order Taylor expansion \cite{adali2010adaptiveSignalProcessing},
so that we have
\begin{align}
\sum_{q=1}^{Q}v_{\mathrm{out},q}^{2} & \geq\sum_{q=1}^{Q}v_{\mathrm{out},q}^{\left(i-1\right)}\left(\beta_{2}t_{q,0}+4\Re\left\{ \mathbf{t}_{q}^{\left(i-1\right)H}\mathbf{A}_{0}\mathbf{t}_{q}\right\} \right)\nonumber \\
 & -\sum_{q=1}^{Q}v_{\mathrm{out},q}^{\left(i-1\right)}\left(v_{\mathrm{out},q}^{\left(i-1\right)}+2\mathbf{t}_{q}^{\left(i-1\right)H}\mathbf{A}_{0}\mathbf{t}_{q}^{\left(i-1\right)}\right),\label{eq:approximateSumVoutSquare}
\end{align}
where the right hand side is the linear approximated objective function
with $v_{\mathrm{out},q}^{\left(i-1\right)}=\frac{1}{2}\beta_{2}t_{q,0}^{\left(i-1\right)}+\mathbf{t}_{q}^{\left(i-1\right)H}\mathbf{A}_{0}\mathbf{t}_{q}^{\left(i-1\right)}$.
Then, ignoring the constant in the linear approximated objective function,
we can equivalently formulate the approximate problem (AP) at iteration
$i$ as
\begin{align}
\underset{t_{q,k},\mathbf{\mathbf{X}}\succeq0}{\mathrm{max}}\;\;\; & \sum_{q=1}^{Q}v_{\mathrm{out},q}^{\left(i-1\right)}\left(\beta_{2}t_{q,0}+4\Re\left\{ \mathbf{t}_{q}^{\left(i-1\right)H}\mathbf{A}_{0}\mathbf{t}_{q}\right\} \right)\label{eq:OPT3-object}\\
\mathsf{\mathrm{s.t.}}\;\;\;\,\;\; & t_{q,k}=\mathrm{Tr}\left(\mathbf{M}_{q,k}\mathbf{X}\right),\:\forall q,k,\label{eq:OPT3-constraint-1}\\
 & \mathrm{Tr}\left(\mathbf{X}\right)\leq2P,\label{eq:OPT3-constraint-2}
\end{align}
which is a Semidefinite Programming (SDP). Substituting \eqref{eq:OPT3-constraint-1}
into \eqref{eq:OPT3-object}, we can rewrite the problem \eqref{eq:OPT3-object}-\eqref{eq:OPT3-constraint-2}
in an equivalent compact form as
\begin{align}
\underset{\mathbf{\mathbf{X}\succeq0}}{\mathrm{max}}\;\;\; & \left\{ \mathrm{Tr}\left(\mathbf{A}_{1}\mathbf{X}\right)\colon\mathrm{Tr}\left(\mathbf{X}\right)\leq2P\right\} ,\label{eq:OPT4-object}
\end{align}
where $\mathbf{A}_{1}=\mathbf{C}_{1}+\mathbf{C}_{1}^{H}$ is Hermitian
and $\mathbf{C}_{1}$ is given by 
\begin{align}
\mathbf{C}_{1}=\sum_{q=1}^{Q}v_{\mathrm{out},q}^{\left(i-1\right)} & \left(\frac{2\beta_{2}+3\beta_{4}t_{q,0}^{\left(i-1\right)}}{4}\mathbf{M}_{q,0}\right.\nonumber \\
 & \left.\:\:\:+\frac{3}{2}\beta_{4}\sum_{k=1}^{N-1}t_{q,k}^{\left(i-1\right)*}\mathbf{M}_{q,k}\right).\label{eq:Matrix C1}
\end{align}
According to \cite{2017_TSP_WPT_Bruno_Yang_Large}, \cite{DanielSDP_rank},
the problem \eqref{eq:OPT4-object} has a rank-1 global optimal solution
$\mathbf{X}^{\star}$ given by
\begin{align}
\mathbf{X}^{\star} & =\mathbf{s}^{\star}\mathbf{s}^{\star H},\label{eq:EVD-1}\\
\mathbf{s}^{\star} & =\sqrt{2P}\left[\mathbf{U}_{\mathbf{A}_{1}}\right]_{\mathrm{max}},\label{eq:EVD-2}
\end{align}
where $\left[\mathbf{U}_{\mathbf{A}_{1}}\right]_{\mathrm{max}}$ is
the eigenvector of $\mathbf{A}_{1}$ corresponding to the maximum
eigenvalue. Therefore, at each iteration, we perform Eigenvalue Decomposition
(EVD) for $\mathbf{A}_{1}$ by the QR algorithm \cite{QR_algorithm}
with a computational complexity of $\mathcal{O}\left(M^{3}N^{3}\right)$
to find a rank-1 global optimal solution of the AP \eqref{eq:OPT3-object}-\eqref{eq:OPT3-constraint-2},
and we repeat the iterations till convergence. The SCA guarantees
to converge to a stationary point of the relaxed problem \eqref{eq:OP2-object}-\eqref{eq:OPT2-constraint-2}.
In addition, because such stationary point is guaranteed to be rank-1
(achieved by \eqref{eq:EVD-1} and \eqref{eq:EVD-2}), it is also
a stationary point of the original problem \eqref{eq:OP2-object}-\eqref{eq:OPT2-constraint-4}.
The initialization of the SCA is important and affects the convergence
speed. Specifically, using singular value decomposition (SVD), we
decompose the channel matrix as $\mathbf{H}_{n}=\mathbf{U}_{n}\mathbf{\Sigma}_{n}\mathbf{V}_{n}^{H}$
where $\mathbf{U}_{n}$ is a $Q\times Q$ unitary matrix, $\mathbf{V}_{n}$
is a $M\times M$ unitary matrix, and $\mathbf{\Sigma}_{n}$ is a
$Q\times M$ diagonal matrix. We choose a good initial point as $\mathbf{s}_{n}^{\left(0\right)}=\sigma_{n}\sqrt{2P/\sum_{n=1}^{N}\sigma_{n}^{2}}\left[\mathbf{V}_{n}\right]_{\mathrm{max}}$
where $\sigma_{n}$ is the maximum singular value of $\mathbf{H}_{n}$
and $\left[\mathbf{V}_{n}\right]_{\mathrm{max}}$ is the vector in
$\mathbf{V}_{n}$ corresponding to $\sigma_{n}$. Accordingly, we
have that $\mathbf{s}^{\left(0\right)}=\left[\mathbf{s}_{1}^{\left(0\right)T},\mathbf{s}_{2}^{\left(0\right)T},\ldots,\mathbf{s}_{N}^{\left(0\right)T}\right]^{T}$,
$\mathbf{X}^{\left(0\right)}=\mathbf{s}^{\left(0\right)}\mathbf{s}^{\left(0\right)H}$,
$t_{q,k}^{\left(0\right)}=\mathrm{Tr}\left(\mathbf{M}_{q,k}\mathbf{X}^{\left(0\right)}\right)$,
and $v_{\mathrm{out},q}^{\left(0\right)}=\frac{1}{2}\beta_{2}t_{q,0}^{\left(0\right)}+\mathbf{t}_{q}^{\left(0\right)H}\mathbf{A}_{0}\mathbf{t}_{q}^{\left(0\right)}$.

Algorithm \ref{alg:DC-Combining-Optimization.} summarizes the overall
algorithm for jointly optimizing the waveform and beamforming with
DC combining\footnote{Algorithm \ref{alg:DC-Combining-Optimization.} proposed in this paper
is different from the algorithm for beamforming only design with DC
combining in \cite{ShanpuShen2020_TWC_MIMO_WPT_SingleTone} in three
aspects: 1) The algorithm in \cite{ShanpuShen2020_TWC_MIMO_WPT_SingleTone}
cannot theoretically guarantee finding a stationary point, but Algorithm
\ref{alg:DC-Combining-Optimization.} can; 2) The algorithm in \cite{ShanpuShen2020_TWC_MIMO_WPT_SingleTone}
is only for a continuous sinewave, but Algorithm \ref{alg:DC-Combining-Optimization.}
is for multi-sine wave which is more general and challenging to optimize;
3) In each iteration, the algorithm in \cite{ShanpuShen2020_TWC_MIMO_WPT_SingleTone}
needs to solve a convex problem but Algorithm \ref{alg:DC-Combining-Optimization.}
only needs to perform EVD which has lower computation complexity.
In addition, Algorithm \ref{alg:DC-Combining-Optimization.} proposed
in this paper is different from Algorithm 2 in \cite{2017_TSP_WPT_Bruno_Yang_Large}
in three aspects: 1) The objective function optimized by Algorithm
\ref{alg:DC-Combining-Optimization.} is the total output DC power
$\sum_{q=1}^{Q}v_{\mathrm{out},q}^{2}/R_{L}$ in single user MIMO
WPT systems, which is an octic polynomial of $\mathbf{s}$. However,
the objective function optimized by Algorithm 2 in \cite{2017_TSP_WPT_Bruno_Yang_Large}
is the weighted sum output DC voltage $\sum_{q=1}^{Q}w_{q}v_{\mathrm{out},q}$
in multi-user MISO WPT systems, which is a quartic polynomial of $\mathbf{s}$;
2) Algorithm \ref{alg:DC-Combining-Optimization.} uses SCA to approximate
$\sum_{q=1}^{Q}v_{\mathrm{out},q}^{2}/R_{L}$ as a linear function
of $\mathbf{t}_{q}$ while Algorithm 2 in \cite{2017_TSP_WPT_Bruno_Yang_Large}
uses SCA to approximate $\sum_{q=1}^{Q}w_{q}v_{\mathrm{out},q}$ as
a linear function; 3) Algorithm \ref{alg:DC-Combining-Optimization.}
considers $v_{\mathrm{out},q}^{\left(i-1\right)}$ at iteration $i$
but Algorithm 2 in \cite{2017_TSP_WPT_Bruno_Yang_Large} does not
consider $v_{\mathrm{out},q}^{\left(i-1\right)}$. The matrix $\mathbf{A}_{1}$
in Algorithm \ref{alg:DC-Combining-Optimization.} has a different
definition from the matrix $\mathbf{A}_{1}$ in Algorithm 2 in \cite{2017_TSP_WPT_Bruno_Yang_Large}.}. It solves a stationary point of the problem \eqref{eq:OP1-MaxPoutStP-1}.

\begin{algorithm}[t]
\begin{enumerate}
\item \textbf{Initialize}: $i=0$, $\mathbf{s}^{\left(0\right)}$, $\mathbf{X}^{\left(0\right)}$,
$\mathbf{t}_{q}^{\left(0\right)}$, and $v_{\mathrm{out},q}^{\left(0\right)}$
for $\forall q$;
\item \textbf{do}
\item ~~~~$i=i+1$;
\item ~~~~$\mathbf{A}_{1}=\mathbf{C}_{1}+\mathbf{C}_{1}^{H}$ where
$\mathbf{C}_{1}$ is computed by \eqref{eq:Matrix C1};
\item ~~~~Update $\mathbf{s}^{\left(i\right)}=\sqrt{2P}\left[\mathbf{U}_{\mathbf{A}_{1}}\right]_{\mathrm{max}}$;
$\mathbf{X}^{\left(i\right)}=\mathbf{s}^{\left(i\right)}\mathbf{s}^{\left(i\right)H}$;
\item ~~~~Update $t_{q,k}^{\left(i\right)}=\mathrm{Tr}\left(\mathbf{M}_{q,k}\mathbf{X}^{\left(i\right)}\right)$,
$\forall q,k$;
\item ~~~~Update $v_{\mathrm{out},q}^{\left(i\right)}=\frac{1}{2}\beta_{2}t_{q,0}^{\left(i\right)}+\mathbf{t}_{q}^{\left(i\right)H}\mathbf{A}_{0}\mathbf{t}_{q}^{\left(i\right)}$,
$\forall q$;
\item \textbf{until} $\left\Vert \mathbf{s}^{\left(i\right)}-\mathbf{s}^{\left(i-1\right)}\right\Vert /\left\Vert \mathbf{s}^{\left(i\right)}\right\Vert \leq\epsilon$
or $i=i_{\mathrm{max}}$
\item Set $\mathbf{s}^{\star}=\mathbf{s}^{\left(i\right)}$;
\end{enumerate}
\caption{\label{alg:DC-Combining-Optimization.}Joint Waveform and Beamforming
Optimization with DC Combining.}
\end{algorithm}

\section{Joint Waveform and Beamforming Optimization with RF Combining}

\begin{figure}[t]
\begin{centering}
\includegraphics[width=8.5cm]{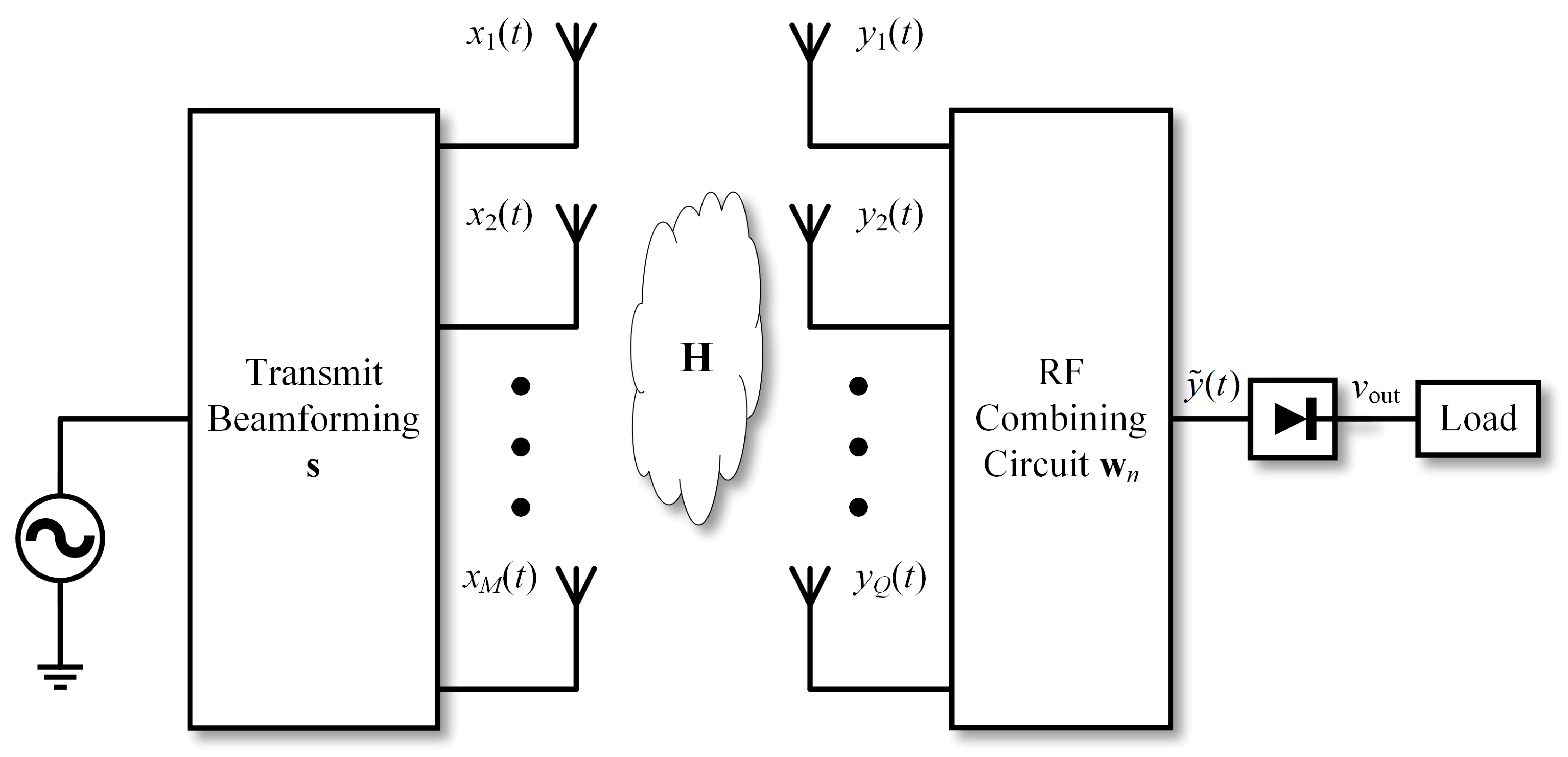}
\par\end{centering}
\caption{\label{fig:RF combinging}Schematic of the multi-sine MIMO WPT system
with RF combining in the receiver.}
\end{figure}

Consider the RF combining scheme for the multiple antennas as shown
in Fig. \ref{fig:RF combinging}. All receive antennas are connected
to an RF combining circuit such as an RF power combiner. The received
signals at all receive antennas are combined together so that the
RF combined signal $\widetilde{y}\left(t\right)$ can be expressed
as
\begin{align}
\widetilde{y}\left(t\right) & =\Re\left\{ \sum_{n=1}^{N}\mathbf{w}_{n}^{H}\mathbf{H}_{n}\mathbf{s}_{n}e^{j\omega_{n}t}\right\} ,
\end{align}
where $\mathbf{w}_{n}$ denotes the receive beamformer at the $n$th
angular frequency. Using the nonlinear rectenna model \eqref{eq:VoutRectennaModel}
with the truncation order $n_{0}=4$, the output DC voltage is given
by
\begin{align}
v_{\mathrm{out}} & =\beta_{2}\mathscr{\mathcal{E}}\left\{ \widetilde{y}\left(t\right)^{2}\right\} +\beta_{4}\mathscr{\mathcal{E}}\left\{ \widetilde{y}\left(t\right)^{4}\right\} ,\label{eq:vout in RF combining}
\end{align}
where $\mathscr{\mathcal{E}}\left\{ \widetilde{y}\left(t\right)^{2}\right\} $
and $\mathscr{\mathcal{E}}\left\{ \widetilde{y}\left(t\right)^{4}\right\} $
are given by
\begin{align}
\mathscr{\mathcal{E}}\left\{ \widetilde{y}\left(t\right)^{2}\right\}  & =\frac{1}{2}\sum_{n=1}^{N}\mathbf{s}_{n}^{H}\mathbf{H}_{n}^{H}\mathbf{w}_{n}\mathbf{w}_{n}^{H}\mathbf{H}_{n}\mathbf{s}_{n}\label{eq:y2 in RF combining}\\
\mathscr{\mathcal{E}}\left\{ \widetilde{y}\left(t\right)^{4}\right\}  & =\frac{3}{8}\sum_{\substack{n_{1},n_{2},n_{3},n_{4}\\
n_{1}+n_{2}=n_{3}+n_{4}
}
}\left(\mathbf{s}_{n_{3}}^{H}\mathbf{H}_{n_{3}}^{H}\mathbf{w}_{n_{3}}\mathbf{w}_{n_{1}}^{H}\mathbf{H}_{n_{1}}\mathbf{s}_{n_{1}}\right.\nonumber \\
 & \left.\qquad\qquad\qquad\quad\;\;\,\cdot\mathbf{s}_{n_{4}}^{H}\mathbf{H}_{n_{4}}^{H}\mathbf{w}_{n_{4}}\mathbf{w}_{n_{2}}^{H}\mathbf{H}_{n_{2}}\mathbf{s}_{n_{2}}\right).\label{eq:y4 in RF combining}
\end{align}

\subsection{General Receive Beamforming}

We first consider an RF combining scheme with general receive beamforming
satisfying the constraint that $\left\Vert \mathbf{w}_{n}\right\Vert \leq1$
$\forall n$. This constraint results from the fact that the output
power of the passive RF combining circuit should be equal or less
than the input power \cite{ShanpuShen2020_TWC_MIMO_WPT_SingleTone}.

We aim to maximize the total output DC power subject to the transmit
power constraint and the general receive beamforming constraint. Maximizing
the output DC power $P_{\mathrm{out}}=v_{\mathrm{out}}^{2}/R_{L}$
is equivalent to maximizing $v_{\mathrm{out}}$, so we can formulate
the equivalent problem as 
\begin{align}
\underset{\mathbf{s},\mathbf{w}_{n}}{\mathrm{max}}\;\;\; & \left\{ v_{\mathrm{out}}\colon\frac{1}{2}\left\Vert \mathbf{s}\right\Vert ^{2}\leq P,\:\left\Vert \mathbf{w}_{n}\right\Vert \leq1,\:\forall n\right\} .\label{eq:OPT5-object}
\end{align}
To handle this, we introduce an auxiliary multiple-input single-output
(MISO) channel $\widetilde{\mathbf{h}}_{n}=\mathbf{w}_{n}^{H}\mathbf{H}_{n}$
$\forall n$, which is an effective MISO channel obtained by concatenating
the MIMO channel $\mathbf{H}_{n}$ with the RF combining $\mathbf{w}_{n}$.
For the joint beamforming and waveform optimization in the equivalent
multi-sine MISO WPT system, it is shown in \cite{2016_TSP_WPT_Bruno_Waveform}
and \cite{2017_TSP_WPT_Bruno_Yang_Large} that the optimal $\mathbf{s}_{n}$
is actually a matched beamformer of the form
\begin{equation}
\mathbf{s}_{n}=\xi_{n}\frac{\widetilde{\mathbf{h}}_{n}^{H}}{\left\Vert \widetilde{\mathbf{h}}_{n}\right\Vert }=\xi_{n}\frac{\:\:\left(\mathbf{w}_{n}^{H}\mathbf{H}_{n}\right)^{H}}{\left\Vert \mathbf{w}_{n}^{H}\mathbf{H}_{n}\right\Vert },\label{eq:optimal beamfommer sn}
\end{equation}
where $\xi_{n}>0$ without loss of generality and $\xi_{n}^{2}$ denotes
the power allocated to the sinewave at the $n$th angular frequency.
We group all $\xi_{n}$ into a vector $\mathbf{p}=\left[\xi_{1},\xi_{2},\ldots,\xi_{N}\right]^{T}$
such that $\mathbf{p}$ describes the power allocation to the different
sinewaves and $\frac{1}{2}\left\Vert \mathbf{p}\right\Vert ^{2}\leq P$.
With the optimal $\mathbf{s}_{n}$ \eqref{eq:optimal beamfommer sn},
the problem \eqref{eq:OPT5-object} can be equivalently converted
to the following problem
\begin{align}
\underset{\mathbf{p},\mathbf{w}_{n}}{\mathrm{max}}\;\;\; & \left\{ v_{\mathrm{out}}\colon\frac{1}{2}\left\Vert \mathbf{p}\right\Vert ^{2}\leq P,\:\left\Vert \mathbf{w}_{n}\right\Vert \leq1,\:\forall n\right\} ,\label{eq:OPT6-object}
\end{align}
where the terms $\mathscr{\mathcal{E}}\left\{ \widetilde{y}\left(t\right)^{2}\right\} $
and $\mathscr{\mathcal{E}}\left\{ \widetilde{y}\left(t\right)^{4}\right\} $
in the objective function \eqref{eq:OPT6-object} can be achieved
by substituting \eqref{eq:optimal beamfommer sn} into \eqref{eq:y2 in RF combining}
and \eqref{eq:y4 in RF combining}, i.e. 
\begin{align}
\mathscr{\mathcal{E}}\left\{ \widetilde{y}\left(t\right)^{2}\right\}  & =\frac{1}{2}\left[\sum_{n=1}^{N}\left\Vert \widetilde{\mathbf{h}}_{n}\right\Vert ^{2}\xi_{n}^{2}\right],\label{eq:y2 in MISO}\\
\mathscr{\mathcal{E}}\left\{ \widetilde{y}\left(t\right)^{4}\right\}  & =\frac{3}{8}\left[\sum_{\substack{n_{1},n_{2},n_{3},n_{4}\\
n_{1}+n_{2}=n_{3}+n_{4}
}
}\left[\prod_{j=1}^{4}\left\Vert \widetilde{\mathbf{h}}_{n_{j}}\right\Vert \xi_{n_{j}}\right]\right].\label{eq:y4 in MISO}
\end{align}
From \eqref{eq:y2 in MISO} and \eqref{eq:y4 in MISO}, we can find
that $v_{\mathrm{out}}$ increases with $\left\Vert \widetilde{\mathbf{h}}_{n}\right\Vert $
given any $\xi_{n}$. Therefore, the optimal receive beamformer $\mathbf{w}_{n}^{\star}$
maximizing the output DC power is given by $\mathbf{w}_{n}^{\star}=\underset{\left\Vert \mathbf{w}_{n}\right\Vert \leq1}{\mathrm{argmax}}\left\Vert \widetilde{\mathbf{h}}_{n}\right\Vert .$
A closed form solution for $\mathbf{w}_{n}^{\star}$ can be obtained
by using SVD $\mathbf{H}_{n}=\mathbf{U}_{n}\mathbf{\Sigma}_{n}\mathbf{V}_{n}^{H}$.
Then, the optimal receive beamformer $\mathbf{w}_{n}^{\star}$ is
given by 
\begin{equation}
\mathbf{w}_{n}^{\star}=\left[\mathbf{U}_{n}\right]_{\mathrm{max}},\label{eq:optimal wn}
\end{equation}
where $\left[\mathbf{U}_{n}\right]_{\mathrm{max}}$ refers to the
vector in $\mathbf{U}_{n}$ corresponding to $\sigma_{n}$ which is
the maximum singular value of $\mathbf{H}_{n}$. Therefore, the maximum
value of $\left\Vert \widetilde{\mathbf{h}}_{n}\right\Vert $ is $\sigma_{n}$.

With the optimal $\mathbf{s}_{n}$ \eqref{eq:optimal beamfommer sn}
and the optimal receive beamformer $\mathbf{w}_{n}^{\star}$ \eqref{eq:optimal wn},
the joint waveform and beamforming optimization for the multi-sine
MIMO WPT system with the general receive beamforming can be equivalently
converted to the waveform optimization for the multi-sine single-input
single-output (SISO) WPT system. Namely, the problem \eqref{eq:OPT6-object}
is equivalent to
\begin{align}
\underset{\mathbf{p}}{\mathrm{max}}\;\;\; & \left\{ v_{\mathrm{out}}\colon\frac{1}{2}\left\Vert \mathbf{p}\right\Vert ^{2}\leq P\right\} ,\label{eq:OPT7-object}
\end{align}
which finds the optimal power allocation across sinewaves to maximize
the output DC voltage. The terms $\mathscr{\mathcal{E}}\left\{ \widetilde{y}\left(t\right)^{2}\right\} $
and $\mathscr{\mathcal{E}}\left\{ \widetilde{y}\left(t\right)^{4}\right\} $
in the objective function \eqref{eq:OPT7-object} are given by
\begin{align}
\mathscr{\mathcal{E}}\left\{ \widetilde{y}\left(t\right)^{2}\right\}  & =\frac{1}{2}\left[\sum_{n=1}^{N}\sigma_{n}^{2}\xi_{n}^{2}\right],\label{eq:y(t)2}\\
\mathscr{\mathcal{E}}\left\{ \widetilde{y}\left(t\right)^{4}\right\}  & =\frac{3}{8}\left[\sum_{\substack{n_{1},n_{2},n_{3},n_{4}\\
n_{1}+n_{2}=n_{3}+n_{4}
}
}\left[\prod_{j=1}^{4}\sigma_{n_{j}}\xi_{n_{j}}\right]\right].\label{eq:y(t)4}
\end{align}
Leveraging \eqref{eq:y(t)2} and \eqref{eq:y(t)4}, the objective
function $v_{\mathrm{out}}$ in \eqref{eq:OPT7-object} writes as
a posynomial, and can be written in the compact form $v_{\mathrm{out}}=\sum_{k=1}^{K}g_{k}\left(\mathbf{\mathbf{p}}\right)$
where $K$ denotes the number of monomials in the posynomial and $g_{k}\left(\mathbf{\mathbf{p}}\right)$
denotes the $k$th monomial. $g_{k}\left(\mathbf{\mathbf{p}}\right)$
can be defined as follows. We first exhaustively search $n_{1}=1,\ldots,N$,
$n_{2}=1,\ldots,N$, $n_{3}=1,\ldots,N$, and $n_{4}=1,\ldots,N$
to find all the combinations of $n_{1}$, $n_{2}$, $n_{3}$, $n_{4}$
which satisfy $n_{1}+n_{2}=n_{3}+n_{4}$. We denote the $k$th combination
of $n_{1}$, $n_{2}$, $n_{3}$, $n_{4}$ satisfying $n_{1}+n_{2}=n_{3}+n_{4}$
as $n_{1}^{\left(k\right)}$, $n_{2}^{\left(k\right)}$, $n_{3}^{\left(k\right)}$,
$n_{4}^{\left(k\right)}$, and there are in total $\bar{K}$ such
combinations. Therefore, we can define $g_{k}\left(\mathbf{\mathbf{p}}\right)$
as 
\begin{equation}
g_{k}\left(\mathbf{\mathbf{p}}\right)=\begin{cases}
\frac{3\beta_{4}}{8}\prod_{j=1}^{4}\sigma_{n_{j}^{\left(k\right)}}\xi_{n_{j}^{\left(k\right)}} & ,k=1,\ldots,\bar{K}\\
\frac{\beta_{2}}{2}\sigma_{k-\bar{K}}^{2}\xi_{k-\bar{K}}^{2} & ,k=\bar{K}+1,\ldots,\bar{K}+N
\end{cases},\label{eq:gk(p)}
\end{equation}
so that we can write $v_{\mathrm{out}}=\sum_{k=1}^{K}g_{k}\left(\mathbf{\mathbf{p}}\right)$
where $K=\bar{K}+N$. The problem \eqref{eq:OPT7-object} aims to
maximize a posynomial subject to a power constraint (the power is
also a posynomial), which is not a standard Geometric Programming
(GP). 
\begin{algorithm}[t]
\begin{enumerate}
\item \textbf{Initialize}: $i=0$ and $\mathbf{p}^{\left(0\right)}$;
\item \textbf{do}
\item ~~~~$i=i+1$;
\item ~~~~$\gamma_{k}=g_{k}\left(\mathbf{p}^{\left(i-1\right)}\right)/\sum_{k=1}^{K}g_{k}\left(\mathbf{p}^{\left(i-1\right)}\right)$,
$\forall k$;
\item ~~~~Update $\mathbf{p}^{\left(i\right)}$ by solving GP \eqref{eq:OPT9-object}-\eqref{eq:OPT9-constraint-2};
\item \textbf{until} $\left\Vert \mathbf{p}^{\left(i\right)}-\mathbf{p}^{\left(i-1\right)}\right\Vert /\left\Vert \mathbf{p}^{\left(i\right)}\right\Vert \leq\epsilon$
or $i=i_{\mathrm{max}}$
\item Set $\mathbf{p}^{\star}=\left[\xi_{1}^{\star},\xi_{2}^{\star},\ldots,\xi_{N}^{\star}\right]^{T}=\mathbf{p}^{\left(i\right)}$;
\item Set $\mathbf{w}_{n}^{\star}=\left[\mathbf{U}_{n}\right]_{\mathrm{max}}$,
$\forall n$;
\item Set $\mathbf{s}_{n}^{\star}=\xi_{n}^{\star}\frac{\left(\mathbf{w}_{n}^{\star H}\mathbf{H}_{n}\right)^{H}}{\left\Vert \mathbf{w}_{n}^{\star H}\mathbf{H}_{n}\right\Vert }$,
$\forall n$;
\end{enumerate}
\caption{\label{alg:RF-Combining-Optimization-General}Joint Waveform and Beamforming
Optimization with RF Combining using General Receive Beamforming.}
\end{algorithm}
To handle this, we introduce an auxiliary variable $\zeta_{0}>0$
and equivalently rewrite the problem \eqref{eq:OPT7-object} as
\begin{align}
\underset{\mathbf{p},\zeta_{0}}{\mathrm{min}}\;\;\; & \frac{1}{\zeta_{0}}\label{eq:OPT8-object}\\
\mathsf{\mathrm{s.t.}}\;\;\; & \frac{1}{2}\left\Vert \mathbf{p}\right\Vert ^{2}\leq P,\label{eq:OPT8-contraint-1}\\
 & \frac{\zeta_{0}}{\sum_{k=1}^{K}g_{k}\left(\mathbf{\mathbf{p}}\right)}\leq1.\label{eq:OPT8-constraint-2}
\end{align}
However, $\zeta_{0}/\sum_{k=1}^{K}g_{k}\left(\mathbf{p}\right)$ is
not a posynomial, which prevents the use of standard GP tools. Therefore,
we use SCA to approximate $\zeta_{0}/\sum_{k=1}^{K}g_{k}\left(\mathbf{p}\right)$
as a monomial and iteratively solve the approximated problem. Particularly,
at iteration $i$, $\zeta_{0}/\sum_{k=1}^{K}g_{k}\left(\mathbf{p}\right)$
is approximated at $\mathbf{p}^{\left(i-1\right)}$, which is the
optimal $\mathbf{p}$ solved at iteration $\left(i-1\right)$, as
a monomial based on the fact that an arithmetic mean (AM) is greater
or equal to the geometric mean (GM) \cite{2007_TWC_GP_Daniel}, so
we have
\begin{equation}
\frac{\zeta_{0}}{\sum_{k=1}^{K}g_{k}\left(\mathbf{p}\right)}\leq\frac{\zeta_{0}}{\prod{}_{k=1}^{K}\left(\frac{g_{k}\left(\mathbf{p}\right)}{\gamma_{k}}\right)^{\gamma_{k}}},\label{eq:AM-GM-bound}
\end{equation}
where $\gamma_{k}=g_{k}\left(\mathbf{p}^{\left(i-1\right)}\right)/\sum_{k=1}^{K}g_{k}\left(\mathbf{p}^{\left(i-1\right)}\right)>0$
and $\sum_{k=1}^{K}\gamma_{k}=1$. We replace the constraint $\zeta_{0}/\sum_{k=1}^{K}g_{k}\left(\mathbf{p}\right)\leq1$
with $\zeta_{0}\prod{}_{k=1}^{K}\left(\frac{g_{k}\left(\mathbf{p}\right)}{\gamma_{k}}\right)^{-\gamma_{k}}\leq1$
in a conservative way and then the problem \eqref{eq:OPT8-object}-\eqref{eq:OPT8-constraint-2}
can be approximated as a standard GP
\begin{align}
\underset{\mathbf{p},\zeta_{0}}{\mathsf{\mathrm{min}}}\;\;\; & \frac{1}{\zeta_{0}}\label{eq:OPT9-object}\\
\mathsf{\mathrm{s.t.}}\;\;\; & \frac{1}{2}\left\Vert \mathbf{p}\right\Vert ^{2}\leq P,\label{eq:OPT9-constraint-1}\\
 & \zeta_{0}\stackrel[k=1]{K}{\prod}\left(\frac{g_{k}\left(\mathbf{p}\right)}{\gamma_{k}}\right)^{-\gamma_{k}}\leq1,\label{eq:OPT9-constraint-2}
\end{align}
which can be solved by existing software using interior point methods,
e.g. CVX \cite{grant2008cvx}. Therefore, at each iteration of SCA,
we solve the standard GP \eqref{eq:OPT9-object}-\eqref{eq:OPT9-constraint-2}
for an updated set of $\left\{ \gamma_{k}\right\} $ by using interior
point methods which have provably polynomial time complexity \cite{2007_TWC_GP_Daniel},
and we repeat the iterations till convergence. The SCA guarantees
to converge to a stationary point of the problem \eqref{eq:OPT8-object}-\eqref{eq:OPT8-constraint-2}
\cite{2007_TWC_GP_Daniel}. Let $\mathbf{p}^{\star}$ denotes the
stationary point of the problem \eqref{eq:OPT8-object}-\eqref{eq:OPT8-constraint-2}.
Given the optimal $\mathbf{w}_{n}^{\star}$ \eqref{eq:optimal wn}
and $\mathbf{p}^{\star}$, the optimal $\mathbf{s}_{n}^{\star}$ can
be found by \eqref{eq:optimal beamfommer sn}. The initialization
of the SCA is important and affects the convergence speed. Herein,
we choose a good initial point for the SCA as $\mathbf{p}^{\left(0\right)}=\sqrt{2P/\sum_{n=1}^{N}\sigma_{n}^{2}}\left[\sigma_{1},\sigma_{2},\ldots,\sigma_{N}\right]^{T}.$

Algorithm \ref{alg:RF-Combining-Optimization-General} summarizes
the overall algorithm for jointly optimizing the waveform and beamforming
with RF combining using general receive beamforming\footnote{In beamforming only design \cite{ShanpuShen2020_TWC_MIMO_WPT_SingleTone},
the general receive beamforming has a simple closed form solution
but only valid for a continuous sinewave. However, when it comes to
joint waveform and beamforming design, the optimization is more challenging
as the waveform needs to be particularly optimized. Therefore we propose
Algorithms \ref{alg:RF-Combining-Optimization-General} in this paper
for more general multi-sine wave to increase the output DC power.}. It solves a stationary point of the problem \eqref{eq:OPT5-object}.

\subsection{Analog Receive Beamforming}

\begin{figure}[t]
\begin{centering}
\includegraphics[width=8.5cm]{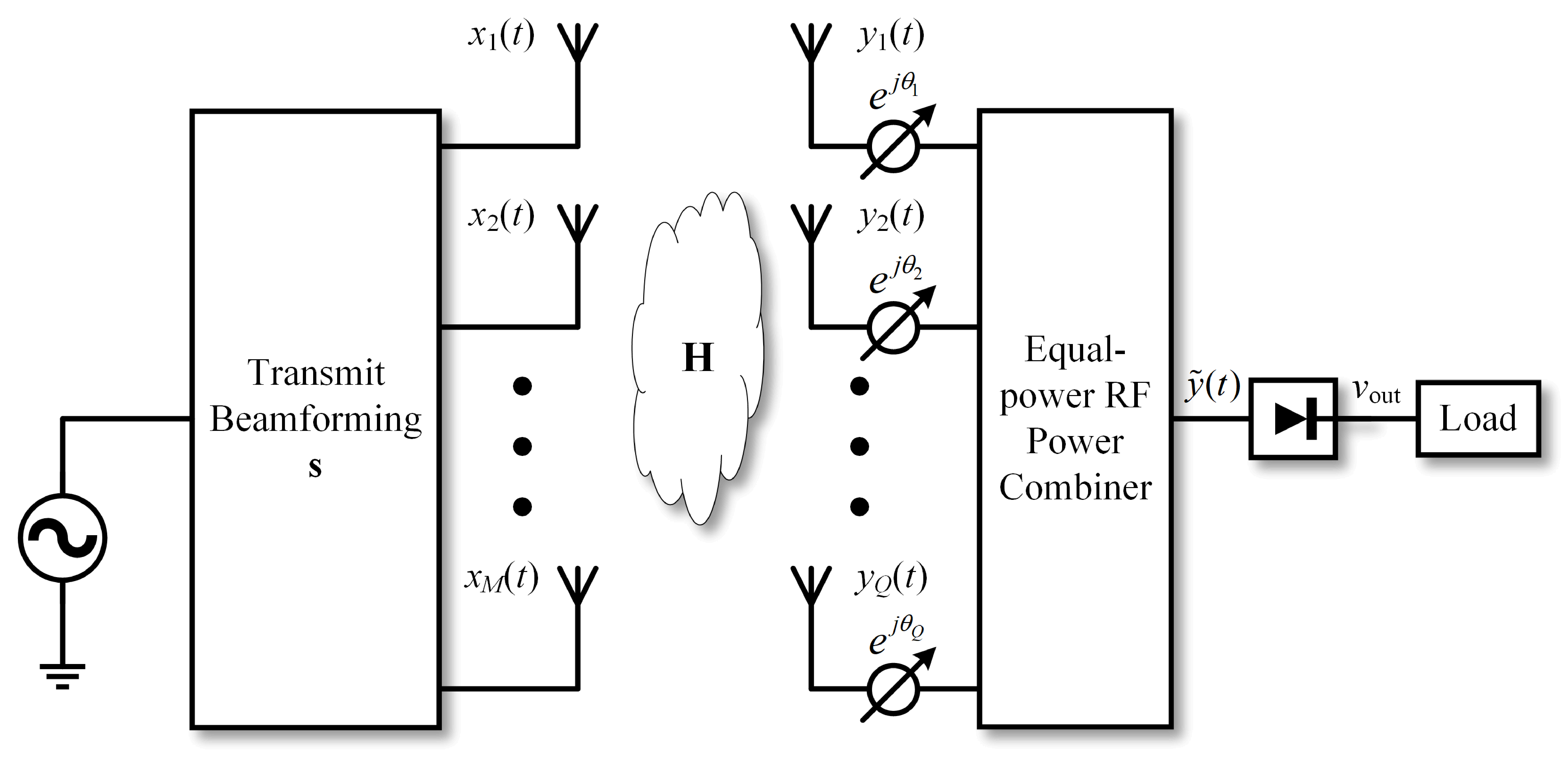}
\par\end{centering}
\caption{\label{fig:RF combinging-ABF}Schematic of the multi-sine MIMO WPT
system with RF combining using analog receive beamforming in the receiver.}
\end{figure}

We also consider an RF combining scheme using a practical RF combining
circuit as shown in Fig. \ref{fig:RF combinging-ABF}. We refer to
it as analog receive beamforming. It consists of an equal-power RF
power combiner and $Q$ phase shifters. Each receive antenna is connected
to a phase shifter and the outputs of the $Q$ phase shifters are
connected to the RF power combiner. Phase shifters have been widely
used in hybrid precoding for massive MIMO communications and the phase
shifts for multiple carrier frequencies are usually modeled to be
identical \cite{2016_JSTSP_HybridPrecoding_Xianghao}, \cite{2017_JSAC_HybridPrecoding_OFDM}.
Therefore, the analog receive beamformers for the $N$ sinewaves can
be modeled to be identical, i.e.
\begin{equation}
\mathbf{w}_{n}=\frac{1}{\sqrt{Q}}\left[e^{-j\theta_{1}},e^{-j\theta_{2}},...,e^{-j\theta_{Q}}\right]^{T},\:\forall n,\label{eq:ABF constraint-1}
\end{equation}
\begin{equation}
-\pi\leq\theta_{q}<\pi,\:\forall q,\label{eq:ABF constraint-2}
\end{equation}
where $\theta_{q}$ denotes the $q$th phase shift for $q=1,\ldots,Q$.

We aim to maximize the output DC voltage $v_{\mathrm{out}}$ (equivalently
maximize the output DC power) subject to the transmit power constraint
and the analog receive beamforming constraints \eqref{eq:ABF constraint-1}
and \eqref{eq:ABF constraint-2}, so we can formulate the problem
as
\begin{align}
\underset{\mathbf{s},\mathbf{w}_{n},\theta_{q}}{\mathrm{max}}\;\;\; & v_{\mathrm{out}}\label{eq:OPT11-object}\\
\mathsf{\mathrm{s.t.}\;\;\,\;\;} & \frac{1}{2}\left\Vert \mathbf{s}\right\Vert ^{2}\leq P,\label{eq:OPT11-constraint-1}\\
 & \mathbf{w}_{n}=\frac{1}{\sqrt{Q}}\left[e^{-j\theta_{1}},e^{-j\theta_{2}},...,e^{-j\theta_{Q}}\right]^{T},\:\forall n,\label{eq:OPT11-constraint-2}\\
 & -\pi\leq\theta_{q}<\pi,\:\forall q,\label{eq:OPT11-constraint-3}
\end{align}
where $v_{\mathrm{out}}$ is given by \eqref{eq:vout in RF combining},
\eqref{eq:y2 in RF combining}, and \eqref{eq:y4 in RF combining}.
The constraints \eqref{eq:OPT11-constraint-2} and \eqref{eq:OPT11-constraint-3}
are more restrictive than the general receive beamforming constraint
$\left\Vert \mathbf{w}_{n}\right\Vert \leq1$ $\forall n$, so the
analog receive beamforming provides a lower output DC voltage than
the general receive beamforming.

The problem \eqref{eq:OPT11-object}-\eqref{eq:OPT11-constraint-3}
can be simplified in three steps: 1) introducing auxiliary MISO channels
$\widetilde{\mathbf{h}}_{n}=\mathbf{w}_{n}^{H}\mathbf{H}_{n}$ $\forall n$
so that the optimal $\mathbf{s}_{n}$ is provided by \eqref{eq:optimal beamfommer sn}
as shown in \cite{2016_TSP_WPT_Bruno_Waveform} and \cite{2017_TSP_WPT_Bruno_Yang_Large};
2) introducing an auxiliary variable $\mathbf{w}=\mathbf{w}_{n}$
$\forall n$ so that the constraints \eqref{eq:OPT11-constraint-2}
and \eqref{eq:OPT11-constraint-3} are equivalent to $\left|\left[\mathbf{w}\right]_{q}\right|=\frac{1}{\sqrt{Q}}$
$\forall q$; and 3) introducing auxiliary variables $r_{n}=\left\Vert \widetilde{\mathbf{h}}_{n}\right\Vert =\left\Vert \mathbf{w}^{H}\mathbf{H}_{n}\right\Vert >0$
$\forall n$. Hence we can equivalently simplify problem \eqref{eq:OPT11-object}-\eqref{eq:OPT11-constraint-3}
as 
\begin{align}
\underset{\mathbf{p},\mathbf{w},r_{n}}{\mathrm{max}}\;\;\; & v_{\mathrm{out}}\label{eq:OPT10-object}\\
\mathsf{\mathrm{s.t.}\;\;\;\;} & \frac{1}{2}\left\Vert \mathbf{p}\right\Vert ^{2}\leq P,\label{eq:OPT10-constraint-1}\\
 & \left|\left[\mathbf{w}\right]_{q}\right|=\frac{1}{\sqrt{Q}},\:\forall q,\label{eq:OPT10-constraint-2}\\
 & r_{n}^{2}=\mathbf{w}^{H}\mathbf{H}_{n}\mathbf{H}_{n}^{H}\mathbf{w},\:\forall n,\label{eq:OPT10-constraint-3}
\end{align}
where the terms $\mathscr{\mathcal{E}}\left\{ \widetilde{y}\left(t\right)^{2}\right\} $
and $\mathscr{\mathcal{E}}\left\{ \widetilde{y}\left(t\right)^{4}\right\} $
in the objective function \eqref{eq:OPT10-object} are given by
\begin{align}
\mathscr{\mathcal{E}}\left\{ \widetilde{y}\left(t\right)^{2}\right\}  & =\frac{1}{2}\left[\sum_{n=1}^{N}r_{n}^{2}\xi_{n}^{2}\right],\label{eq:ABF-Y2}\\
\mathscr{\mathcal{E}}\left\{ \widetilde{y}\left(t\right)^{4}\right\}  & =\frac{3}{8}\left[\sum_{\substack{n_{1},n_{2},n_{3},n_{4}\\
n_{1}+n_{2}=n_{3}+n_{4}
}
}\left[\prod_{j=1}^{4}r_{n_{j}}\xi_{n_{j}}\right]\right].\label{eq:ABF-Y4}
\end{align}
Leveraging \eqref{eq:ABF-Y2} and \eqref{eq:ABF-Y4}, the objective
function $v_{\mathrm{out}}$ \eqref{eq:OPT10-object} is a posynomial
so that it can be written in a compact form that $v_{\mathrm{out}}=\sum_{k=1}^{K^{\prime}}g_{k}^{\prime}\left(\mathbf{\mathbf{p}},\mathbf{r}\right)$
where $K^{\prime}$ denotes the number of monomials in the posynomial
and $g_{k}^{\prime}\left(\mathbf{\mathbf{p}},\mathbf{r}\right)$ denotes
the $k$th monomial with $\mathbf{r}=\left[r_{1},r_{2},\ldots,r_{N}\right]^{T}$.
$g_{k}^{\prime}\left(\mathbf{\mathbf{p}},\mathbf{r}\right)$ can be
defined in a similar way to \eqref{eq:gk(p)}. The problem \eqref{eq:OPT10-object}-\eqref{eq:OPT10-constraint-3}
is more difficult than the problem \eqref{eq:OPT6-object} (the general
receive beamforming) because we cannot decouple the optimization of
$\mathbf{p}$ and $\mathbf{w}$ to provide a closed-form solution
for the optimal $\mathbf{w}$. Therefore, we need to jointly optimize
$\mathbf{p}$ and $\mathbf{w}$. To that end, we first replace the
constraint $\left|\left[\mathbf{w}\right]_{q}\right|=\frac{1}{\sqrt{Q}}$
with $\left|\left[\mathbf{w}\right]_{q}\right|\leq\frac{1}{\sqrt{Q}}$
without affecting the optimal solution of the problem \eqref{eq:OPT10-object}-\eqref{eq:OPT10-constraint-3}.
The reason is that the objective function \eqref{eq:OPT10-object}
monotonically increases with $r_{n}$ and $r_{n}=\left\Vert \mathbf{w}^{H}\mathbf{H}_{n}\right\Vert $.
Hence, the optimal $\mathbf{w}$ that maximizes the objective function
must satisfy the equality even though $\left|\left[\mathbf{w}\right]_{q}\right|\leq\frac{1}{\sqrt{Q}}$.
We also replace the constraint $r_{n}^{2}=\mathbf{w}^{H}\mathbf{H}_{n}\mathbf{H}_{n}^{H}\mathbf{w}$
with $r_{n}^{2}\leq\mathbf{w}^{H}\mathbf{H}_{n}\mathbf{H}_{n}^{H}\mathbf{w}$
without affecting the optimal solution of the problem \eqref{eq:OPT10-object}-\eqref{eq:OPT10-constraint-3}
since the objective function \eqref{eq:OPT10-object} monotonically
increases with $r_{n}$. Hence, the optimal $r_{n}$ that maximizes
the objective function must satisfy the equality even though $r_{n}^{2}\leq\mathbf{w}^{H}\mathbf{H}_{n}\mathbf{H}_{n}^{H}\mathbf{w}$.
In addition, we introduce an auxiliary variable $\zeta_{1}>0$ and
then equivalently rewrite the problem \eqref{eq:OPT10-object}-\eqref{eq:OPT10-constraint-3}
as
\begin{align}
\underset{\mathbf{p},\mathbf{w},r_{n},\zeta_{1}}{\mathsf{\mathrm{min}}}\;\;\; & \frac{1}{\zeta_{1}}\label{eq:OPT12-object}\\
\mathsf{\mathrm{s.t.}}\;\;\;\;\;\; & \frac{1}{2}\left\Vert \mathbf{p}\right\Vert ^{2}\leq P,\label{eq:OPT12-constraint-1}\\
 & \left|\left[\mathbf{w}\right]_{q}\right|\leq\frac{1}{\sqrt{Q}},\:\forall q,\label{eq:OPT12-constraint-2}\\
 & r_{n}^{2}\leq\mathbf{w}^{H}\mathbf{H}_{n}\mathbf{H}_{n}^{H}\mathbf{w},\:\forall n,\label{eq:OPT12-constraint-3}\\
 & \frac{\zeta_{1}}{\sum_{k=1}^{K^{\prime}}g_{k}^{\prime}\left(\mathbf{\mathbf{p}},\mathbf{r}\right)}\leq1.\label{eq:OPT12-constraint-5}
\end{align}
However, \eqref{eq:OPT12-constraint-3} is not convex, and $\zeta_{1}/\sum_{k=1}^{K^{\prime}}g_{k}^{\prime}\left(\mathbf{\mathbf{p}},\mathbf{r}\right)$
is not a posynomial which prevents the transformation to a convex
constraint.

To solve the nonconvex problem \eqref{eq:OPT12-object}-\eqref{eq:OPT12-constraint-5},
we use SCA to approximate \eqref{eq:OPT12-constraint-3} and \eqref{eq:OPT12-constraint-5}
as convex constraints and iteratively solve the approximated problem.
Particularly, at iteration $i$, $r_{n}^{2}\leq\mathbf{w}^{H}\mathbf{H}_{n}\mathbf{H}_{n}^{H}\mathbf{w}$
$\forall n$ is approximated at $\mathbf{w}^{\left(i-1\right)}$,
which is the optimal $\mathbf{w}$ solved at iteration $\left(i-1\right)$,
as a convex constraint 
\begin{equation}
r_{n}^{2}\leq2\Re\left\{ \mathbf{w}^{\left(i-1\right)H}\mathbf{H}_{n}\mathbf{H}_{n}^{H}\mathbf{w}\right\} -\mathbf{w}^{\left(i-1\right)H}\mathbf{H}_{n}\mathbf{H}_{n}^{H}\mathbf{w}^{\left(i-1\right)},\:\forall n\text{,}\label{eq:rn2<linear}
\end{equation}
based on the first-order Taylor expansion \cite{adali2010adaptiveSignalProcessing},
while $\zeta_{1}/\sum_{k=1}^{K^{\prime}}g_{k}^{\prime}\left(\mathbf{\mathbf{p}},\mathbf{r}\right)$
is approximated at $\mathbf{p}^{\left(i-1\right)}$ and $\mathbf{r}^{\left(i-1\right)}$,
which is optimal $\mathbf{\mathbf{p}}$ and $\mathbf{r}$ solved at
iteration $\left(i-1\right)$, as a monomial based on the AM-GM inequality
\cite{2007_TWC_GP_Daniel}, i.e.
\begin{equation}
\frac{\zeta_{1}}{\sum_{k=1}^{K^{\prime}}g_{k}^{\prime}\left(\mathbf{\mathbf{p}},\mathbf{r}\right)}\leq\frac{\zeta_{1}}{\prod{}_{k=1}^{K^{\prime}}\left(\frac{g_{k}^{\prime}\left(\mathbf{\mathbf{p}},\mathbf{r}\right)}{\gamma_{k}^{\prime}}\right)^{\gamma_{k}^{\prime}}},
\end{equation}
where $\gamma_{k}^{\prime}=g_{k}^{\prime}\left(\mathbf{p}^{\left(i-1\right)},\mathbf{r}^{\left(i-1\right)}\right)/\sum_{k=1}^{K^{\prime}}g_{k}^{\prime}\left(\mathbf{p}^{\left(i-1\right)},\mathbf{r}^{\left(i-1\right)}\right)$
$\forall k$ and $\sum_{k=1}^{K^{\prime}}\gamma_{k}^{\prime}=1$.
We replace \eqref{eq:OPT12-constraint-3} with \eqref{eq:rn2<linear}
and replace \eqref{eq:OPT12-constraint-5} with $\zeta_{1}\prod{}_{k=1}^{K^{\prime}}\left(\frac{g_{k}^{\prime}\left(\mathbf{\mathbf{p}},\mathbf{r}\right)}{\gamma_{k}^{\prime}}\right)^{-\gamma_{k}^{\prime}}\leq1$
both in a conservative way, so that the problem \eqref{eq:OPT12-object}-\eqref{eq:OPT12-constraint-5}
can be approximated as
\begin{align}
\underset{\mathbf{p},\mathbf{w},r_{n},\zeta_{1}}{\mathsf{\mathrm{min}}}\;\;\; & \frac{1}{\zeta_{1}}\label{eq:OPT13-object}\\
\mathsf{\mathrm{s.t.}}\;\;\;\;\;\; & \frac{1}{2}\left\Vert \mathbf{p}\right\Vert ^{2}\leq P,\label{eq:OPT13-constraint-1}\\
 & \left|\left[\mathbf{w}\right]_{q}\right|\leq\frac{1}{\sqrt{Q}},\:\forall q,\label{eq:OPT13-constraint-2}\\
 & r_{n}^{2}\leq2\Re\left\{ \mathbf{w}^{\left(i-1\right)H}\mathbf{H}_{n}\mathbf{H}_{n}^{H}\mathbf{w}\right\} \nonumber \\
 & \qquad-\mathbf{w}^{\left(i-1\right)H}\mathbf{H}_{n}\mathbf{H}_{n}^{H}\mathbf{w}^{\left(i-1\right)},\:\forall n\text{,}\label{eq:OPT13-constraint-3}\\
 & \zeta_{1}\stackrel[k=1]{K^{\prime}}{\prod}\left(\frac{g_{k}^{\prime}\left(\mathbf{\mathbf{p}},\mathbf{r}\right)}{\gamma_{k}^{\prime}}\right)^{-\gamma_{k}^{\prime}}\leq1,\label{eq:OPT13-constraint-5}
\end{align}
which can be equivalently transformed to a convex problem by using
a logarithmic transformation. To see the details, we first rewrite
the monomial term as $\zeta_{1}\prod{}_{k=1}^{K^{\prime}}\left(\frac{g_{k}^{\prime}\left(\mathbf{\mathbf{p}},\mathbf{r}\right)}{\gamma_{k}^{\prime}}\right)^{-\gamma_{k}^{\prime}}=c_{1}\zeta_{1}\prod{}_{n=1}^{N}\xi_{n}^{a_{n}}r_{n}^{b_{n}}$
where $c_{1}$, $a_{n}$, $b_{n}$ $\forall n$ are constants. We
then introduce auxiliary variables $\tilde{\zeta}_{1}=\log\zeta_{1}$,
$\tilde{\xi}_{n}=\log\xi_{n}$, $\tilde{r}_{n}=\log r_{n}$ $\forall n$,
so that $e^{\tilde{\zeta}_{1}}=\zeta_{1}$, $e^{\tilde{\xi}_{n}}=\xi_{n}$,
$e^{\tilde{r}_{n}}=r_{n}$ $\forall n$. Using the logarithmic transformation
for the objective function \eqref{eq:OPT13-object} and the constraints
\eqref{eq:OPT13-constraint-1}, \eqref{eq:OPT13-constraint-3}, and
\eqref{eq:OPT13-constraint-5}, we can equivalently transform the
problem \eqref{eq:OPT13-object}-\eqref{eq:OPT13-constraint-5} as
\begin{align}
\underset{\tilde{\xi}_{n},\mathbf{w},\tilde{r}_{n},\tilde{\zeta}_{1}}{\mathsf{\mathrm{min}}}\;\;\; & -\tilde{\zeta}_{1}\label{eq:OPT14-Object}\\
\mathsf{\mathrm{s.t.}}\;\;\;\;\;\,\; & \sum_{n=1}^{N}e^{2\tilde{\xi}_{n}}\leq2P,\label{eq:OPT14-consraint-1}\\
 & \left|\left[\mathbf{w}\right]_{q}\right|\leq\frac{1}{\sqrt{Q}},\:\forall q,\\
 & e^{2\tilde{r}_{n}}\leq2\Re\left\{ \mathbf{w}^{\left(i-1\right)H}\mathbf{H}_{n}\mathbf{H}_{n}^{H}\mathbf{w}\right\} \nonumber \\
 & \qquad\;\;\,-\mathbf{w}^{\left(i-1\right)H}\mathbf{H}_{n}\mathbf{H}_{n}^{H}\mathbf{w}^{\left(i-1\right)},\:\forall n\text{,}\\
 & \log c_{1}+\tilde{\zeta}_{1}+\sum_{n=1}^{N}a_{n}\tilde{\xi}_{n}+\sum_{n=1}^{N}b_{n}\tilde{r}_{n}\leq0,\label{eq:OPT14-constraint-4}
\end{align}
which is a convex problem that can be solved by existing software
using interior point methods, e.g. CVX. Therefore, at each iteration
of SCA, we solve the problem \eqref{eq:OPT13-object}-\eqref{eq:OPT13-constraint-5}
by solving the equivalent convex problem \eqref{eq:OPT14-Object}-\eqref{eq:OPT14-constraint-4}
with interior point methods which have polynomial time complexity,
and we repeat the iteration till convergence. The SCA guarantees to
converge to a stationary point of the problem \eqref{eq:OPT12-object}-\eqref{eq:OPT12-constraint-5}.
Let $\mathbf{p}^{\star}$ and $\mathbf{w}^{\star}$ denote the stationary
point of the problem \eqref{eq:OPT12-object}-\eqref{eq:OPT12-constraint-5}.
Given $\mathbf{p}^{\star}$ and $\mathbf{w}^{\star}$, the optimal
$\mathbf{s}_{n}^{\star}$ can be found by \eqref{eq:optimal beamfommer sn}.
The initialization of SCA is important and affects the convergence
speed. Herein, we choose a good initial point as $\mathbf{w}^{\left(0\right)}=\frac{1}{\sqrt{Q}}e^{j\mathrm{arg}\left(\left[\mathbf{U}_{\bar{n}}\right]_{\mathrm{max}}\right)}$
where $\bar{n}=\arg\max_{n}\sigma_{n}$ is the strongest channel using
SVD $\mathbf{H}_{n}=\mathbf{U}_{n}\mathbf{\Sigma}_{n}\mathbf{V}_{n}^{H}$
and $\left[\mathbf{U}_{\bar{n}}\right]_{\mathrm{max}}$ is the vector
in $\mathbf{U}_{\bar{n}}$ corresponding to the maximum singular value
of $\mathbf{H}_{\bar{n}}$. Besides $\mathbf{p}^{\left(0\right)}=\sqrt{2P/\sum_{n=1}^{N}\sigma_{n}^{2}}\left[\sigma_{1},\sigma_{2},\ldots,\sigma_{N}\right]^{T}$,
$r_{n}^{\left(0\right)}=\left\Vert \mathbf{w}^{\left(0\right)H}\mathbf{H}_{n}\right\Vert $,
and $\zeta_{1}^{\left(0\right)}=0$.

Algorithm \ref{alg:RF-Combining-Optimization-Analog} summarizes the
overall algorithm for jointly optimizing the waveform and beamforming
with RF combining using analog receive beamforming\footnote{Algorithms \ref{alg:RF-Combining-Optimization-Analog} proposed in
this paper is different from the algorithm for beamforming only design
using analog receive beamforming in \cite{ShanpuShen2020_TWC_MIMO_WPT_SingleTone}
in two aspects: 1) The algorithm in \cite{ShanpuShen2020_TWC_MIMO_WPT_SingleTone}
cannot theoretically guarantee finding a stationary point, but Algorithm
\ref{alg:RF-Combining-Optimization-Analog} can; 2) The algorithm
in \cite{ShanpuShen2020_TWC_MIMO_WPT_SingleTone} is only for a continuous
sinewave, but Algorithm \ref{alg:RF-Combining-Optimization-Analog}
is for multi-sine wave which is more general and more challenging
to optimize.}. It solves a stationary point of the problem \eqref{eq:OPT11-object}-\eqref{eq:OPT11-constraint-3}.

\begin{algorithm}[t]
\begin{enumerate}
\item \textbf{Initialize}: $i=0$, $\mathbf{p}^{\left(0\right)}$, $\mathbf{w}^{\left(0\right)}$,
$\mathbf{r}^{\left(0\right)}$, and $\zeta_{1}^{\left(0\right)}$;
\item \textbf{do}
\item ~~~~$i=i+1$;
\item ~~~~$\gamma_{k}^{\prime}=g_{k}^{\prime}\left(\mathbf{p}^{\left(i-1\right)},\mathbf{r}^{\left(i-1\right)}\right)/\sum_{k=1}^{K^{\prime}}g_{k}^{\prime}\left(\mathbf{p}^{\left(i-1\right)},\mathbf{r}^{\left(i-1\right)}\right)$;
\item ~~~~Update $\mathbf{p}^{\left(i\right)}$, $\mathbf{w}^{\left(i\right)}$,
$\mathbf{r}^{\left(i\right)}$, $\zeta_{1}^{\left(i\right)}$ by solving
\eqref{eq:OPT13-object}-\eqref{eq:OPT13-constraint-5};
\item \textbf{until} $\left|\zeta_{1}^{\left(i\right)}-\zeta_{1}^{\left(i-1\right)}\right|/\left|\zeta_{1}^{\left(i\right)}\right|<\epsilon$
or $i=i_{\mathrm{max}}$
\item Set $\mathbf{p}^{\star}=\left[\xi_{1}^{\star},\xi_{2}^{\star},\ldots,\xi_{N}^{\star}\right]^{T}=\mathbf{p}^{\left(i\right)}$;
\item Set $\mathbf{w}^{\star}=\mathbf{w}^{\left(i\right)}$;
\item Set $\mathbf{s}_{n}^{\star}=\xi_{n}^{\star}\frac{\left(\mathbf{w}^{\star H}\mathbf{H}_{n}\right)^{H}}{\left\Vert \mathbf{w}^{\star H}\mathbf{H}_{n}\right\Vert }$,
$\forall n$;
\end{enumerate}
\caption{\label{alg:RF-Combining-Optimization-Analog}Joint Waveform and Beamforming
Optimization with RF combining using Analog Receive Beamforming.}
\end{algorithm}

\section{Performance Evaluations}

We consider two types of performance evaluations. The first one is
based on the simplified and tractable nonlinear rectenna model \eqref{eq:VoutRectennaModel}
as introduced in Section III, while the second one relies on an accurate
and realistic rectenna modeling in the circuit simulation solver Advanced
Design System (ADS).

\subsection{Nonlinear Model-Based Performance Evaluations}

The first type of evaluations consider the output DC power averaged
over channel realizations of the multi-sine MIMO WPT system with DC
and RF combinings. The evaluation is performed in a scenario representative
of a WiFi-like environment at a center frequency of 5.18 GHz with
a 36 dBm transmit power and 66 dB path loss in a large open space
environment with a NLOS channel power delay profile with 18 taps obtained
from model B \cite{Hiperlan2Channel}. Taps are modeled as i.i.d.
circularly symmetric complex Gaussian random variables, each with
an average power $\rho_{l}$. The multipath response is normalized
such that $\sum_{l=1}^{18}\rho_{l}=1$. With one transmit antenna,
this leads to an average received power of \textminus 30 dBm (1 $\mu\mathrm{W}$).
Equivalently, this system model can be viewed as a transmission over
the aforementioned normalized multipath channel with an average transmit
power fixed to \textminus 30 dBm. The $N$ sinewaves are centered
around 5.18 GHz with a uniform frequency gap $\Delta_{\omega}=2\pi\Delta_{f}$
where $\Delta_{f}=B/N$ and the bandwidth $B=10$ MHz. For the parameters
of the rectifier, we assume $v_{t}=25.86$ mV, $n_{i}=1.05$, and
$R_{L}=10\;\mathrm{k}\Omega$.

For DC combining, we evaluate the adaptive optimized (OPT) waveform
and transmit beamforming using Algorithm \ref{alg:DC-Combining-Optimization.}
versus a benchmark: a waveform and transmit beamforming design based
on adaptive single sinewave (ASS) strategy \cite{2016_TSP_WPT_Bruno_Waveform}.
Specifically, we obtain $\sigma_{n}$, which is the maximum singular
value of $\mathbf{H}_{n}$, by using SVD $\mathbf{H}_{n}=\mathbf{U}_{n}\mathbf{\Sigma}_{n}\mathbf{V}_{n}^{H}$,
and then find the strongest channel $\bar{n}=\arg\max_{n}\sigma_{n}$.
Therefore, the transmit beamformer $\mathbf{s}_{n}^{\mathrm{ASS}}$
is given by
\begin{equation}
\mathbf{s}_{n}^{\mathrm{ASS}}=\begin{cases}
\sqrt{2P}\left[\mathbf{V}_{n}\right]_{\mathrm{max}} & ,n=\bar{n}\\
\boldsymbol{0} & ,n\neq\bar{n}
\end{cases}\label{eq:TX - ASS}
\end{equation}
where $\left[\mathbf{V}_{n}\right]_{\mathrm{max}}$ refers to the
vector in $\mathbf{V}_{n}$ corresponding to the maximum singular
value of $\mathbf{H}_{n}$. Such transmit beamformer is optimal for
maximizing the output DC power when the linear rectenna model (having
a constant RF-to-DC conversion efficiency) is considered \cite{2013_TWC_SWIPT_RZhang}.

For RF combining, we evaluate the adaptive optimized (OPT) waveform
and transmit beamforming with the general receive beamforming using
Algorithm \ref{alg:RF-Combining-Optimization-General} and the adaptive
optimized waveform and transmit beamforming with the analog receive
beamforming (ABF) using Algorithm \ref{alg:RF-Combining-Optimization-Analog}.
For comparison, we also consider a benchmark: a waveform and transmit
beamforming with the general receive beamforming based on ASS strategy.
Specifically, we still use SVD for the channel matrix and find the
strongest channel $\bar{n}=\arg\max_{n}\sigma_{n}$. Therefore, the
transmit beamformer $\mathbf{s}_{n}^{\mathrm{ASS}}$ is given by \eqref{eq:TX - ASS}
and the general receive beamformer is given by \eqref{eq:optimal wn}.
Interestingly, as shown in \cite{ShanpuShen2020_TWC_MIMO_WPT_SingleTone},
RF combining has the same performance as DC combining when the linear
rectenna model is considered. Namely, ASS based RF combining is also
optimal considering the linear rectenna model.

\begin{figure*}[t]
\begin{centering}
\includegraphics[scale=0.3]{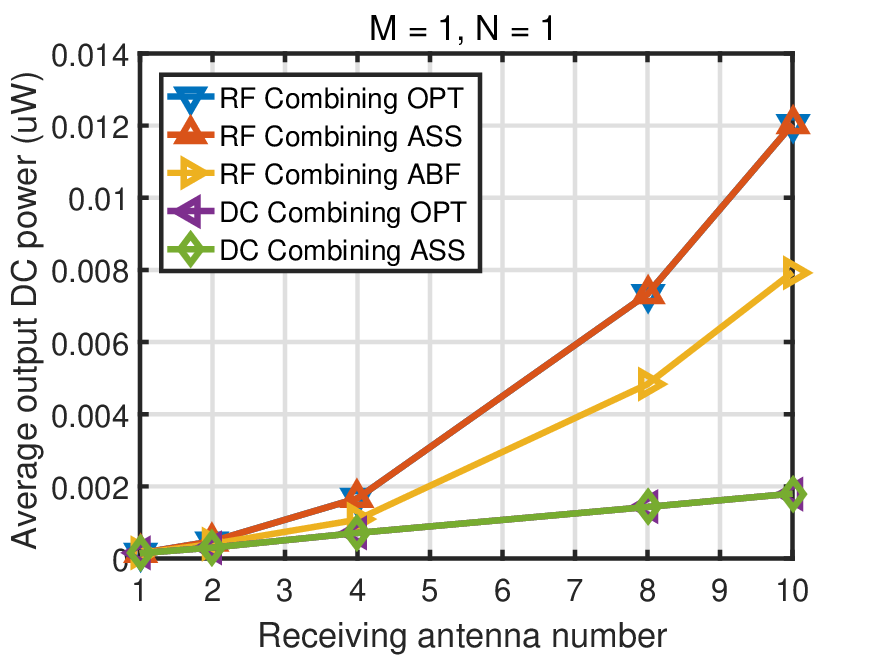}\includegraphics[scale=0.3]{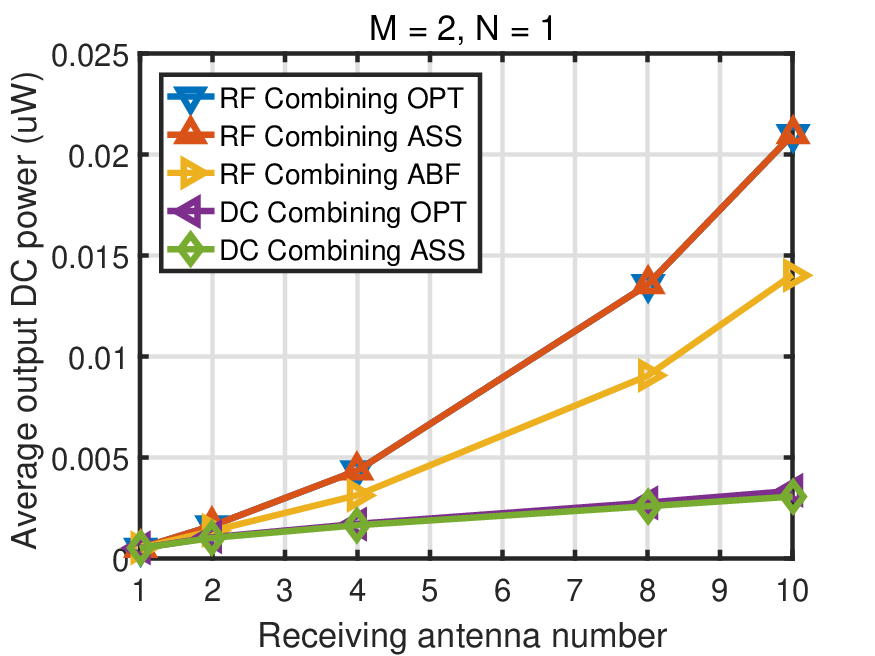}\includegraphics[scale=0.3]{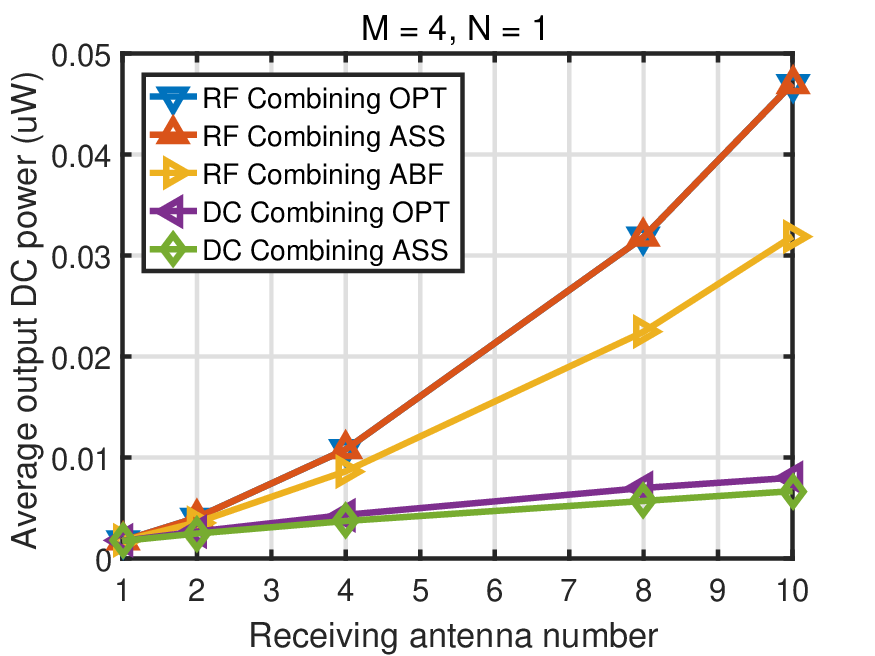}\includegraphics[scale=0.3]{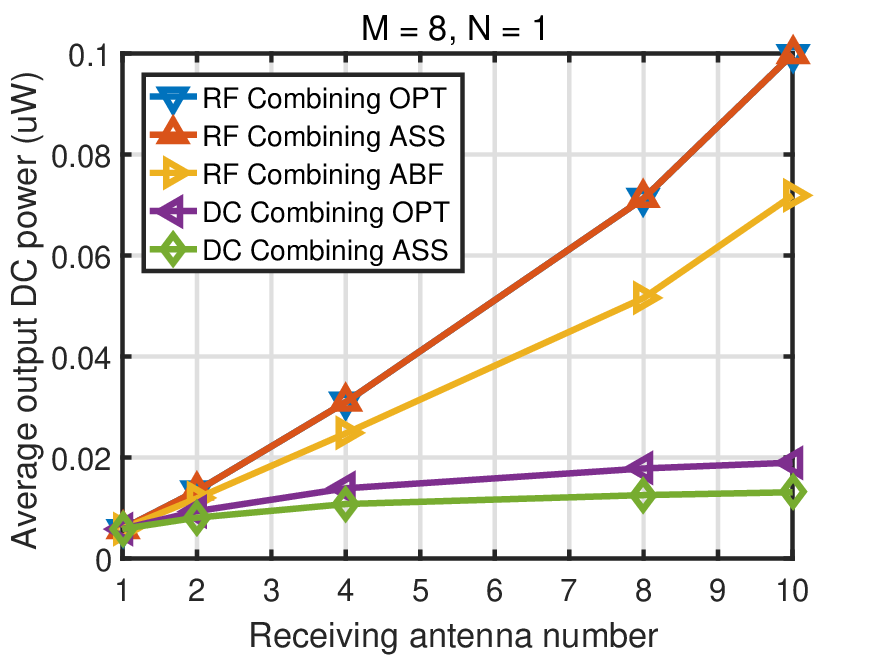}
\par\end{centering}
\begin{centering}
\includegraphics[scale=0.3]{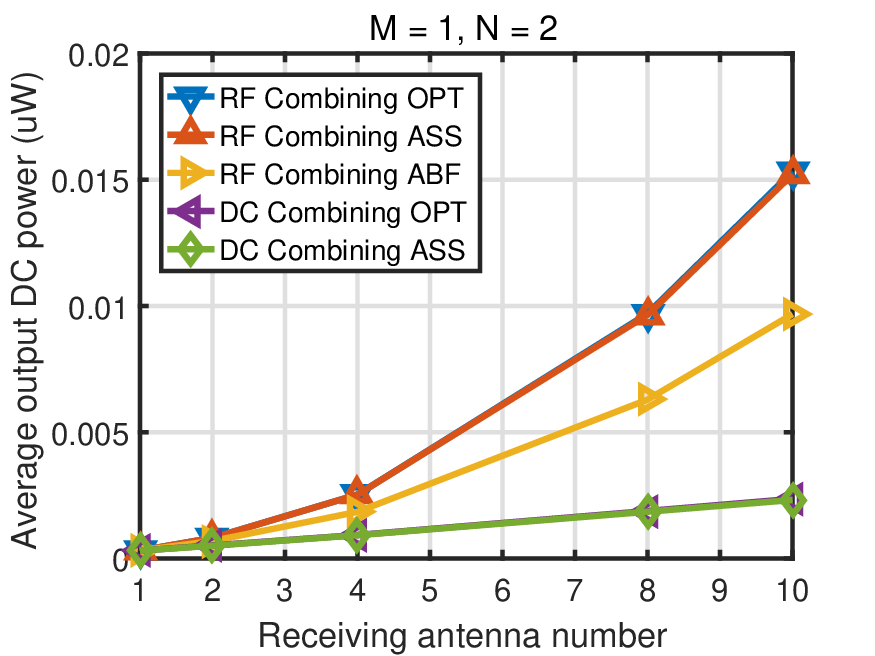}\includegraphics[scale=0.3]{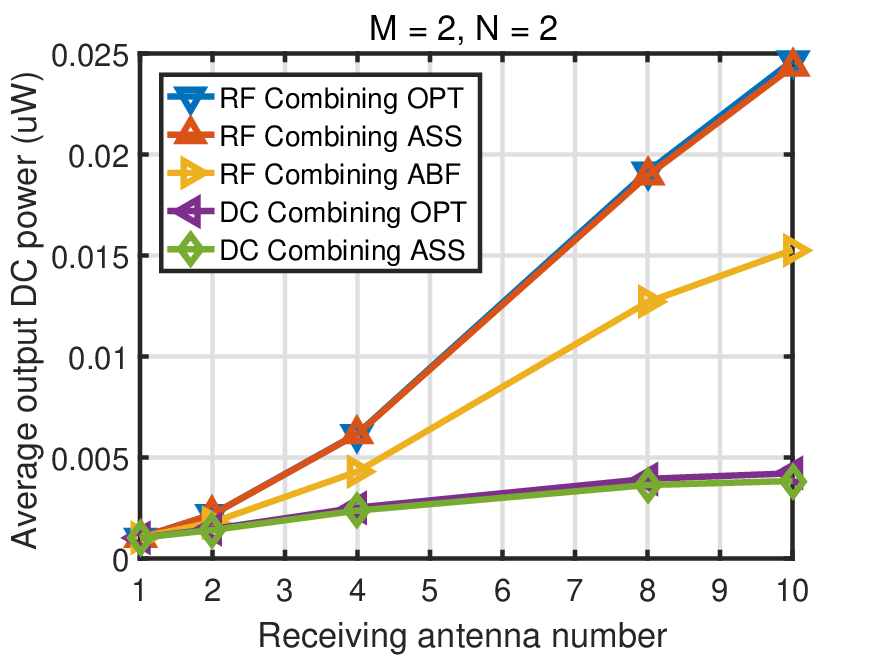}\includegraphics[scale=0.3]{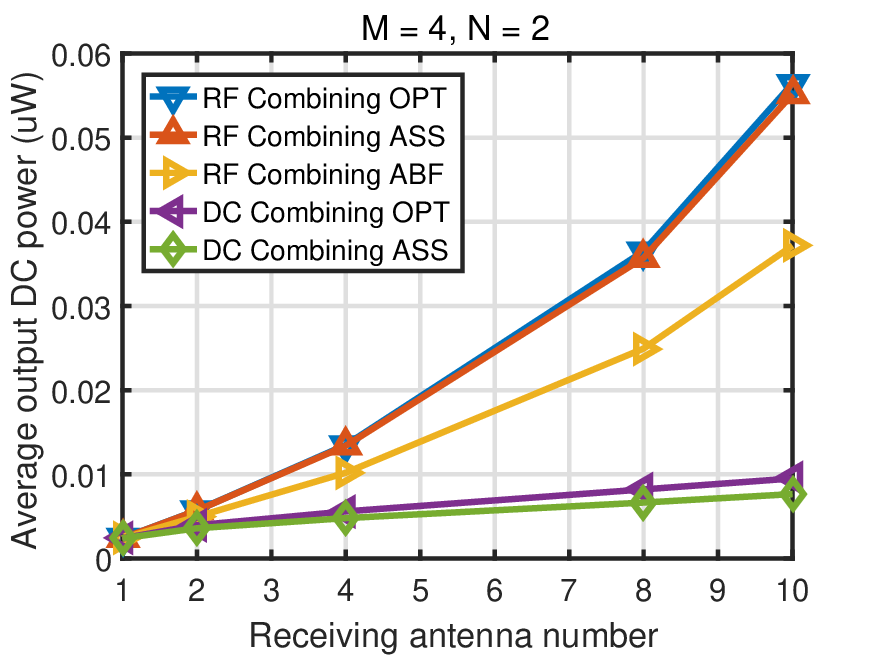}\includegraphics[scale=0.3]{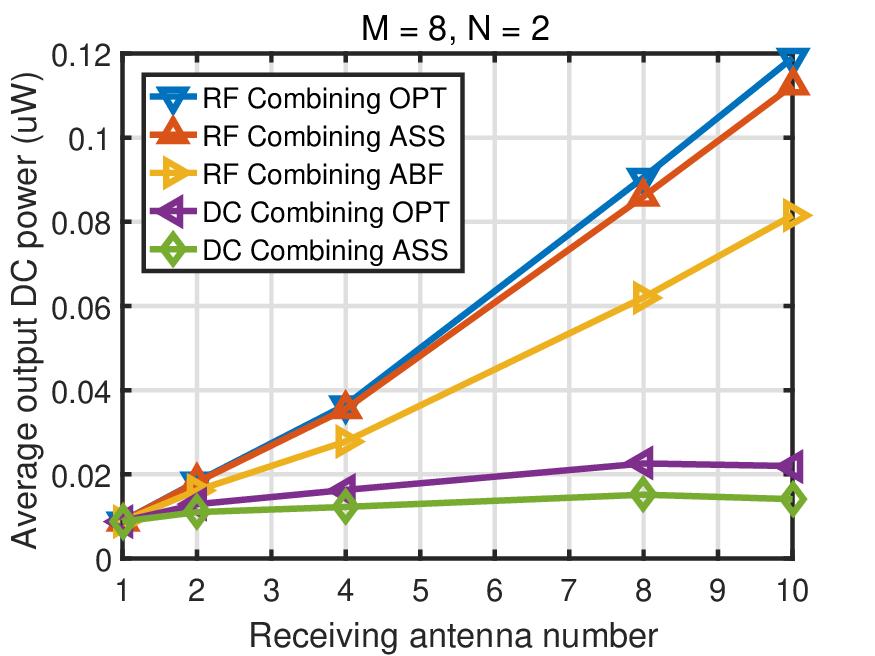}
\par\end{centering}
\begin{centering}
\includegraphics[scale=0.3]{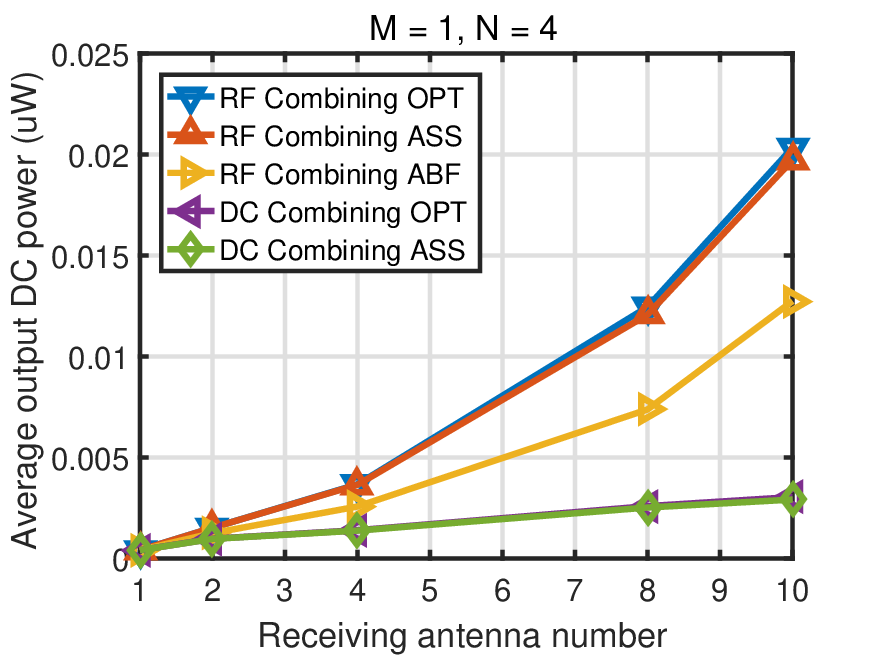}\includegraphics[scale=0.3]{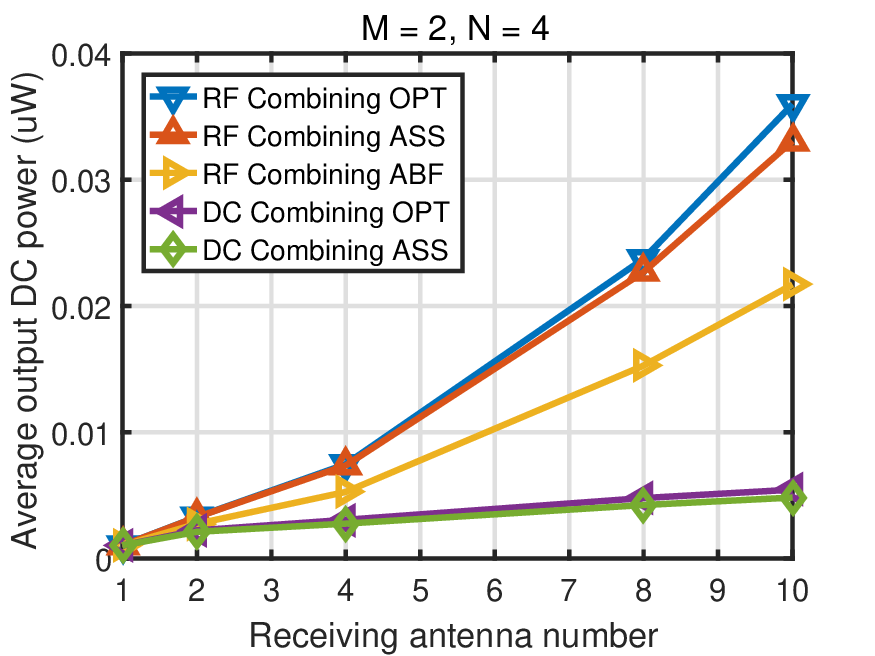}\includegraphics[scale=0.3]{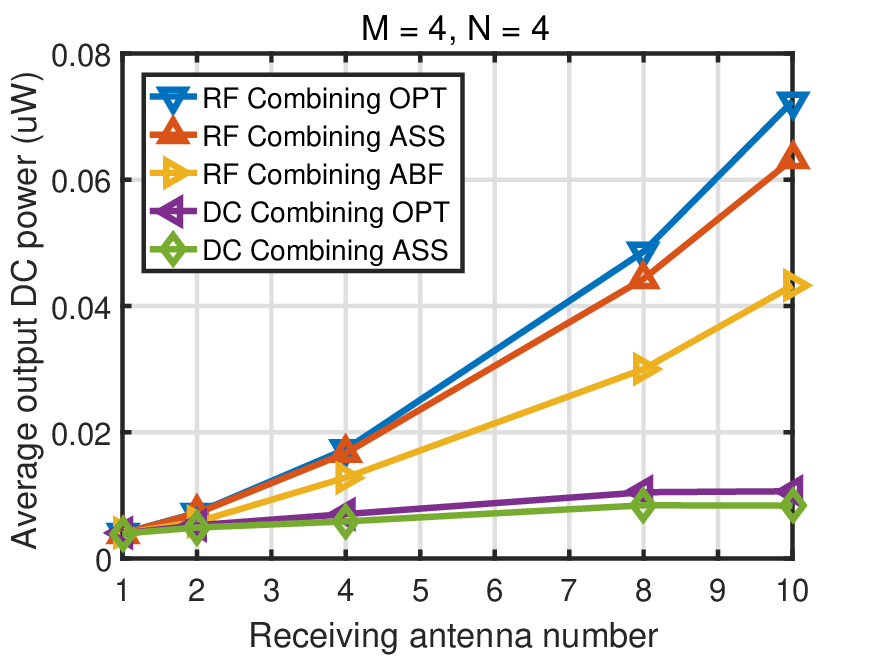}\includegraphics[scale=0.3]{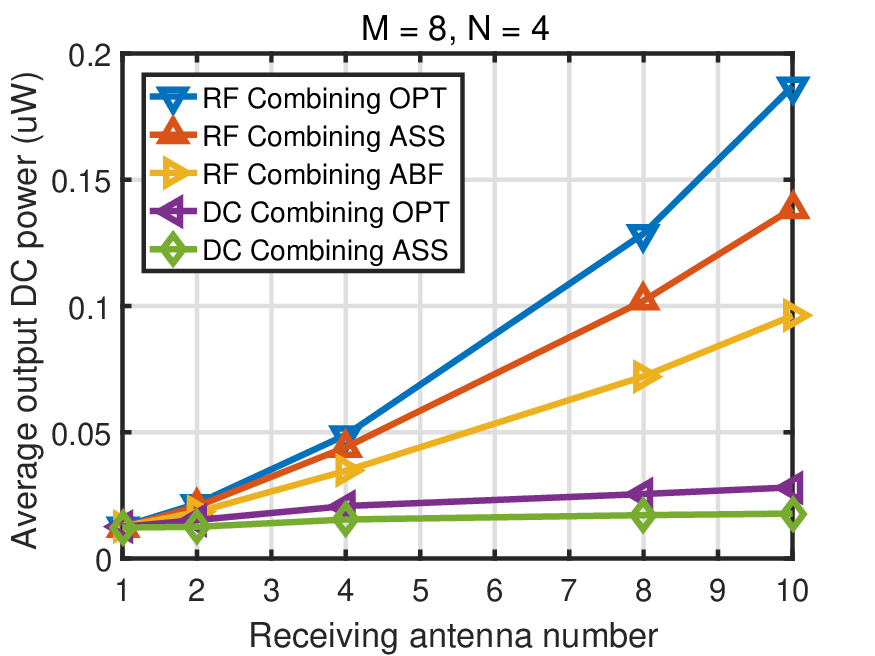}
\par\end{centering}
\begin{centering}
\includegraphics[scale=0.3]{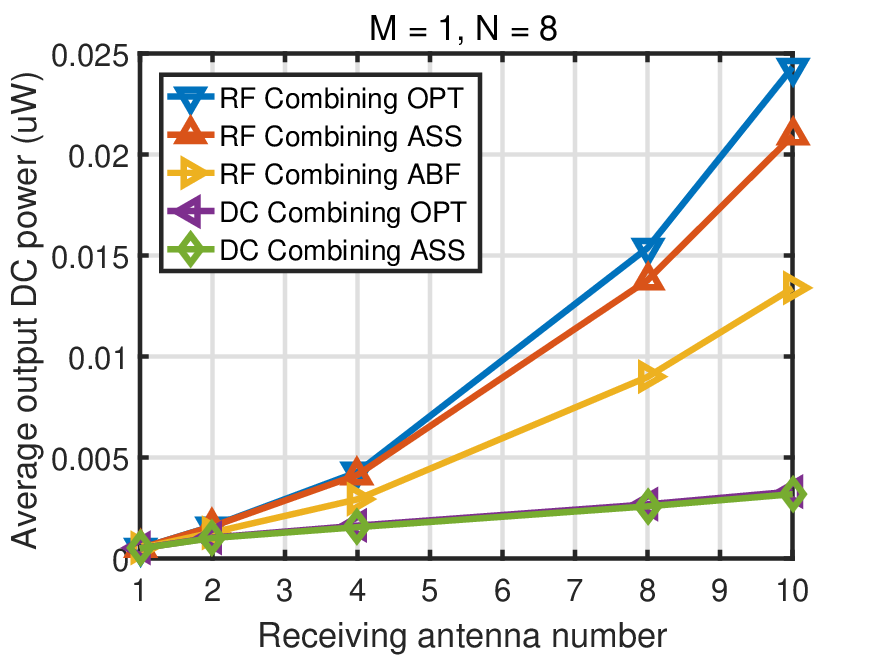}\includegraphics[scale=0.3]{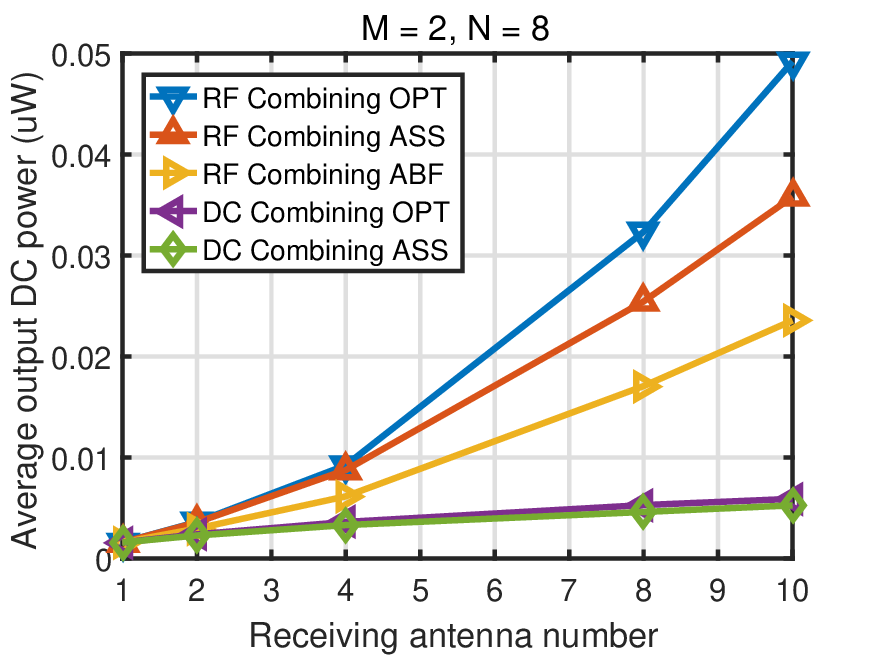}\includegraphics[scale=0.3]{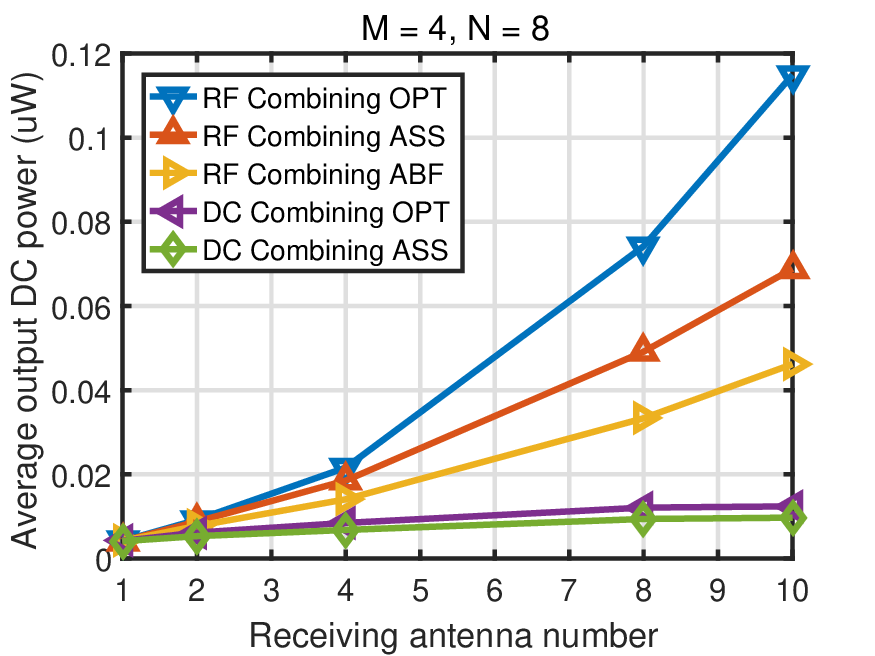}\includegraphics[scale=0.3]{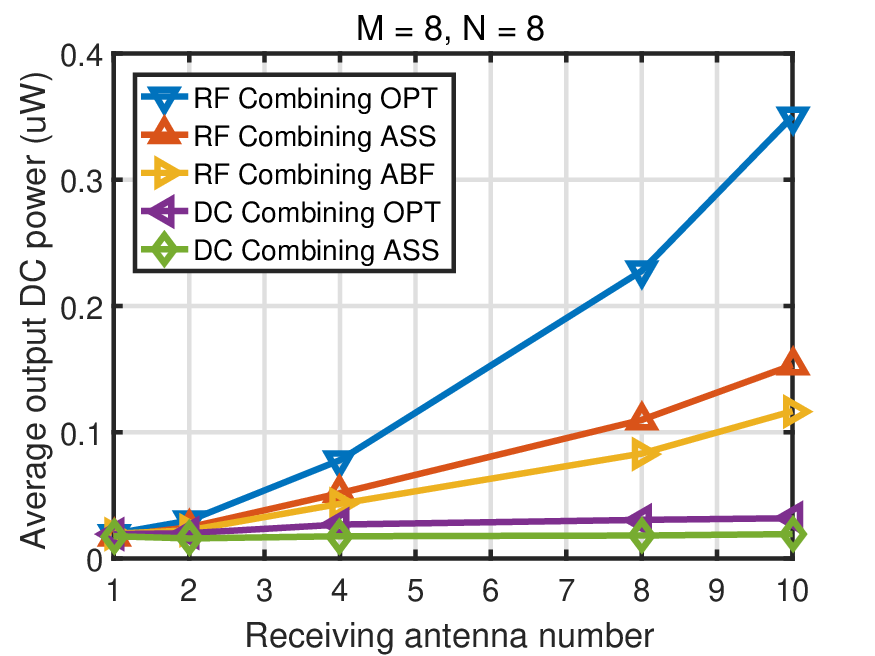}
\par\end{centering}
\begin{centering}
\includegraphics[scale=0.3]{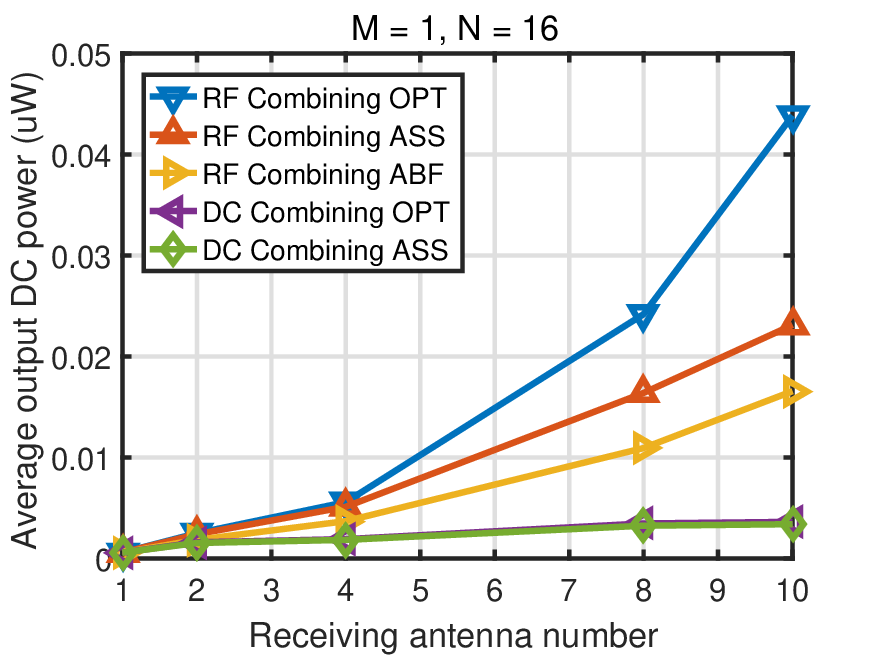}\includegraphics[scale=0.3]{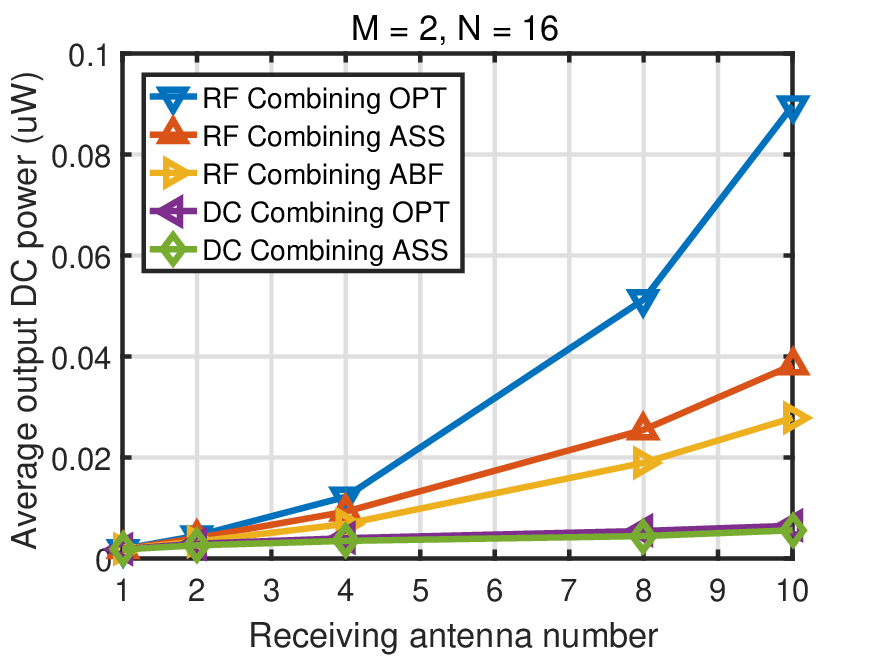}\includegraphics[scale=0.3]{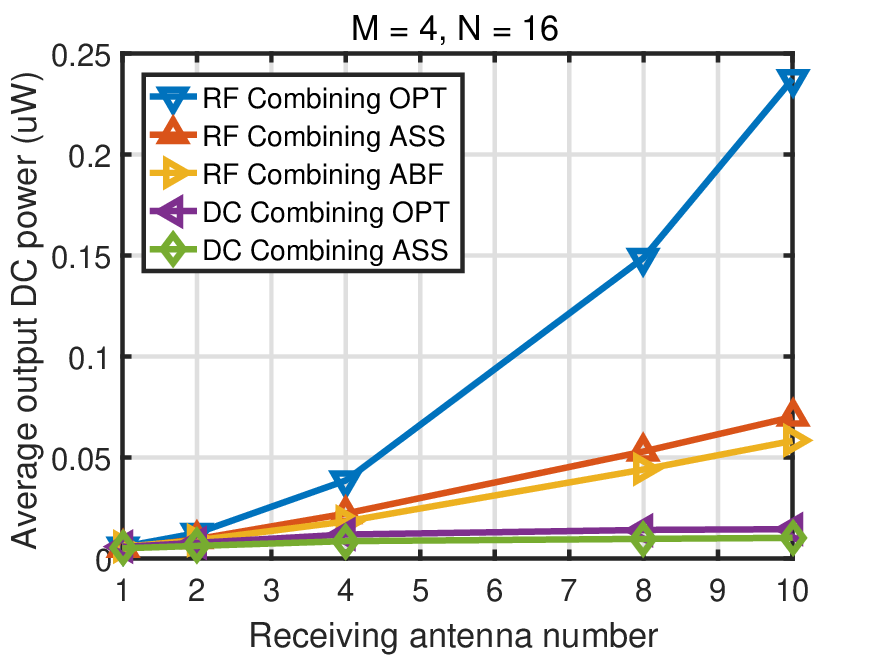}\includegraphics[scale=0.3]{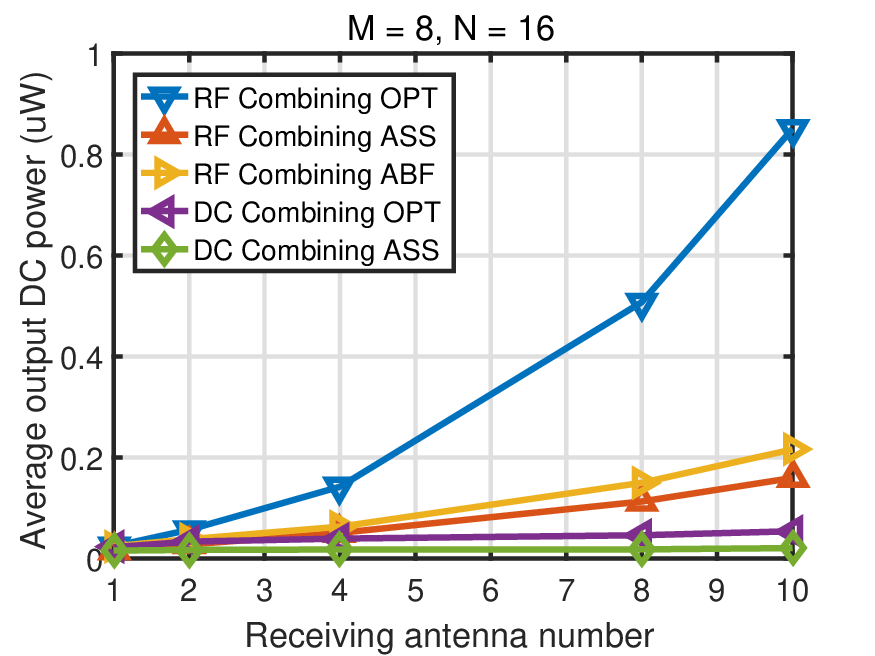}
\par\end{centering}
\centering{}\caption{\label{fig:Average-output-DC-Rectenna Model}Average output DC power
versus the number of receive antennas $Q$ for different numbers of
transmit antennas $M$ and different numbers of frequencies $N$ based
on the nonlinear rectenna model.}
\end{figure*}

Fig. \ref{fig:Average-output-DC-Rectenna Model} displays the output
DC power averaged over channel realizations versus the number of receive
antennas $Q$ for different numbers of transmit antennas $M$ and
different numbers of frequencies $N$. We make the following observations.

\textit{First}, the output DC power increases with the number of transmit
antennas and also the number of receive antennas for the five waveform
and beamforming designs using DC or RF combinings, showing that the
output DC power can be effectively increased by leveraging the transmit
or receive beamforming gain.

\textit{Second}, the output DC power increases with the number of
frequencies for the five waveform and beamforming designs using DC
or RF combinings. Compared with the beamforming only design ($N=1$),
the jointly waveform and beamforming design ($N>1$) can provide a
higher output DC power, showing the benefit of jointly optimizing
the waveform and beamforming over beamforming only. For the ASS based
waveform and beamforming, the increase of the output DC power comes
from the frequency diversity gain in the frequency selective channel.
For the other designs optimized with the nonlinear rectenna model,
the increase not only comes from the frequency diversity gain but
also the rectenna nonlinearity. Overall, it shows that the output
DC power can be effectively increased by leveraging the frequency
diversity gain or rectenna nonlinearity through using multi-sine waveform.

\textit{Third}, for DC combining, the OPT waveform and beamforming
achieves higher output DC power than the ASS based waveform and beamforming.
This is because the OPT waveform and beamforming leverages the rectenna
nonlinearity while the ASS based waveform and beamforming (which is
optimized for the inaccurate linear rectenna model) ignores the rectenna
nonlinearity. Recall that the rectenna nonlinearity is responsible
for the RF-to-DC conversion efficiency to be a function of the input
waveform \cite{2017_TOC_WPT_YZeng_Bruno_RZhang}, \cite{2016_TSP_WPT_Bruno_Waveform}.
Ignoring the nonlinearity results in assuming the RF-to-DC conversion
efficiency to be constant, which is again demonstrated in this paper
to be inaccurate and to lead to suboptimal designs.

\textit{Fourth}, for RF combining, the OPT general receive beamforming
achieves higher output DC power than the ASS based general receive
beamforming. Again, this is because the OPT general receive beamforming
leverages the rectenna nonlinearity while the ASS based general receiving
beamforming ignores the nonlinearity. In addition, the general receive
beamforming outperforms the analog receive beamforming. This is because
the constraints of the analog receive beamforming \eqref{eq:ABF constraint-1}
and \eqref{eq:ABF constraint-2} is more restrictive than that of
the general receive beamforming.

\textit{Fifth}, RF combining outperforms DC combining, especially
when the number of receive antennas goes large. This is because RF
combining leverages the rectenna nonlinearity more efficiently than
DC combining. Indeed, the rectenna has a nonlinear characteristics
such that the RF-to-DC conversion efficiency increases with the input
RF power. RF combining inputs the combined RF signal (having higher
RF power) into a single rectifier while DC combining inputs each RF
signal to each rectifier. Therefore, RF combining has a higher RF-to-DC
conversion efficiency and output DC power. This observation was made
in \cite{ShanpuShen2020_TWC_MIMO_WPT_SingleTone} and is shown here
to also hold in the presence of more complex input waveform.

\begin{figure}[t]
\begin{centering}
\includegraphics[scale=0.3]{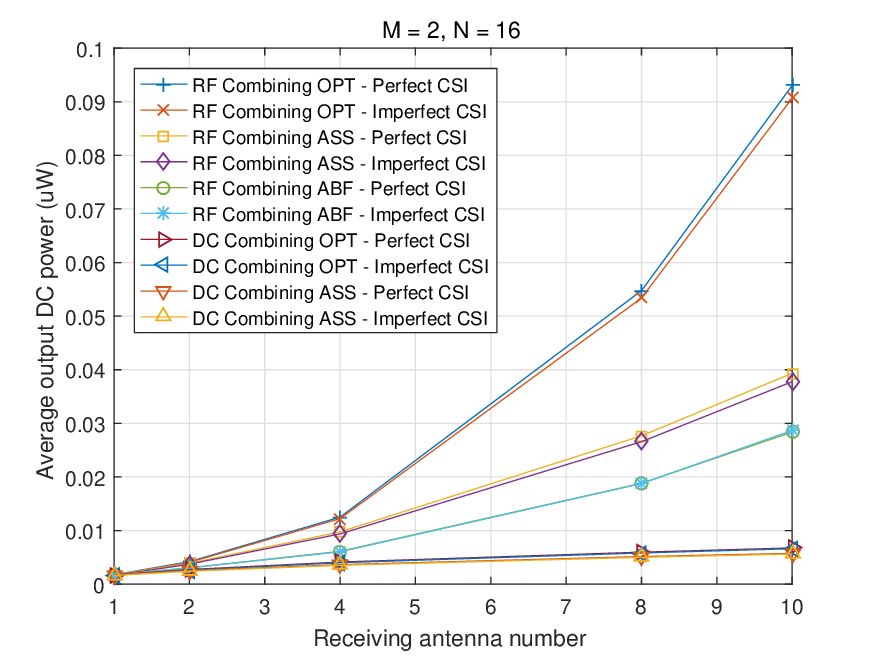}\includegraphics[scale=0.3]{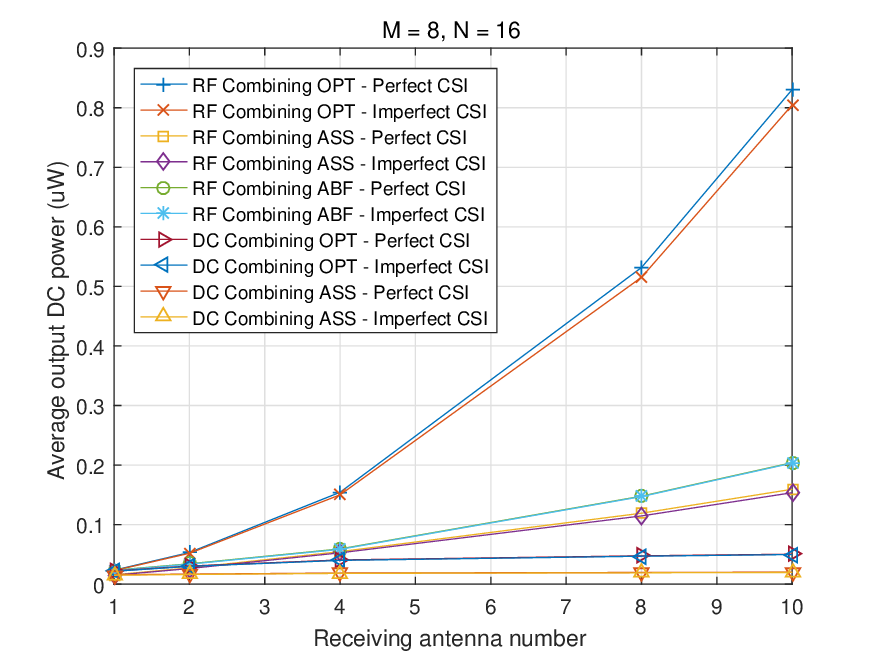}
\par\end{centering}
\centering{}\caption{\label{fig:imperfect_CSI}Average output DC power with perfect CSI
and imperfect CSI based on the nonlinear rectenna model.}
\end{figure}

It is worth noting that the proposed algorithms for DC and RF combinings
are robust to the imperfect CSI case. To show that, we evaluate the
average output DC power for the proposed Algorithms 1, 2, and 3 and
the benchmark algorithms with perfect CSI and imperfect CSI. Particularly,
the imperfect CSI is modeled as 
\begin{equation}
\hat{\mathbf{H}}_{n}=\sqrt{1-\tau^{2}}\mathbf{H}_{n}+\tau\tilde{\mathbf{H}}_{n}
\end{equation}
where $\mathbf{H}_{n}$ denotes the perfect CSI and $\tilde{\mathbf{H}}_{n}$
denotes the estimation error with each entry following i.i.d. complex
Gaussian distribution with zero mean and unit variance. $\tau\in\left[0,1\right]$
indicates the inaccuracy of the CSI and we set $\tau=0.1$ in the
evaluation. The evaluation results with perfect and imperfect CSI
are shown in Fig. \ref{fig:imperfect_CSI}. We can find that the performance
gap between the perfect and imperfect CSI cases are very small for
all the proposed algorithms and benchmarks at different numbers of
antennas and numbers of frequencies, which shows that the proposed
algorithms are robust to the imperfect CSI. Therefore, all the observations
and conclusions drawn from the perfect CSI case still hold for the
imperfect CSI case.

\begin{figure}[t]
\begin{centering}
\includegraphics[scale=0.3]{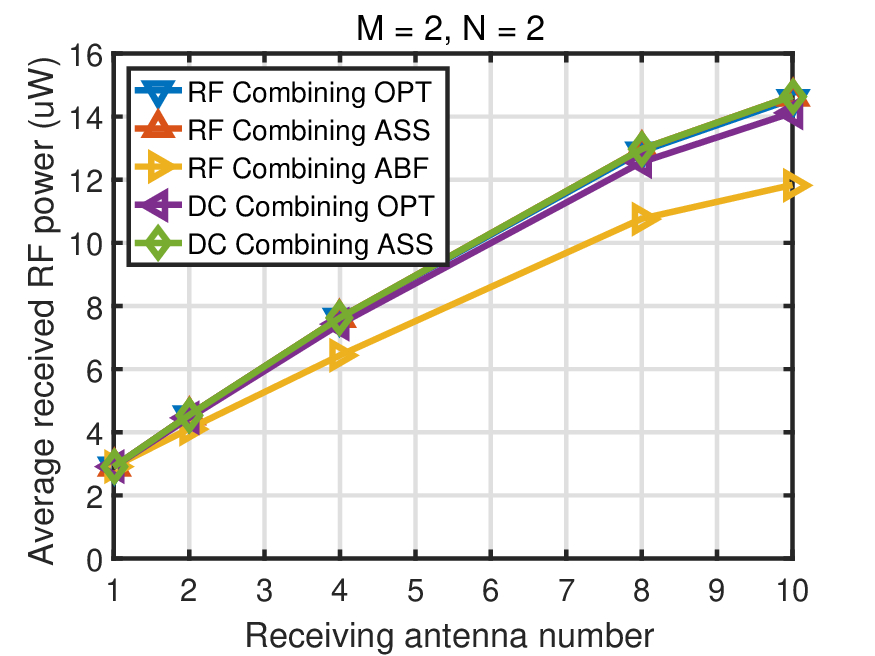}\includegraphics[scale=0.3]{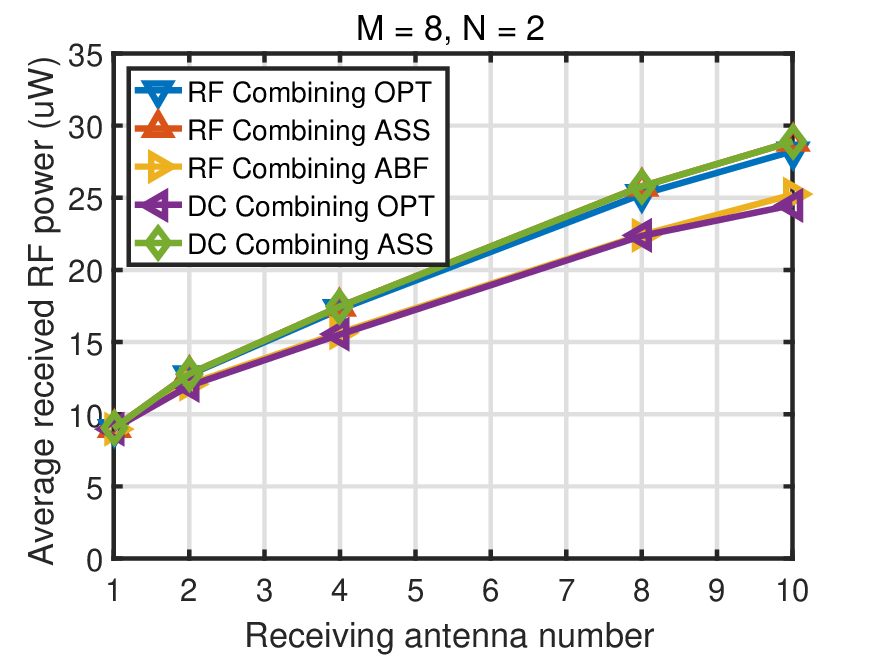}
\par\end{centering}
\centering{}\includegraphics[scale=0.3]{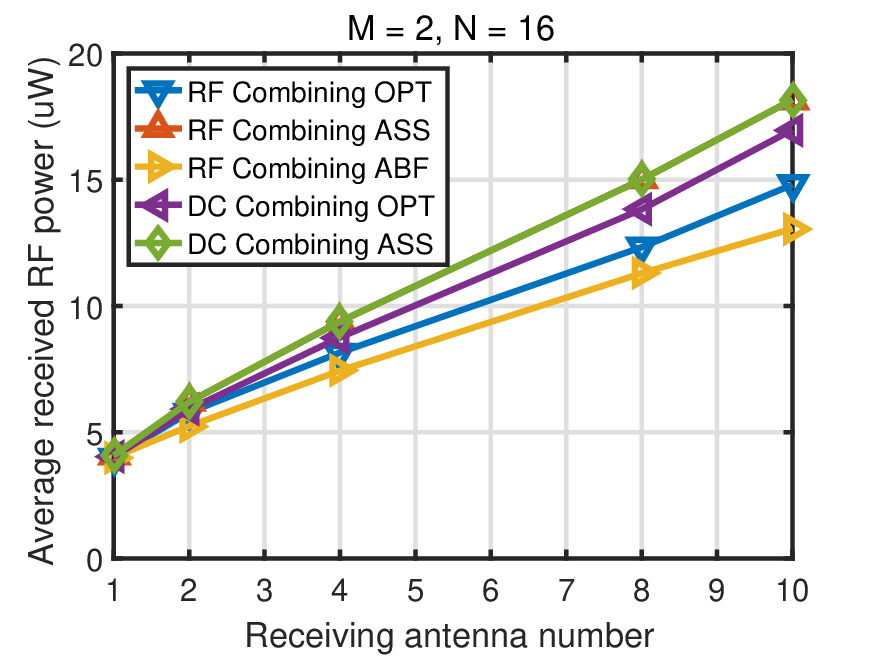}\includegraphics[scale=0.3]{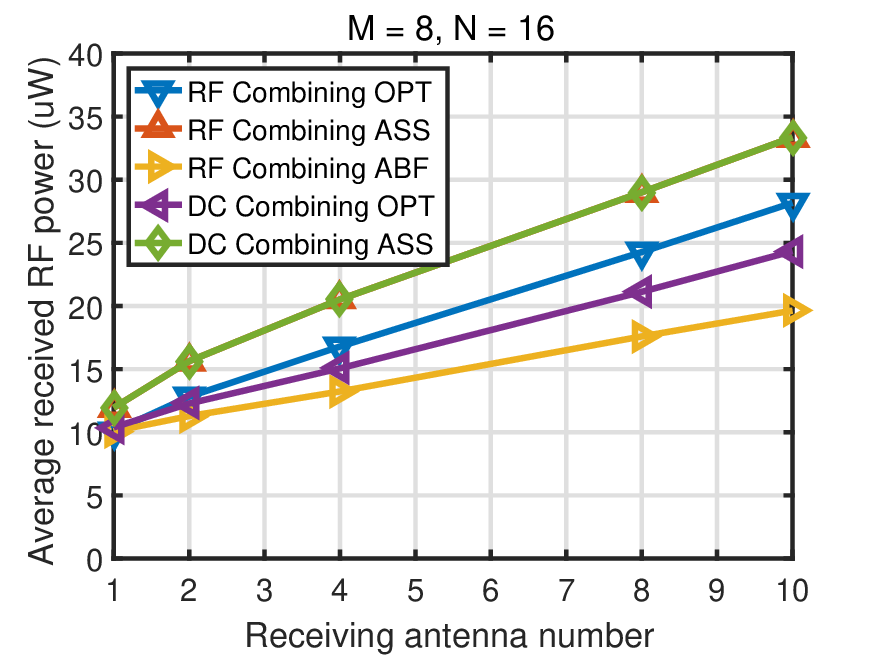}\caption{\label{fig:Average-received-RF}Average received RF power versus the
number of receive antennas $Q$ for different numbers of transmit
antennas $M$ and different numbers of frequencies $N$}
\end{figure}

It is also worthwhile evaluating the received RF power for the different
waveform and beamforming designs to understand the crucial role played
by the rectenna nonlinearity. Recall again that the linear rectenna
model assumes a constant RF-to-DC conversion efficiency, for which
maximizing the output DC power is equivalent to maximizing the received
RF power. Fig. \ref{fig:Average-received-RF} displays the received
RF power averaged over channel realizations versus the number of receive
antennas $Q$ for different numbers of transmit antennas $M$ and
different numbers of frequencies $N$. We make the following observations.
\textit{First}, the ASS based DC combining has the same received RF
power as the ASS based RF combining. This is because they all use
the SVD of channel matrix $\mathbf{H}_{n}$ and choose the strongest
channel. This is also consistent with the conclusion in \cite{ShanpuShen2020_TWC_MIMO_WPT_SingleTone}
that DC combining has the same performance as RF combining if the
linear rectenna model is considered.\textsl{ }\textit{Second}, the
OPT waveform and beamforming based DC combining (or RF combining)
has less received RF power than the ASS based DC combining (or RF
combining). This is because OPT waveform and beamforming based DC
or RF combining is optimized with the nonlinear rectenna model while
the ASS based DC or RF combining is optimal for the linear rectenna
model.\textsl{ }\textit{Third}, the analog receive beamforming has
less received RF power than the general receive beamforming which
is because the constraints of analog receive beamforming \eqref{eq:ABF constraint-1}
and \eqref{eq:ABF constraint-2} is more restrictive.

Based on the above observations from Fig. \ref{fig:Average-output-DC-Rectenna Model}
and Fig. \ref{fig:Average-received-RF}, we can find that maximizing
the received RF power does not mean maximizing the output DC power
due to the rectenna nonlinearity. Therefore, we should consider and
leverage the rectenna nonlinearity in WPT to increase the output DC
power. This behavior has been extensively emphasized in \cite{2016_TSP_WPT_Bruno_Waveform},
\cite{2018_MM_WPT_Bruno_1GWPT}, \cite{2019_JSAC_WIPT_Bruno_RZhang_RSchober_DIKim_HVPoor}
but finds further consequences in the multi-sine MIMO WPT. To conclude,
the rectenna nonlinearity can be leveraged by using multi-sine waveform
with DC and RF combinings while RF combining can leverage the rectenna
nonlinearity more efficiently.

\subsection{Accurate and Realistic Performance Evaluations}

The second type of evaluations uses the circuit simulation solver
ADS to accurately model the rectenna so as to validate the joint waveform
and beamforming optimization with the DC and RF combinings and the
rectenna nonlinearity model.

\begin{figure}[t]
\begin{centering}
\includegraphics[width=7cm]{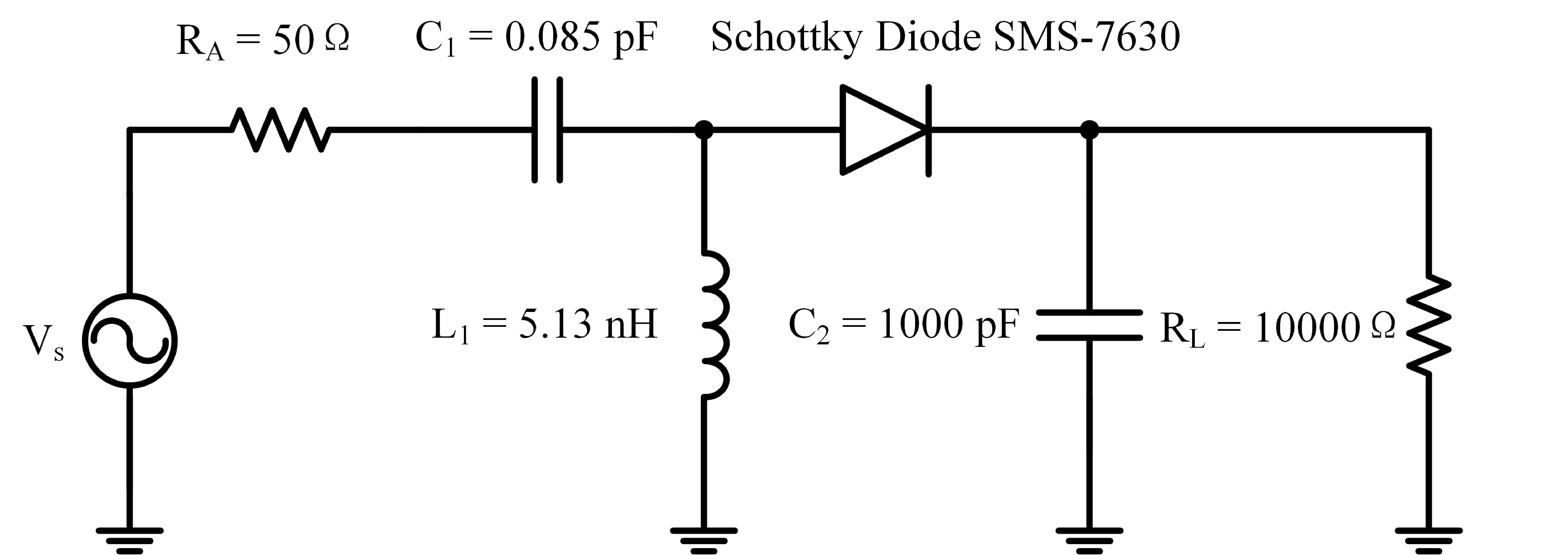}
\par\end{centering}
\caption{\label{fig:Rectenna-ADS}Rectenna with a single diode and L-matching
network used for circuit evaluation in ADS.}
\end{figure}

To that end, in DC combining, for a given channel realization, we
first optimize the waveform and beamforming in Matlab so that we can
find the RF signal received by each receive antenna. Then, in ADS,
we input the RF signal received by each receive antenna to a realistic
rectifier as shown in Fig. \ref{fig:Rectenna-ADS}. Hence $Q$ rectifiers
as shown in Fig. \ref{fig:Rectenna-ADS} are used. The output DC power
for each rectifier will be solved by ADS so that the total output
DC power can be computed. In RF combining, we compute the output DC
power in a similar way but only one rectifier as shown in Fig. \ref{fig:Rectenna-ADS}
is used to rectify the combined RF signal in ADS. The rectenna circuit
contains a voltage source, an antenna impedance, an L-matching network,
a Schottky diode SMS-7630, a capacitor as low-pass filter, and a load
resistor. The L-matching network is used to guarantee a good matching
between the rectifier and the antenna and to minimize the impedance
mismatch due to variations in frequency and input RF power level.
With the SPICE model of SMS-7630, the values of the capacitor $C_{1}$
and the inductor $L_{1}$ in the matching network are optimized in
ADS to achieve a good impedance matching. The output capacitor is
chosen as $C_{2}=1000$ pF and the load resistor is chosen as $R_{L}=10\;\mathrm{k}\Omega$.

\begin{figure*}[t]
\begin{centering}
\includegraphics[scale=0.3]{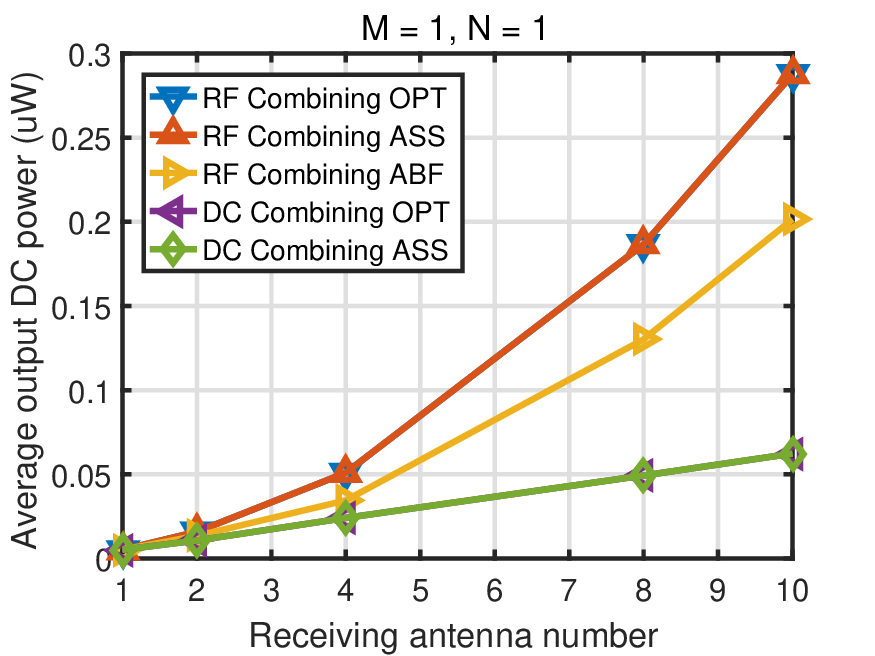}\includegraphics[scale=0.3]{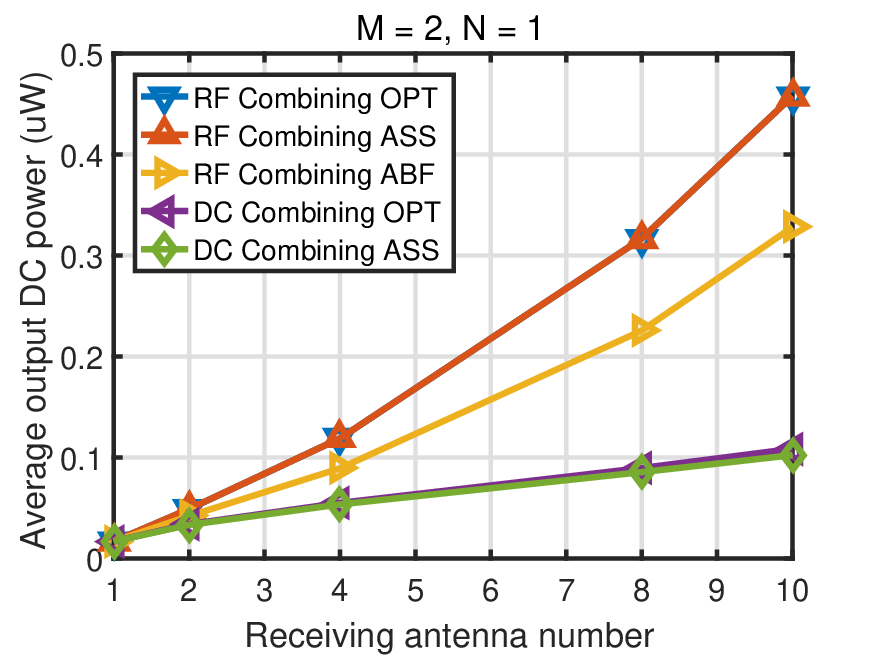}\includegraphics[scale=0.3]{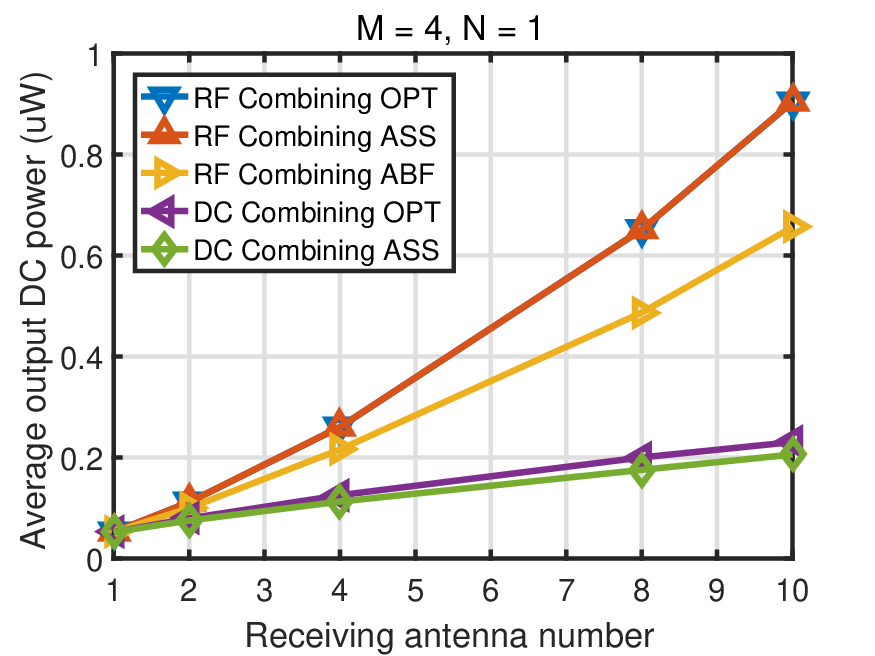}\includegraphics[scale=0.3]{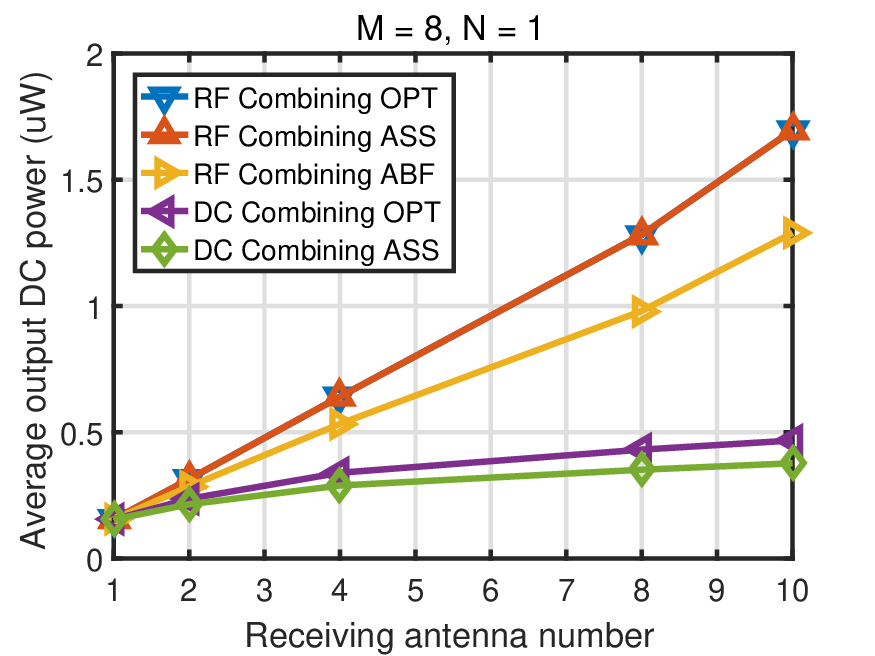}
\par\end{centering}
\begin{centering}
\includegraphics[scale=0.3]{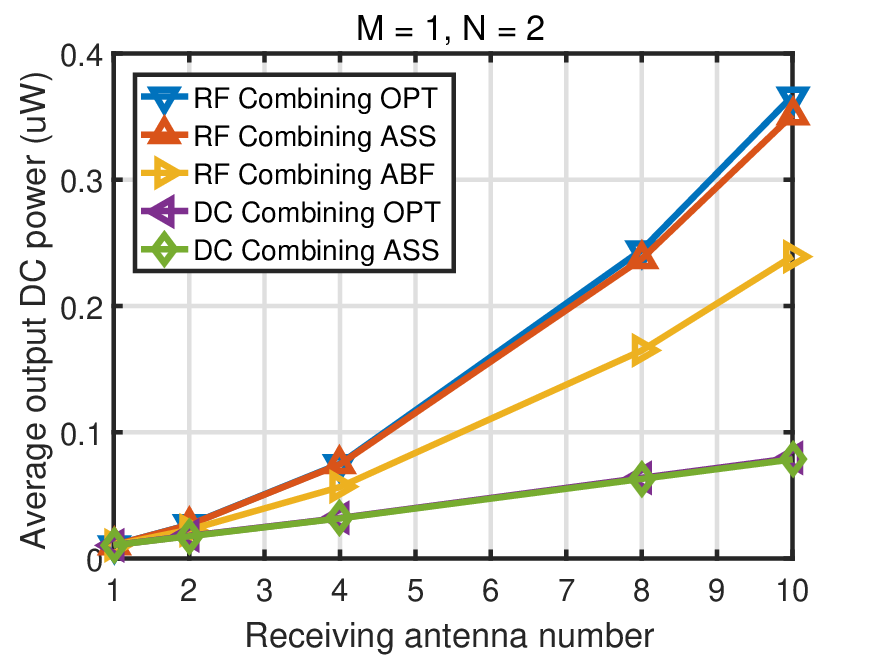}\includegraphics[scale=0.3]{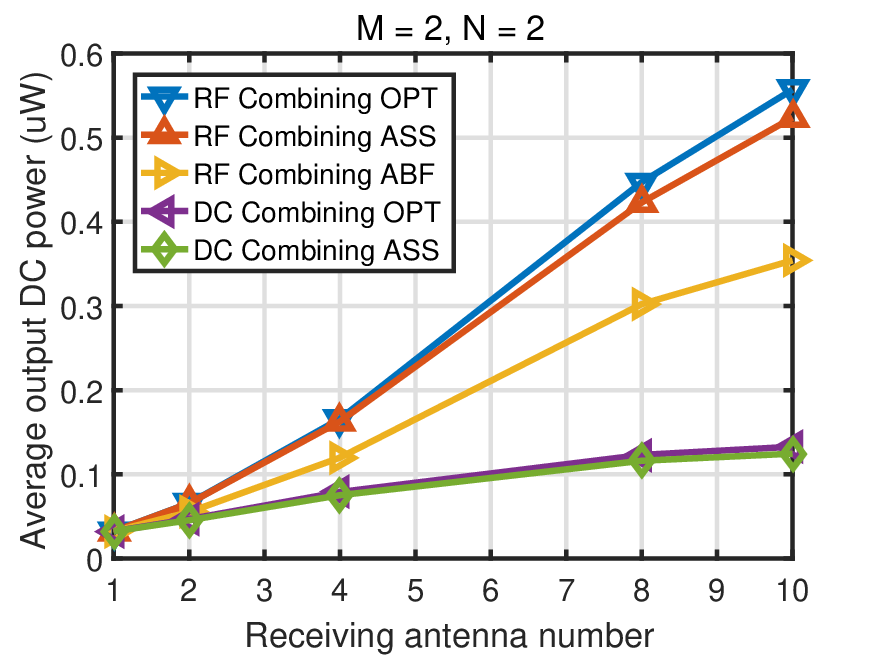}\includegraphics[scale=0.3]{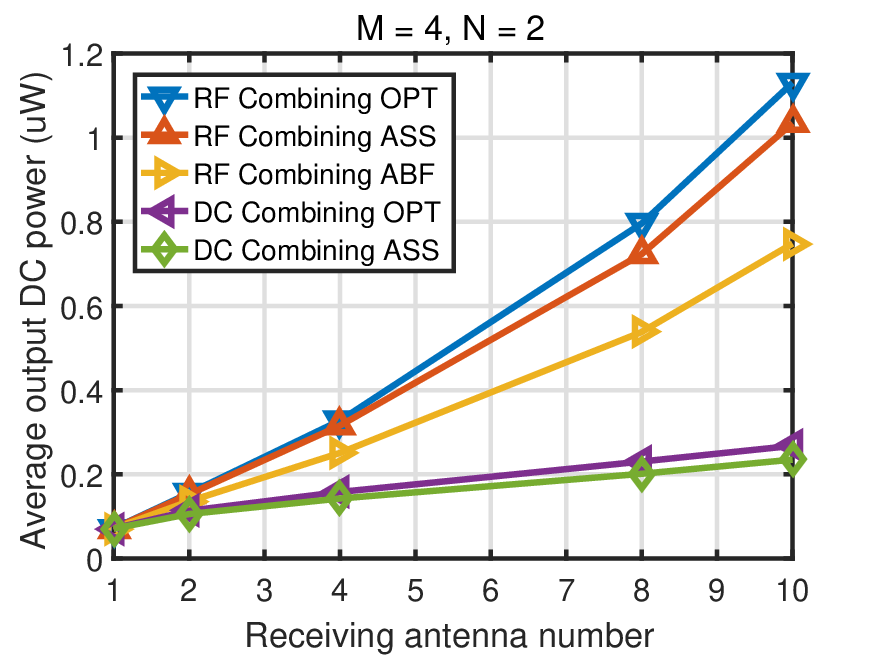}\includegraphics[scale=0.3]{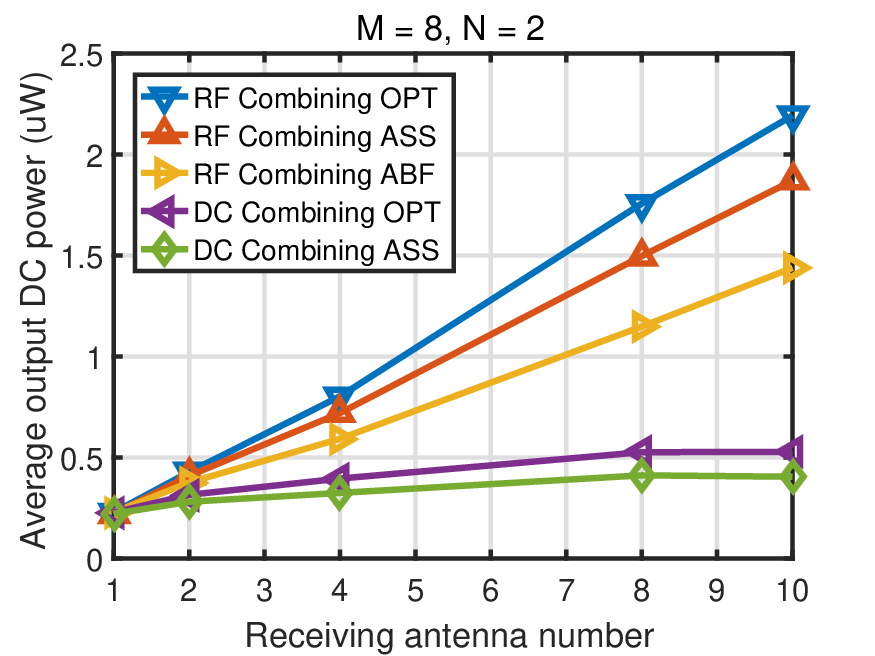}
\par\end{centering}
\begin{centering}
\includegraphics[scale=0.3]{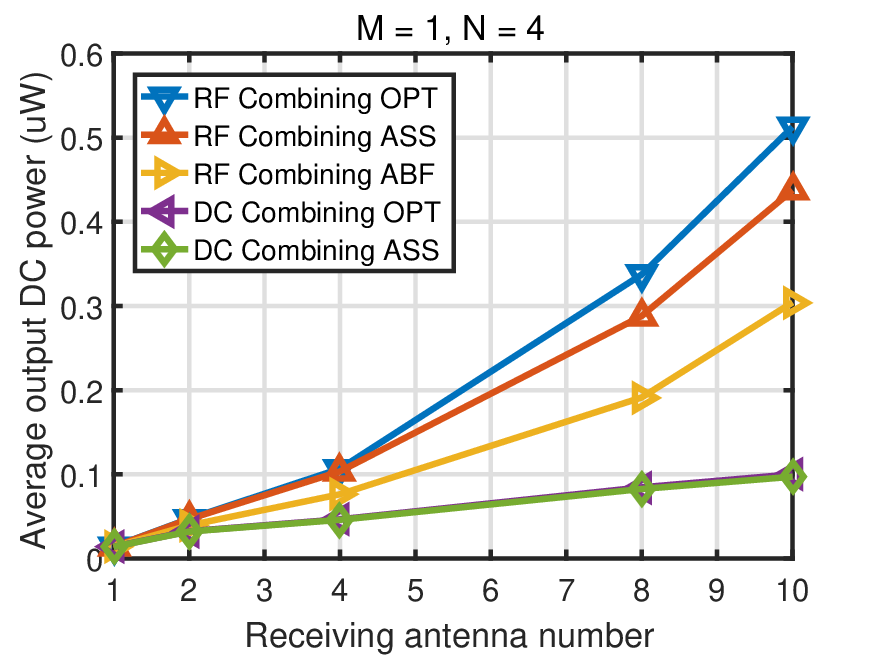}\includegraphics[scale=0.3]{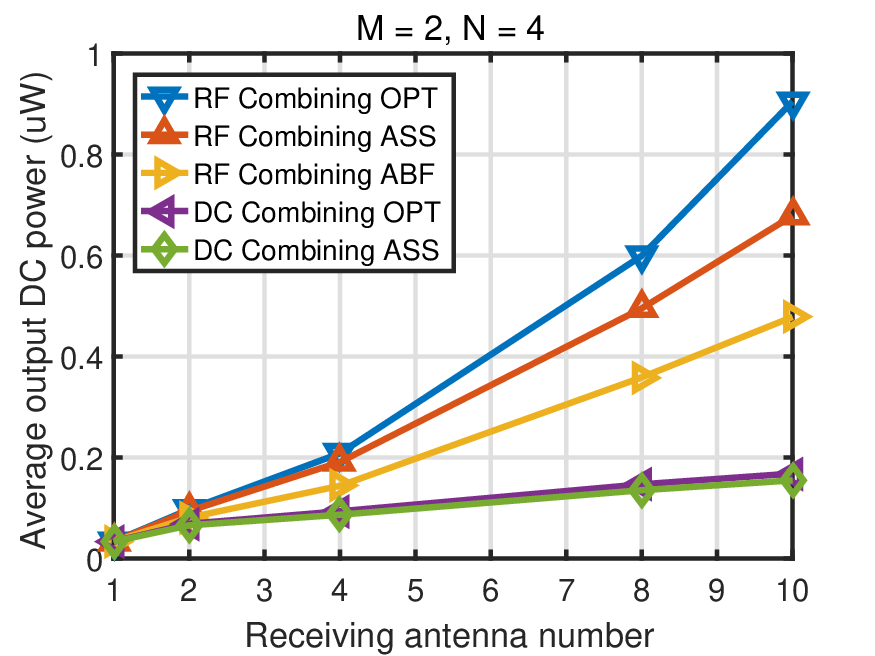}\includegraphics[scale=0.3]{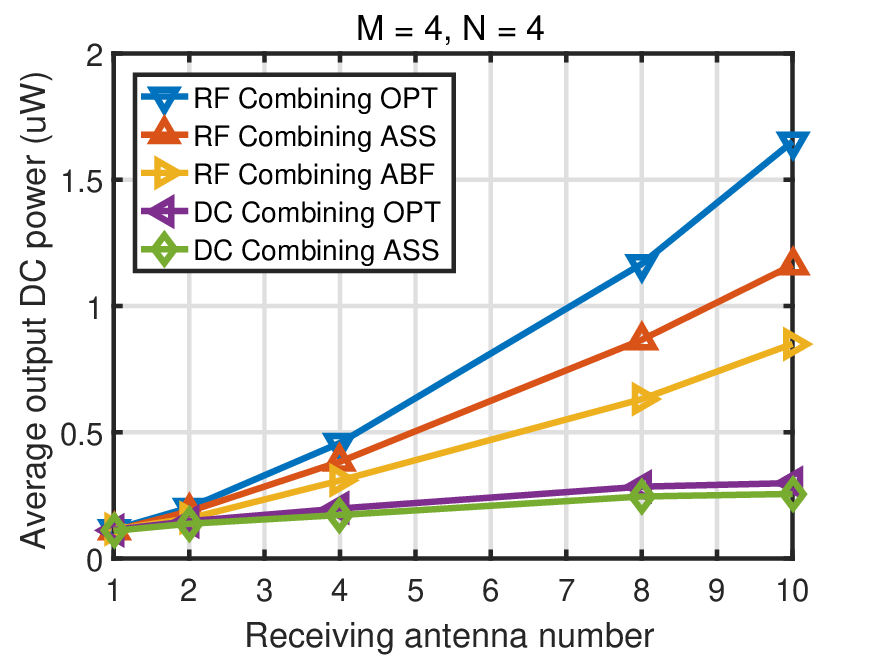}\includegraphics[scale=0.3]{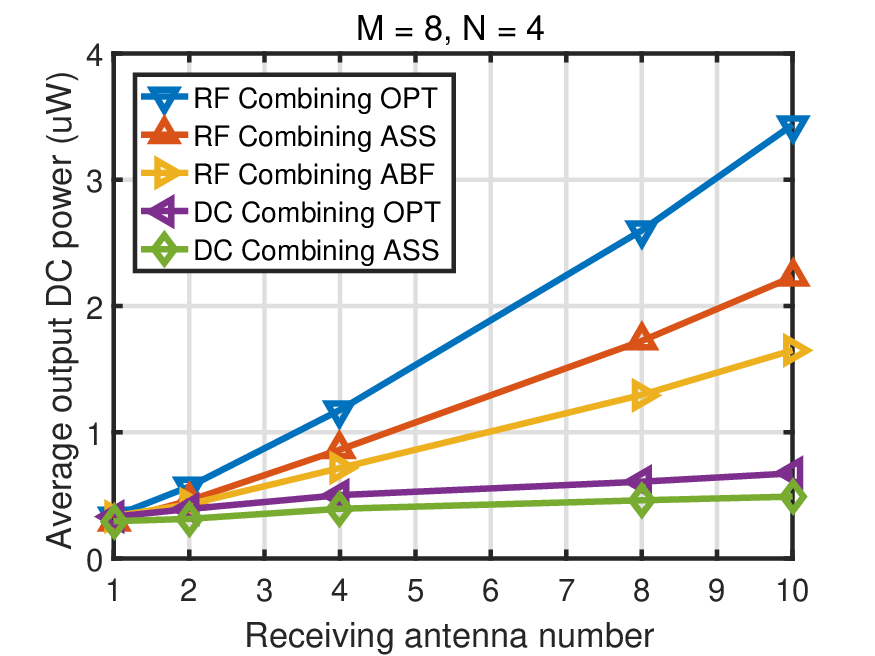}
\par\end{centering}
\begin{centering}
\includegraphics[scale=0.3]{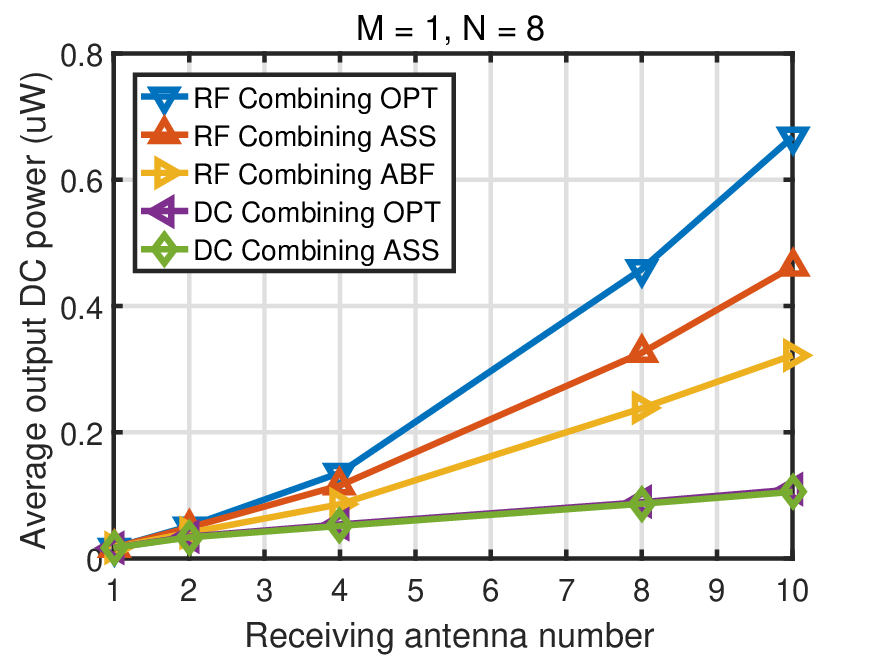}\includegraphics[scale=0.3]{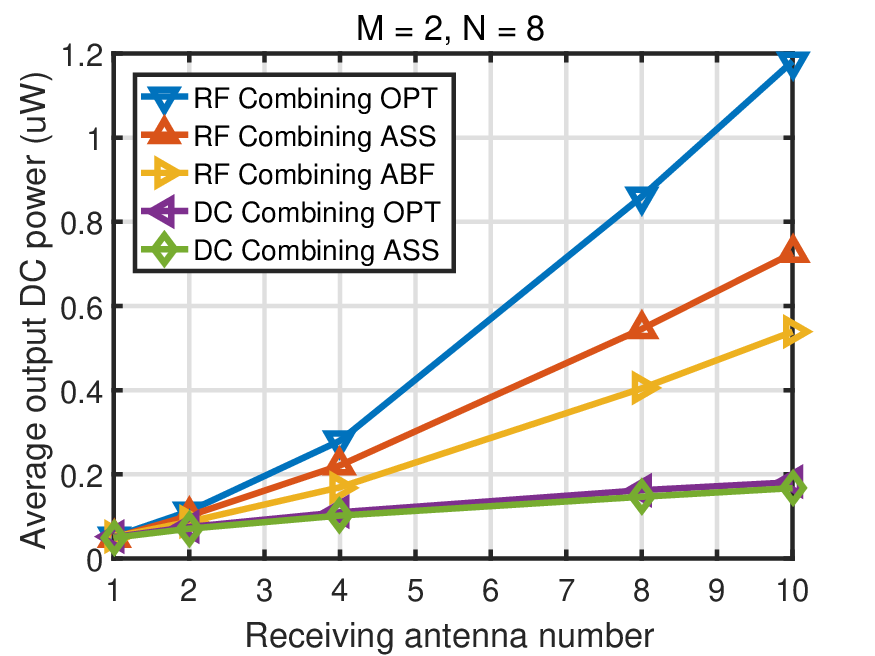}\includegraphics[scale=0.3]{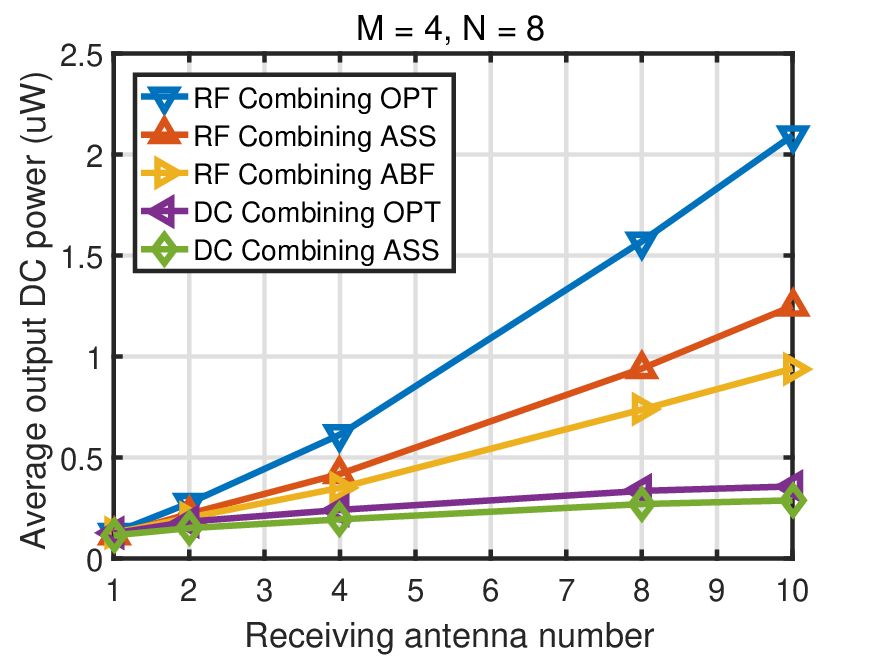}\includegraphics[scale=0.3]{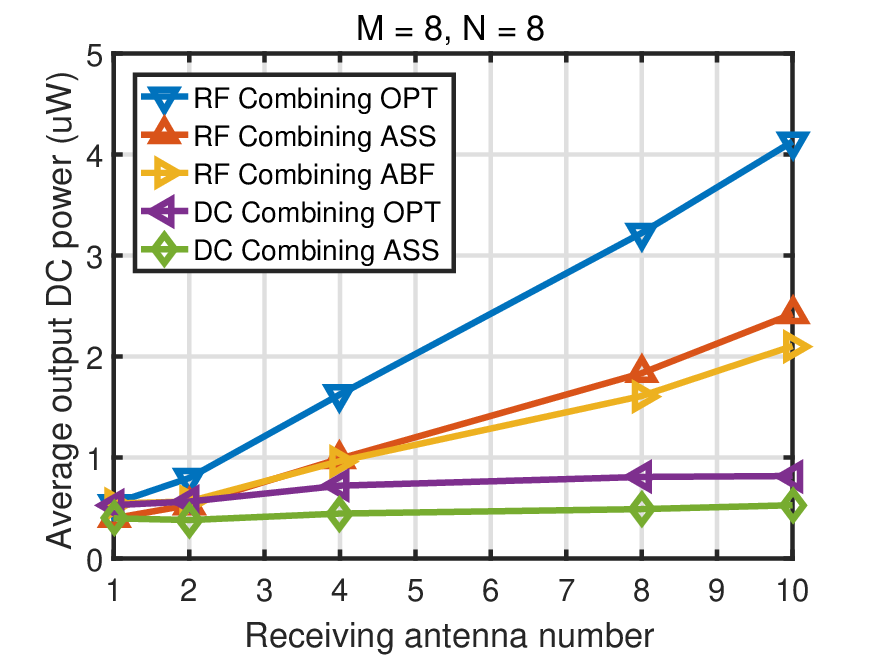}
\par\end{centering}
\begin{centering}
\includegraphics[scale=0.3]{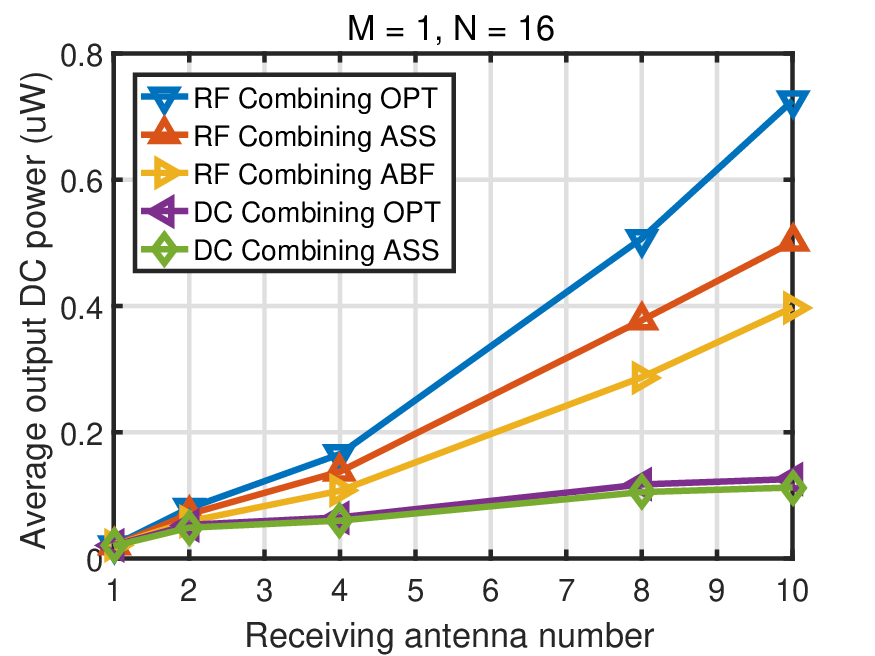}\includegraphics[scale=0.3]{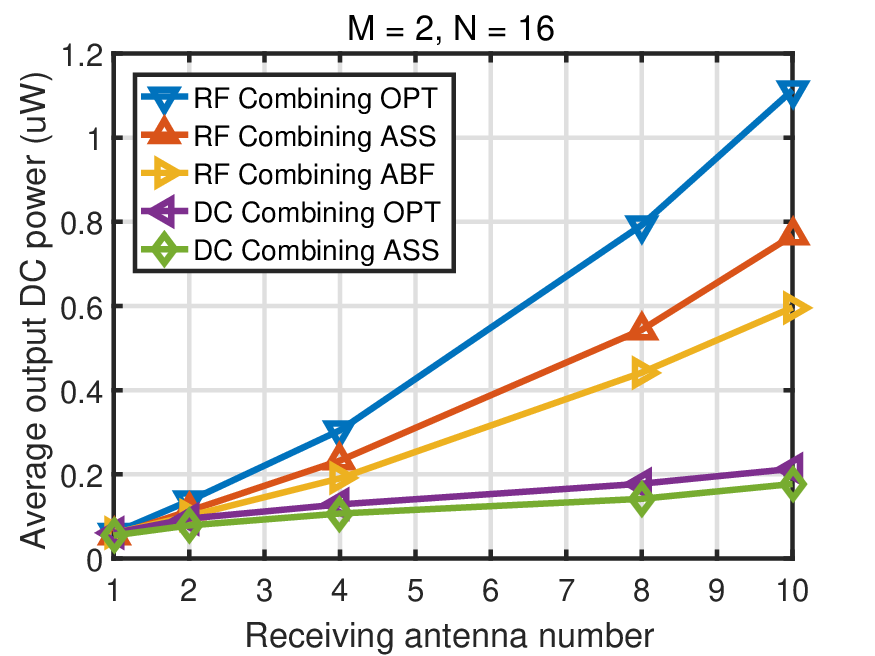}\includegraphics[scale=0.3]{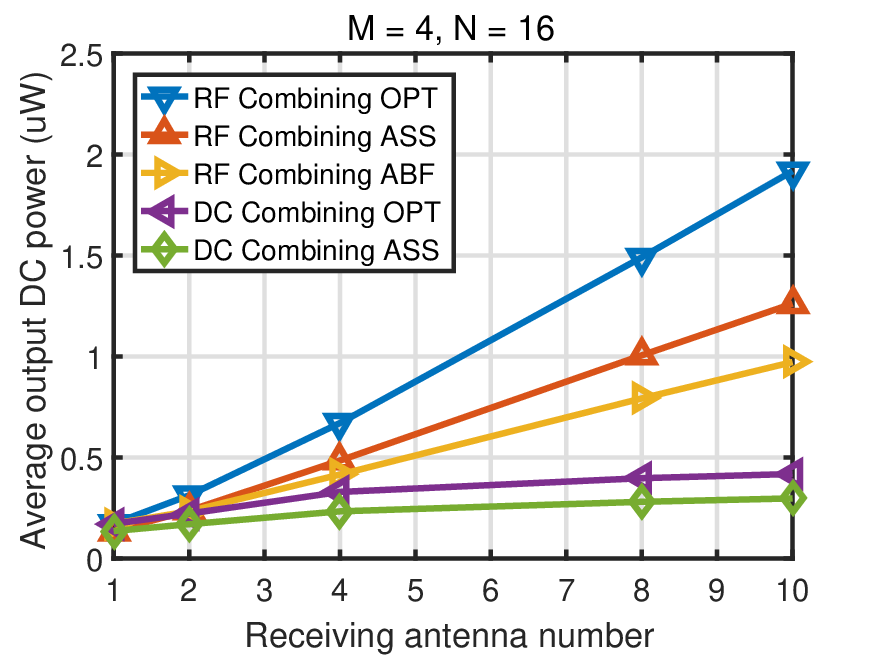}\includegraphics[scale=0.3]{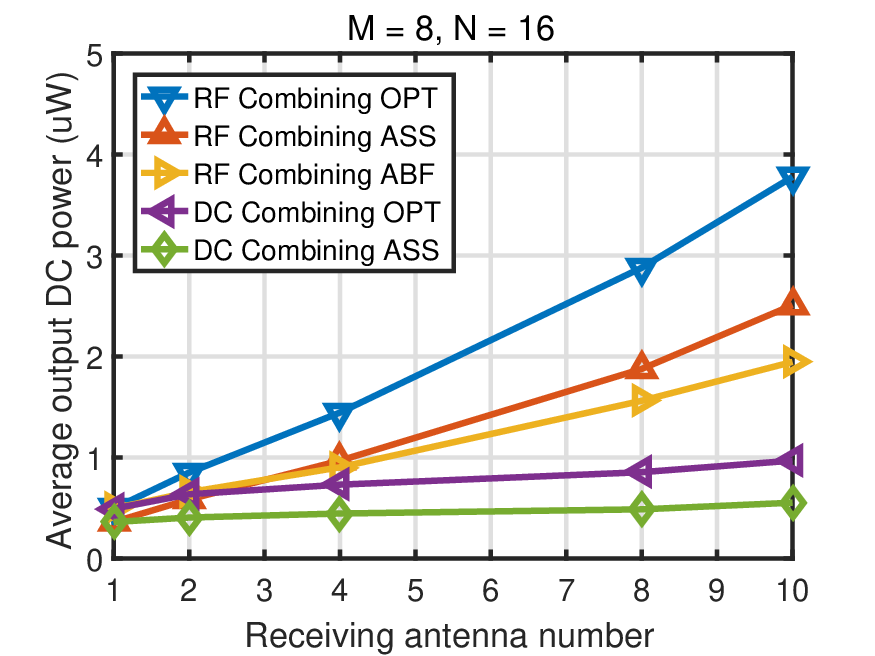}
\par\end{centering}
\caption{\label{fig:Average-output-DC-SPICE}Average output DC power versus
the number of receive antennas $Q$ for different numbers of transmit
antennas $M$ and different numbers of frequencies $N$ based on the
accurate and realistic circuit simulation.}
\end{figure*}

We now evaluate the performance of the multi-sine MIMO WPT system
with DC and RF combinings using the accurate rectenna modeling in
ADS. Again, we consider two DC combinings (based on OPT and ASS) and
three RF combinings (based on OPT, ASS, and ABF). Fig. \ref{fig:Average-output-DC-SPICE}
displays the output DC power averaged over channel realizations versus
the number of receive antennas $Q$ for different numbers of transmit
antennas $M$ and different numbers of frequencies $N$ based on ADS.
We can make the following observations which are similar to the observations
in Fig. \ref{fig:Average-output-DC-Rectenna Model}.

\textit{First}, the output DC power increases with the number of transmit
antennas and the number of receive antennas in both DC and RF combinings.

\textit{Second}, the output DC power increases with the number of
frequencies. Compared with the beamforming only design ($N=1$), the
jointly waveform and beamforming design ($N>1$) is shown to provide
a higher output DC power. Specifically, for OPT DC combining, the
relative gain of the joint waveform and beamforming design over the
beamforming only design can exceed 100\% when $N=16$ and can reach
to 180\% when $M=2$, $N=16$, and $Q=2$, while for OPT RF combining,
the relative gain can exceed 100\% when $N\geq8$ and can reach to
180\% when $M=2$, $N=16$, and $Q=2$.

\textit{Third}, for DC combining, the OPT waveform and beamforming
achieves higher output DC power than the ASS based waveform and beamforming.
The relative gain of OPT DC combining versus ASS DC combining can
be up to 75\% when $M=8$, $N=16$, and $Q=10$.

\textit{Fourth}, for RF combining, the OPT general receive beamforming
achieves higher output DC power than the ASS based general receive
beamforming. The relative gain of OPT RF combining versus ASS RF combining
can be up to 71\% when $M=8$, $N=8$, and $Q=10$. In addition, the
general receive beamforming outperforms the analog receive beamforming.

\textit{Fifth}, RF combinings lead to higher output DC power than
DC combinings, especially when the number of receive antennas goes
large. The relative gain of OPT RF combining versus OPT DC combining
increases with $N$ until $N>8$ due to the breakdown effect of the
diode \cite{2016_TSP_WPT_Bruno_Waveform} and it can exceed 100\%
when $Q\geq4$ and can be up to 550\% when $M=2$, $N=8$, and $Q=10$.

The explanations for these observations can be found in the first
type of evaluations. It is worth noting that the values of average
output DC power shown in Fig. \ref{fig:Average-output-DC-Rectenna Model}
and Fig. \ref{fig:Average-output-DC-SPICE} are different because
the two types of evaluations are calculated by different models. For
the first type of evaluations shown in Fig. \ref{fig:Average-output-DC-Rectenna Model},
the average output DC power is calculated by using the nonlinear rectenna
model \eqref{eq:VoutRectennaModel}. It should be noted that the rectenna
model \eqref{eq:VoutRectennaModel} is derived based on some simplifications
and assumptions (detailed in \cite{2017_TSP_WPT_Bruno_Yang_Large})
so that it can characterize the rectenna nonlinearity and optimize
the waveform and beamforming in a simple and tractable manner. However,
this does not mean that the model in \eqref{eq:VoutRectennaModel}
is accurate enough to predict the rectifier output DC power using
$P_{\mathrm{out}}=v_{\mathrm{out}}^{2}/R_{L}$ where $R_{L}$ refers
to the load resistance. Nevertheless, the model and its benefits in
optimizing waveform and beamforming have been validated by circuit
simulations in \cite{2016_TSP_WPT_Bruno_Waveform}, \cite{2017_AWPL_WPT_Bruno_LowComplexity},
\cite{ShanpuShen2020_TWC_MIMO_WPT_SingleTone}, \cite{2018_TWC_WPT_Bruno_Transmit_Diversity}
and experimentally in \cite{2018_TWC_WPT_Bruno_Transmit_Diversity},
\cite{2019_Junghoon_Prototyping}. On the other hand, for the second
type of evaluations shown in Fig. \ref{fig:Average-output-DC-SPICE},
the average output DC power is calculated by simulating a realistic
rectenna in the circuit simulation solver ADS. Using ADS does not
provide a simple and tractable manner to optimize the waveform and
beamforming but it can accurately simulate the output DC power. Therefore,
we use it to verify the waveform and beamforming optimized by using
the nonlinear rectenna model \eqref{eq:VoutRectennaModel}. The similar
observations in Fig. \ref{fig:Average-output-DC-Rectenna Model} and
Fig. \ref{fig:Average-output-DC-SPICE} confirm the usefulness of
the rectenna nonlinearity model and show the benefit of the joint
waveform and beamforming design which leverages the beamforming gain,
the frequency diversity gain, and the rectenna nonlinearity to increase
the output DC power.

\section{Conclusion and Future Works}

In this paper, we consider the joint design of waveform and beamforming
for MIMO WPT systems, accounting for the rectenna nonlinearity to
increase the output DC power. This is the first paper to jointly optimize
the waveform and beamforming for MIMO WPT systems. DC combining and
RF combining for the multiple rectennas at the receiver are considered.

For DC combining, assuming perfect CSIT and leveraging the nonlinear
rectenna model, the waveform and transmit beamforming are jointly
optimized with guarantee of converging to a stationary point to maximize
the output DC power by using SCA and SDR.

For RF combining, assuming perfect CSIT and CSIR and leveraging the
nonlinear rectenna model, the waveform and transmit and receive beamformings
are jointly optimized to maximize the output DC power. The optimal
transmit and receive beamforming are provided in closed form and the
waveform is optimized with guarantee of converging to a stationary
point by using SCA. A practical RF combining circuit consisting of
phase shifters and an RF power combiner is also considered. The waveform,
transmit beamforming, and analog receive beamforming are jointly optimized
with guarantee of converging to a stationary point by using SCA.

We also provide two types of performance evaluations for the joint
waveform and beamforming design with DC and RF combinings. The first
is based on the nonlinear rectenna model while the second is based
on accurate and realistic circuit simulations in ADS. The two evaluations
agree well with each other, demonstrating the usefulness of the nonlinear
rectenna model, and they show that the output DC power can be increased
by the joint waveform and beamforming design which leverages the beamforming
gain, the frequency diversity gain , and the rectenna nonlinearity.
It is also shown that the joint waveform and beamforming design provides
a higher output DC power than the beamforming only design with a relative
gain exceeding 100\% when $N=16$ and reaching to 180\% when $M=2$,
$N=16$, and $Q=2$. Moreover, RF combining is shown to provide a
higher output DC power than DC combining with a relative gain which
can be up to 550\% when $M=2$, $N=8$, and $Q=10$.

Future research avenues include the following four aspects.

\textit{First}, designing limited feedback MIMO WPT system. In this
paper, we assume the CSI is perfectly known so that the proposed MIMO
WPT system architectures with DC and RF combinings are simplified
and do not contain modules for the CSI acquisition. To acquire the
CSI, we need to extra modules in the architectures such as communication
module and switch module, and we can use forward-link training or
reverse-link training to acquire CSI \cite{2017_TOC_WPT_YZeng_Bruno_RZhang}.
To avoid the requirement of CSI, we can consider designing limited
feedback MIMO WPT systems in future, as shown in related works on
limited feedback MISO WPT systems \cite{2018_TWC_WPT_Bruno_HYang_LimitedFeedback},
\cite{ShanpuShen2020_TIE_WPT_DAS}.

\textit{Second}, unifying the optimizations for DC combining and RF
combining. DC combining and RF combining can be unified as a framework
named as hybrid combining, where the $Q$ receive antennas are divided
into $G$ groups with each group having $Q_{G}=Q/G$ antennas. In
each group, RF combining is used to combine the $Q_{G}$ receive antennas.
Then, DC combining is used to combine the output DC voltage from the
$G$ groups. The hybrid combining becomes DC combining when $G=Q$
and becomes RF combining when $G=1$. Therefore, we can consider developing
efficient algorithms for the hybrid combining to unify the optimizations
for DC combining and RF combining in future.

\textit{Third}, considering MIMO WPT systems for a multi-user scenario.
In the multi-user scenario, we can optimize the transmit waveform
and beamforming and receive beamforming to maximize the weighted sum
output DC power of all users. Therefore, we can consider developing
efficient algorithms for the multi-user scenario in future.

\textit{Fourth}, applying MIMO WPT systems in SWIPT \cite{2013_TWC_SWIPT_RZhang}
and wireless powered communication \cite{2015_ComMag_WirelessPowerComm}.
Particularly, the proposed algorithms for MIMO WPT systems can be
directly applied in SWIPT with time switching scheme \cite{2013_TWC_SWIPT_RZhang}.
However, with power splitting scheme \cite{2013_TWC_SWIPT_RZhang},
we need to jointly optimize the waveform, beamforming, and power splitting
to maximize the energy and rate, which is more difficult and left
as a future work.



\begin{thebibliography}{10}
	\providecommand{\url}[1]{#1}
	\csname url@samestyle\endcsname
	\providecommand{\newblock}{\relax}
	\providecommand{\bibinfo}[2]{#2}
	\providecommand{\BIBentrySTDinterwordspacing}{\spaceskip=0pt\relax}
	\providecommand{\BIBentryALTinterwordstretchfactor}{4}
	\providecommand{\BIBentryALTinterwordspacing}{\spaceskip=\fontdimen2\font plus
		\BIBentryALTinterwordstretchfactor\fontdimen3\font minus
		\fontdimen4\font\relax}
	\providecommand{\BIBforeignlanguage}[2]{{%
			\expandafter\ifx\csname l@#1\endcsname\relax
			\typeout{** WARNING: IEEEtran.bst: No hyphenation pattern has been}%
			\typeout{** loaded for the language `#1'. Using the pattern for}%
			\typeout{** the default language instead.}%
			\else
			\language=\csname l@#1\endcsname
			\fi
			#2}}
	\providecommand{\BIBdecl}{\relax}
	\BIBdecl
	
	\bibitem{2017_TOC_WPT_YZeng_Bruno_RZhang}
	Y.~{Zeng}, B.~{Clerckx}, and R.~{Zhang}, ``Communications and signals design
	for wireless power transmission,'' \emph{IEEE Trans. Commun.}, vol.~65,
	no.~5, pp. 2264--2290, May 2017.
	
	\bibitem{2014_ComMag_SWIPT}
	I.~{Krikidis}, S.~{Timotheou}, S.~{Nikolaou}, G.~{Zheng}, D.~W.~K. {Ng}, and
	R.~{Schober}, ``Simultaneous wireless information and power transfer in
	modern communication systems,'' \emph{IEEE Communications Magazine}, vol.~52,
	no.~11, pp. 104--110, 2014.
	
	\bibitem{2019TAP_WPT_WeiLin}
	W.~{Lin}, R.~W. {Ziolkowski}, and J.~{Huang}, ``Electrically small,
	low-profile, highly efficient, huygens dipole rectennas for wirelessly
	powering internet-of-things devices,'' \emph{IEEE Trans. Antennas Propag.},
	vol.~67, no.~6, pp. 3670--3679, 2019.
	
	\bibitem{2019TMTT_HuchengSun_GridArray}
	Y.~{Hu}, S.~{Sun}, H.~{Xu}, and H.~{Sun}, ``Grid-array rectenna with wide angle
	coverage for effectively harvesting {RF} energy of low power density,''
	\emph{IEEE Trans. Microw. Theory Techn.}, vol.~67, no.~1, pp. 402--413, Jan
	2019.
	
	\bibitem{2009_IntConfRFID}
	M.~S. {Trotter}, J.~D. {Griffin}, and G.~D. {Durgin}, ``Power-optimized
	waveforms for improving the range and reliability of {RFID} systems,'' in
	\emph{IEEE Int. Conf. RFID}, April 2009, pp. 80--87.
	
	\bibitem{2016_TSP_WPT_Bruno_Waveform}
	B.~{Clerckx} and E.~{Bayguzina}, ``Waveform design for wireless power
	transfer,'' \emph{IEEE Trans. Signal Process.}, vol.~64, no.~23, pp.
	6313--6328, Dec 2016.
	
	\bibitem{2017_TSP_WPT_Bruno_Yang_Large}
	Y.~{Huang} and B.~{Clerckx}, ``Large-scale multiantenna multisine wireless
	power transfer,'' \emph{IEEE Trans. Signal Process.}, vol.~65, no.~21, pp.
	5812--5827, Nov 2017.
	
	\bibitem{2017_AWPL_WPT_Bruno_LowComplexity}
	B.~{Clerckx} and E.~{Bayguzina}, ``Low-complexity adaptive multisine waveform
	design for wireless power transfer,'' \emph{IEEE Antennas Wireless Propag.
		Lett.}, vol.~16, pp. 2207--2210, 2017.
	
	\bibitem{2017_IEEE_SPAWC_Waveform}
	M.~R.~V. {Moghadam}, Y.~{Zeng}, and R.~{Zhang}, ``Waveform optimization for
	radio-frequency wireless power transfer,'' in \emph{IEEE Int. Workshop Signal
		Process. Adv. Wireless Commun.}, July 2017, pp. 1--6.
	
	\bibitem{2020_WCNC_WPT}
	B.~A. {Mouris}, H.~{Forssell}, and R.~{Thobaben}, ``A novel low-complexity
	power-allocation algorithm for multi-tone signals for wireless power
	transfer,'' in \emph{2020 IEEE Wireless Communications and Networking
		Conference (WCNC)}, 2020, pp. 1--6.
	
	\bibitem{2018_TWC_WPT_Bruno_HYang_LimitedFeedback}
	Y.~{Huang} and B.~{Clerckx}, ``Waveform design for wireless power transfer with
	limited feedback,'' \emph{IEEE Trans. Wireless Commun.}, vol.~17, no.~1, pp.
	415--429, Jan 2018.
	
	\bibitem{2019_JSAC_Waveform_MultipleDevice}
	K.~{Kim}, H.~{Lee}, and J.~{Lee}, ``Waveform design for fair wireless power
	transfer with multiple energy harvesting devices,'' \emph{IEEE J. Sel. Areas
		Commun.}, vol.~37, no.~1, pp. 34--47, Jan 2019.
	
	\bibitem{2018_TSP_WIPT_Bruno_WIPT}
	B.~{Clerckx}, ``Wireless information and power transfer: Nonlinearity, waveform
	design, and rate-energy tradeoff,'' \emph{IEEE Trans. Signal Process.},
	vol.~66, no.~4, pp. 847--862, Feb 2018.
	
	\bibitem{2019_TOC_SWIPT_nonZeroInput}
	M.~{Varasteh}, B.~{Rassouli}, and B.~{Clerckx}, ``{SWIPT} signaling over
	frequency-selective channels with a nonlinear energy harvester: Non-zero mean
	and asymmetric inputs,'' \emph{IEEE Trans. Commun.}, vol.~67, no.~10, pp.
	7195--7210, 2019.
	
	\bibitem{2019_TWC_WIPT_AsymmetricModulation}
	E.~{Bayguzina} and B.~{Clerckx}, ``Asymmetric modulation design for wireless
	information and power transfer with nonlinear energy harvesting,'' \emph{IEEE
		Trans. Wireless Commun.}, vol.~18, no.~12, pp. 5529--5541, 2019.
	
	\bibitem{2017_CL_WPBackscatterComm}
	B.~{Clerckx}, Z.~{Bayani Zawawi}, and K.~{Huang}, ``Wirelessly powered
	backscatter communications: Waveform design and {SNR}-energy tradeoff,''
	\emph{IEEE Communications Letters}, vol.~21, no.~10, pp. 2234--2237, Oct
	2017.
	
	\bibitem{2019_TWC_MuWPScatteredComm}
	Z.~B. {Zawawi}, Y.~{Huang}, and B.~{Clerckx}, ``Multiuser wirelessly powered
	backscatter communications: Nonlinearity, waveform design, and {SINR}-energy
	tradeoff,'' \emph{IEEE Trans. Wireless Commun.}, vol.~18, no.~1, pp.
	241--253, Jan 2019.
	
	\bibitem{ShanpuShen2016_TAP_Impedancematching}
	S.~Shen and R.~D. Murch, ``Impedance matching for compact multiple antenna
	systems in random {RF} fields,'' \emph{IEEE Trans. Antennas Propag.},
	vol.~64, no.~2, pp. 820--825, Feb. 2016.
	
	\bibitem{ShanpuShen2017_AWPL_DPTB}
	S.~Shen, C.~Y. Chiu, and R.~D. Murch, ``A dual-port triple-band {L}-probe
	microstrip patch rectenna for ambient {RF} energy harvesting,'' \emph{IEEE
		Antennas Wireless Propag. Lett.}, vol.~16, pp. 3071--3074, 2017.
	
	\bibitem{ShanpuShen2018_EuCap_QPDP}
	S.~Shen, Y.~Zhang, C.-Y. Chiu, and R.~D. Murch, ``A compact quad-port
	dual-polarized dipole rectenna for ambient {RF} energy harvesting,'' in
	\emph{2018 12th European Conference on Antennas and Propagation}, London,
	United Kingdom, Apr. 2018.
	
	\bibitem{ShanpuShen2017_TAP_EHPIXEL}
	S.~Shen, C.~Y. Chiu, and R.~D. Murch, ``Multiport pixel rectenna for ambient
	{RF} energy harvesting,'' \emph{IEEE Trans. Antennas Propag.}, vol.~66,
	no.~2, pp. 644--656, Feb. 2018.
	
	\bibitem{ShanpuShen2019_TMTT_Freqdepend}
	S.~Shen, Y.~Zhang, C.-Y. Chiu, and R.~D. Murch, ``An ambient {RF} energy
	harvesting system where the number of antenna ports is dependent on
	frequency,'' \emph{IEEE Trans. Microw. Theory Tech.}, vol.~67, no.~9, pp.
	3821--3832, Sep. 2019.
	
	\bibitem{ShanpuShen2019_TIE_Hybrid_Combining}
	S.~{Shen}, Y.~{Zhang}, C.~Y. {Chiu}, and R.~{Murch}, ``A triple-band high-gain
	multibeam ambient {RF} energy harvesting system utilizing hybrid combining,''
	\emph{IEEE Trans. Ind. Electron.}, vol.~67, no.~11, pp. 9215--9226, 2020.
	
	\bibitem{2011AWPL_EH_InvestRectArray}
	U.~Olgun, C.-C. Chen, and J.~L. Volakis, ``Investigation of rectenna array
	configurations for enhanced {RF} power harvesting,'' \emph{IEEE Antennas
		Wireless Propag. Lett.}, vol.~10, pp. 262--265, 2011.
	
	\bibitem{2016_TWC_RZhang_MIMOWPT_Limitedfeed}
	J.~{Xu} and R.~{Zhang}, ``A general design framework for {MIMO} wireless energy
	transfer with limited feedback,'' \emph{IEEE Trans. Signal Process.},
	vol.~64, no.~10, pp. 2475--2488, May 2016.
	
	\bibitem{2019_SPL_MIMO_WPT}
	G.~{Ma}, J.~{Xu}, Y.~{Zeng}, and M.~R.~V. {Moghadam}, ``A generic receiver
	architecture for {MIMO} wireless power transfer with nonlinear energy
	harvesting,'' \emph{IEEE Signal Processing Letters}, vol.~26, no.~2, pp.
	312--316, Feb 2019.
	
	\bibitem{ShanpuShen2020_TWC_MIMO_WPT_SingleTone}
	S.~{Shen} and B.~{Clerckx}, ``Beamforming optimization for {MIMO} wireless
	power transfer with nonlinear energy harvesting: {RF} combining versus {DC}
	combining,'' \emph{IEEE Trans. Wireless Commun.}, vol.~20, no.~1, pp.
	199--213, 2021.
	
	\bibitem{2015_CL_Nonlinaer_SWIPT}
	E.~{Boshkovska}, D.~W.~K. {Ng}, N.~{Zlatanov}, and R.~{Schober}, ``Practical
	non-linear energy harvesting model and resource allocation for swipt
	systems,'' \emph{IEEE Communications Letters}, vol.~19, no.~12, pp.
	2082--2085, 2015.
	
	\bibitem{2019_JSAC_WIPT_Bruno_RZhang_RSchober_DIKim_HVPoor}
	B.~{Clerckx}, R.~{Zhang}, R.~{Schober}, D.~W.~K. {Ng}, D.~I. {Kim}, and H.~V.
	{Poor}, ``Fundamentals of wireless information and power transfer: From {RF}
	energy harvester models to signal and system designs,'' \emph{IEEE J. Sel.
		Areas Commun.}, vol.~37, no.~1, pp. 4--33, Jan 2019.
	
	\bibitem{KKKKI}
	Y.~H. Lam, W.~H. Ki, and C.~Y. Tsui, ``Single inductor multiple-input
	multiple-output switching converter and method of use,'' Aug.~14 2007, {US}
	Patent 7, 256, 568.
	
	\bibitem{boyd2004convex}
	S.~Boyd and L.~Vandenberghe, \emph{Convex optimization}.\hskip 1em plus 0.5em
	minus 0.4em\relax Cambridge university press, 2004.
	
	\bibitem{adali2010adaptiveSignalProcessing}
	T.~Adali and S.~Haykin, \emph{Adaptive signal processing: next generation
		solutions}.\hskip 1em plus 0.5em minus 0.4em\relax John Wiley \& Sons, 2010,
	vol.~55.
	
	\bibitem{DanielSDP_rank}
	Y.~Huang and D.~P. Palomar, ``Rank-constrained separable semidefinite
	programming with applications to optimal beamforming,'' \emph{IEEE
		Transactions on Signal Processing}, vol.~58, no.~2, pp. 664--678, 2009.
	
	\bibitem{QR_algorithm}
	B.~N. {Parlett}, ``The {QR} algorithm,'' \emph{Computing in Science
		Engineering}, vol.~2, no.~1, pp. 38--42, 2000.
	
	\bibitem{2007_TWC_GP_Daniel}
	M.~{Chiang}, C.~W. {Tan}, D.~P. {Palomar}, D.~{O'neill}, and D.~{Julian},
	``Power control by geometric programming,'' \emph{IEEE Trans. Wireless
		Commun.}, vol.~6, no.~7, pp. 2640--2651, July 2007.
	
	\bibitem{grant2008cvx}
	M.~Grant, S.~Boyd, and Y.~Ye, ``{CVX}: {MATLAB} software for disciplined convex
	programming,'' 2008.
	
	\bibitem{2016_JSTSP_HybridPrecoding_Xianghao}
	X.~{Yu}, J.~{Shen}, J.~{Zhang}, and K.~B. {Letaief}, ``Alternating minimization
	algorithms for hybrid precoding in millimeter wave {MIMO} systems,''
	\emph{IEEE Journal of Selected Topics in Signal Processing}, vol.~10, no.~3,
	pp. 485--500, 2016.
	
	\bibitem{2017_JSAC_HybridPrecoding_OFDM}
	F.~{Sohrabi} and W.~{Yu}, ``Hybrid analog and digital beamforming for mmwave
	{OFDM} large-scale antenna arrays,'' \emph{IEEE Journal on Selected Areas in
		Communications}, vol.~35, no.~7, pp. 1432--1443, 2017.
	
	\bibitem{Hiperlan2Channel}
	J.~Medbo and P.~Schramm, \emph{Channel Models for {HIPERLAN/2} in Different
		Indoor Scenarios}.\hskip 1em plus 0.5em minus 0.4em\relax ETSI EP BRAN
	3ERI085B, Mar 1998.
	
	\bibitem{2013_TWC_SWIPT_RZhang}
	R.~{Zhang} and C.~K. {Ho}, ``{MIMO} broadcasting for simultaneous wireless
	information and power transfer,'' \emph{IEEE Trans. Wireless Commun.},
	vol.~12, no.~5, pp. 1989--2001, May 2013.
	
	\bibitem{2018_MM_WPT_Bruno_1GWPT}
	B.~{Clerckx}, A.~{Costanzo}, A.~{Georgiadis}, and N.~{Borges Carvalho},
	``Toward {1G} mobile power networks: {RF}, signal, and system designs to make
	smart objects autonomous,'' \emph{IEEE Microw. Mag.}, vol.~19, no.~6, pp.
	69--82, Sep. 2018.
	
	\bibitem{2018_TWC_WPT_Bruno_Transmit_Diversity}
	B.~{Clerckx} and J.~{Kim}, ``On the beneficial roles of fading and transmit
	diversity in wireless power transfer with nonlinear energy harvesting,''
	\emph{IEEE Trans. Wireless Commun.}, vol.~17, no.~11, pp. 7731--7743, Nov
	2018.
	
	\bibitem{2019_Junghoon_Prototyping}
	J.~{Kim}, B.~{Clerckx}, and P.~D. {Mitcheson}, ``Signal and system design for
	wireless power transfer : Prototype, experiment and validation,'' \emph{IEEE
		Trans. Wireless Commun.}, pp. 1--1, 2020.
	
	\bibitem{ShanpuShen2020_TIE_WPT_DAS}
	S.~{Shen}, J.~{Kim}, C.~{Song}, and B.~{Clerckx}, ``Wireless power transfer
	with distributed antennas: System design, prototype, and experiments,''
	\emph{IEEE Trans. Ind. Electron.}, pp. 1--1, 2020.
	
	\bibitem{2015_ComMag_WirelessPowerComm}
	S.~{Bi}, C.~K. {Ho}, and R.~{Zhang}, ``Wireless powered communication:
	opportunities and challenges,'' \emph{IEEE Communications Magazine}, vol.~53,
	no.~4, pp. 117--125, April 2015.
	
\end{thebibliography}
\end{document}